\documentclass{aa}

\usepackage{graphicx}
\usepackage{amssymb}
\usepackage{amsmath}
\usepackage{natbib}
\bibpunct{(}{)}{;}{a}{}{,}

\def\void#1{{}}
\begin{document}
   \title{Multi-object spectroscopy of low redshift EIS clusters. III.
	Properties of optically selected clusters \thanks{Based on
	observations made with the Danish1.5-m telescope at ESO, La
	Silla, Chile. Table 3 is only available in electronic
	form at the CDS via anonymous ftp to cdsarc.u-strasbg.fr
	(130.79.128.5) or via
	http://cdsweb.u-strasbg.fr/cgi-bin/qcat?J/A+A/ . Table 5 is
	also available in electronic form at CDS.}}
	\titlerunning{Multi-object spectroscopy of low redshift EIS
	clusters. III.}

   \author{L.F. Olsen\inst{1}
           \and
	   C. Benoist\inst{2}
	  \and
	   L. da Costa\inst{3} 	  
	   \and
           L. Hansen\inst{1}
	   \and
           H.E. J{\o}rgensen\inst{1}
	   }

   \offprints{L.F. Olsen, lisbeth@astro.ku.dk, present address: Observatoire de la C\^{o}te d'Azur, Laboratoire Cassiop\'ee, BP 4229, F-06304 Nice Cedex 4,  France}

   \institute{Copenhagen University Observatory, Juliane Maries Vej 30, DK-2100 Copenhagen, Denmark
	    \and Observatoire de la C\^{o}te d'Azur, Laboratoire Cassiop\'ee, BP 4229, F-06304 Nice Cedex 4,  France
	    \and European Southern Observatory, Karl-Schwartzschild-Str. 2, D-85748 Garching b. M\"{u}nchen, Germany
            }

   \date{Received.....; accepted .....}

   \abstract{ We have carried out an investigation of the properties
   of low redshift EIS clusters using both spectroscopy and imaging
   data.  We present new redshifts for 738 galaxies in 21 ESO Imaging
   Survey (EIS) Cluster fields. We use the ``gap''-technique to search
   for significant overdensities in redshift space and to identify
   groups/clusters of galaxies corresponding to the original EIS
   matched filter cluster candidates. In this way we spectroscopically
   confirm 20 of the 21 cluster candidates with a matched-filter
   estimated redshift $z_{MF}=0.2$. We have now obtained spectroscopic
   redshifts for 34 EIS cluster candidates with $z_{MF}=0.2$
   \citep[see also ][]{hansen02,olsen03}. Of those we
   spectroscopically confirm 32 with redshifts ranging from $z=0.064$
   to $0.283$. We find that: 1) the velocity dispersions of the
   systems range from $\sigma_v\leq130\mathrm{km/s}$ to
   $\sigma_v=1200\mathrm{km/s}$, typical of galaxy groups to rich
   clusters; 2) richnesses corresponding to Abell classes $R \leq1$;
   and 3) concentration indices ranging from $C=0.2$ to $C=1.2$.  From
   the analysis of the colours of the galaxy populations we find that
   53\% of the spectroscopically confirmed systems have a
   ``significant'' red sequence.  These systems are on average richer
   and have higher velocity dispersions. We find that the colour of
   the red sequence galaxies matches passive stellar evolution
   predictions.   }

   \maketitle
%

\section{Introduction}

The evolution of galaxy clusters' properties, as well as that of their
constituent galaxies, are important issues for contemporary cosmology
and astrophysics. The requirement for large samples of clusters of
galaxies covering a large range in redshift has prompted systematic
efforts to assemble catalogues of distant galaxy clusters
\citep[e.g. ][]{gunn86,postman96,scodeggio99,gladders01,gonzalez01,
bahcall03}.  The main goal behind such works is to assemble large
samples of clusters with $z\gtrsim0.5$  because at these redshifts
the evolutionary effects become more significant. However, another
important issue in evolutionary studies is to have a well-defined
comparison sample at lower redshifts. This sample can be taken from
other surveys, but it would be preferable to build it from the same
survey, in order to minimize the differences in selection effects.

During the past decade a number of galaxy cluster catalogues based on
optical imaging data and constructed using objective methods have
become available \cite[e.g. ][]{postman96,gladders01,
postman02,bahcall03, goto02}. Each method uses its own combination of
single passband luminosities, colour indices and galaxy position, and
it is thus of great interest to compare whether the various methods
detect the same systems. The comparison can be carried out along two
tracks. One is to directly compare detections by different methods
over the same area, and the other is to compare the general properties
of the samples created by different detection algorithms
\citep[e.g. ][]{goto02,kim02,bahcall03,lopes04,rizzo04}.  Whatever the
method utilized, spectroscopic follow-up is essential to confirm
that the candidates are physical systems as well as to characterize
their properties.

This work is part of a major on-going confirmation effort to study all
EIS cluster candidates \citep{olsen99a, olsen99b, scodeggio99}.  This
sample consists of 302 cluster candidates with matched filter
estimated redshifts $0.2\leq z_\mathrm{MF}\leq1.3$ and a median
estimated redshift of $z_\mathrm{MF}=0.5$. The cluster candidates were
identified using the matched filter technique originally suggested by
\cite{postman96}. The spectroscopic confirmation of the clusters was
initiated by \cite{ramella00}, who used the multi-object spectroscopy
mode at the ESO 3.6m telescope at La Silla, Chile, to obtain
confirmations of intermediate redshift candidates ($0.5\lesssim
z_\mathrm{MF}\lesssim0.7$).  They targeted six cluster candidates
of which four were confirmed. \cite{benoist02} presented the first
results for the high redshift sample ($z\gtrsim0.8$) with confirmation
of three EIS clusters.  

In this work we report on a systematic spectroscopic follow-up of the
low-redshift EIS cluster candidates having $z_{MF}=0.2$. The sample
was drawn from candidates located in EIS patches~A, B and D
\citep{nonino99} and consists of 68\% (34 systems) of all EIS cluster
candidates at this redshift.  The present work follows that of
Hansen et al. (2002, hereafter Paper I) and Olsen et al. (2003,
hereafter Paper II). In Paper~I we presented the results of a
feasability study confirming five clusters in patch D of which three
have $z_\mathrm{MF}=0.2$ and two have $z_\mathrm{MF}=0.3$.  In
Paper~II we presented the follow-up of candidates in patches~A and B
where 9 out of 10 additional cluster candidates were confirmed.  In
this third paper we present the spectroscopic results for the 21
remaining systems in EIS patch~D.

The paper is structured as follows: Sect.~\ref{sec:compl} gives an
overview of the observations and data reduction.
Sect.~\ref{sec:z_identification} describes the identification of
systems in redshift space as well as the procedure adopted for
associating the redshift groups to the EIS
detections. Sect.~\ref{sec:properties} describes the properties of the
spectroscopically confirmed systems including an analysis of the
colour properties of the galaxy populations.  In
Sect.~\ref{sec:discussion} we discuss our results and relate the
dynamical properties to the colour properties. Finally,
Sect.~\ref{sec:conclusions} summarizes the paper.


\section{Observations and data reduction}
\label{sec:compl}

\begin{table*}
\caption{EIS cluster candidates in patch~D with $z_{MF}=0.2$.}
\label{tab:cl_targets}
\begin{minipage}{\textwidth}
\begin{center}
\begin{tabular}{lllr}
\hline\hline
Field\footnote{Here, and in the rest of this paper, we have added a ``J'' in the name to conform with international standards. The EIS identification is the same except for this ``J''.} & $\alpha_{J2000}$ & $\delta_{J2000}$ & $\Lambda_{cl, org}$ \\
\hline
EISJ0946-2029\footnote{ Reported in Paper I} & 09 46 12.8 & -20 29 49.6 & 59.5 \\
EISJ0946-2133 & 09 46 31.1 & -21 33 24.1 & 30.9 \\
EISJ0947-2120$^b$ & 09 47 06.9 & -21 20 55.6 & 43.4 \\
EISJ0948-2044$^b$ & 09 48 07.9 & -20 44 31.2 & 42.8 \\
EISJ0949-2145 & 09 49 49.4 & -21 45 25.7 & 42.0 \\
EISJ0949-2046 & 09 49 51.5 & -20 46 40.6 & 32.8 \\
EISJ0950-2133 & 09 50 46.1 & -21 33 37.4 & 30.7 \\
EISJ0951-2052 & 09 51 08.3 & -20 52 23.6 & 32.0 \\
EISJ0951-2026 & 09 51 28.9 & -20 26 33.0 & 43.6 \\
EISJ0951-2145 & 09 51 47.3 & -21 45 27.1 & 57.9 \\
EISJ0952-2150 & 09 52 46.8 & -21 50 15.1 & 33.7 \\
EISJ0952-2103 & 09 52 47.6 & -21 03 02.7 & 34.3 \\
EISJ0952-2144 & 09 52 48.6 & -21 44 32.8 & 36.1 \\
EISJ0952-2018 & 09 52 55.3 & -20 18 37.6 & 35.4 \\
EISJ0953-2053 & 09 53 05.9 & -20 53 29.9 & 50.1 \\
EISJ0953-2156 & 09 53 33.8 & -21 56 10.1 & 35.4 \\
EISJ0953-2017 & 09 53 55.5 & -20 17 32.8 & 34.9 \\
EISJ0955-2123 & 09 55 01.3 & -21 23 19.6 & 34.0 \\
EISJ0955-2151 & 09 55 04.1 & -21 51 35.0 & 38.7 \\
EISJ0955-2037 & 09 55 16.9 & -20 37 04.1 & 36.7 \\
EISJ0955-2020 & 09 55 19.8 & -20 20 25.4 & 39.0 \\
EISJ0956-2054 & 09 56 02.7 & -20 54 08.6 & 37.3 \\
EISJ0957-2051 & 09 57 07.2 & -20 51 45.3 & 27.6 \\
EISJ0957-2143 & 09 57 12.4 & -21 43 13.1 & 40.8 \\
\hline
\end{tabular}
\end{center}
\end{minipage}
\end{table*}

The targeted cluster candidates were selected from patch~D with
a matched filter estimated redshift $z_{MF}=0.2$ \citep{scodeggio99}. In
Table~\ref{tab:cl_targets} we list the 24 selected cluster candidates.
Three of these systems were already studied in Paper I as noted in the
table, leaving 21 systems for the present work. The table gives: in
Col.~1 the name of the field referring to the notation adopted by
\cite{scodeggio99}; in Cols.~2 and 3 the matched filter position; and
in Col.~4 the $\Lambda_{cl, org}$-richness. This richness range 
roughly corresponds to the Abell richness classes $\leq 1$
\citep[e.g.][]{postman96}.

The observations were carried out using the Danish Faint Object
Spectrograph and Camera (DFOSC) mounted on the Danish 1.54m telescope
at ESO, La Silla, Chile. With a field of view of $13.7 \times
13.7$~square arcmins corresponding to $2.53\mathrm{Mpc}$ at $z=0.2$ (assuming
$\mathrm{H}_0=75\mathrm{km/s/Mpc}$, $\Omega_\mathrm{m} = 0.3$ and
$\Omega_\Lambda=0.7$), this instrument is well-suited for MOS
observations of moderate redshift clusters.  The effective field that
could be covered with MOS slit masks was typically
$11.0\times5.5$~square arcmins, depending on the exact configuration of
galaxy positions in each field. The slit width was set to $2\arcsec$,
and the slit length varied according to the extent of each galaxy.  We
used grism \#4, giving a dispersion of $220\;\mathrm{{\AA}/mm}$, and
covering a wavelength range from $3800$ to
$7500\;\mathrm{{\AA}}$. However, the useful range for each spectrum
depends on the exact position of the slit with respect to the chip and
the intrinsic galaxy spectrum. The resolution as determined from HeNe
line spectra was found to be $16.6\;\mathrm{{\AA}\;FWHM}$.

\begin{figure}
\resizebox{0.5\textwidth}{!}{\includegraphics{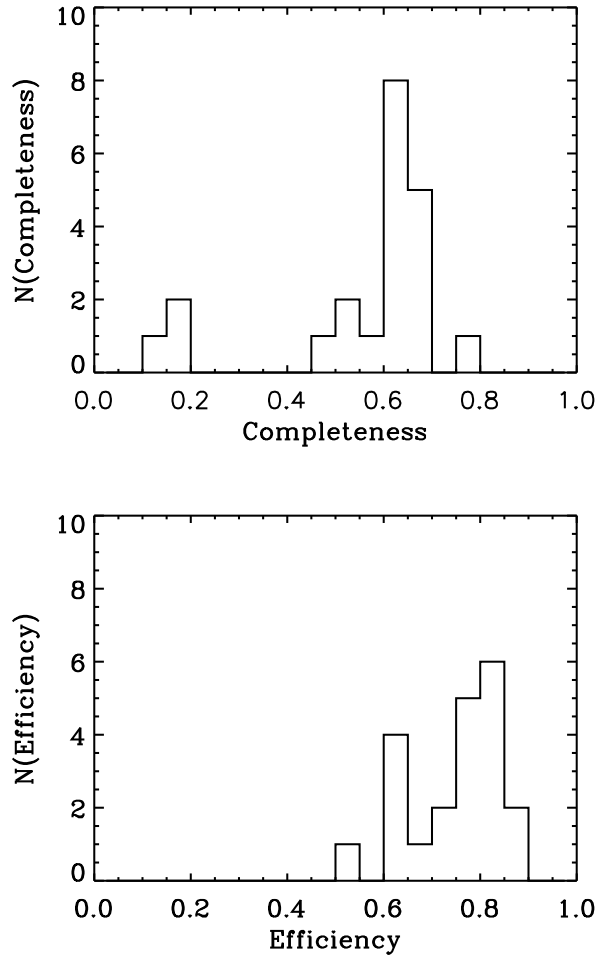}}
\caption{  The distribution of completeness, the fraction of targeted
galaxies to all galaxies (upper panel), and efficiency, the fraction
of spectra that yielded a redshift determination (lower panel) per
field.}
\label{fig:avrg_compl}
\end{figure}

We targeted preferentially the bright galaxies with I-magnitude,
$I\leq19.5$\footnote{All magnitudes are quoted in the EIS magnitude
system as provided by the EIS team, see
\protect\cite{nonino99,prandoni99,benoist99}}.  The Schechter
magnitude at $z=0.2$ is estimated to be $I^*\sim17.5$ using an
absolute Schechter magnitude of $M^*_I=-21.90$ as commonly adopted
\citep[e.g.][]{postman96,olsen99a}. The corresponding apparent
magnitude was computed using the K-correction for an elliptical galaxy
template spectrum from the Kinney library \citep{kinney96}. We thus
estimate our survey to cover galaxies to 2~magnitudes fainter than the
Schechter magnitude.  This procedure was chosen to avoid possible
biases introduced by an additional colour selection of the target
galaxies.  The allocated observing time allowed us to expose two slit
masks for each cluster field.  The exposure time for each mask was in
all cases one hour. We estimate the S/N of the spectra to be in the
range 5 to 15.

The data reduction was performed using the IRAF\footnote{ IRAF is
distributed by the National Optical Astronomy Observatories, which is
operated by AURA Inc. under contract with NSF.} package. The CCD bias
level was determined from overscan regions and subtracted. The
flatfielding was carried out using the two sets of flatfields obtained
immediately before and after each observation.  After the basic
reductions we used standard procedures to extract the spectra and to
obtain redshifts by Fourier cross-correlating our spectra with
standard galaxy spectra templates from \cite{kinney96}.  For the
cross-correlation the template spectra were always redshifted close to
the redshift under consideration.   All the cross-correlation
function results were visually inspected and the reliability of
the peak was evaluated.  Whenever a peak in the correlation function
was accepted as real or possibly real, the observed spectrum was
inspected and compared to the expected positions of the most prominent
spectral features.  We required that some features like the Ca H and K
lines, the 4000\,{\AA} break, or emission lines should be identified
before a determination was accepted as certain. In Paper I the
reduction procedures are described in more detail.

\begin{figure}
\resizebox{0.5\textwidth}{!}{\includegraphics{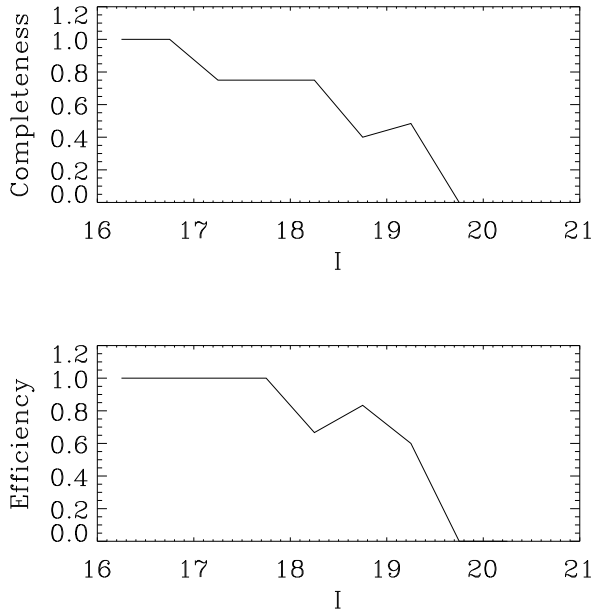}}
\caption{ The completeness (upper panel) and efficiency (lower panel)
as function of magnitude as found for the cluster EISJ0953-2017. The
completeness and efficiency for this cluster as computed in
Table~\protect\ref{tab:spec_compl} correspond to the typical values
as can be seen in Fig.~\protect\ref{fig:avrg_compl}.}
\label{fig:completeness}
\end{figure}

With two slit masks per field regardless of the galaxy density we do
not reach the same level of completeness in all fields, due to the
variations in the local galaxy density.  Therefore, we have
investigated how the completeness varies from field to field. In
Table~\ref{tab:spec_compl} we summarize the spectroscopic results. The
Table lists: in Col.~1 the field name; in Col.~2 the number of target
galaxies; in Col.~3 the number of derived redshifts; in Col.~4 the
completeness as defined below; and in Col.~5 the efficiency of
obtaining redshifts (the ratio between Col.~3 and Col.~2).  The
completeness given in Col.~4 is defined as the ratio of observed to
all galaxies brighter than $I=19.5$ within a rectangular region. The
latter is defined as the smallest rectangle covering all observed
galaxies and is outlined by dashed lines in
Fig.~\ref{fig:spatial_dist}.   

In Fig.~\ref{fig:avrg_compl} we show the distributions of the
completeness and efficiency. One finds that in general the
completeness is $\sim60\%$ except for three fields. This could have
two reasons: (1) the galaxies are distributed such that fewer slits
could fit in or (2) the field is much richer than the average
field. Inspecting Fig.~\ref{fig:spatial_dist} it seems that the low
completeness is probably caused by a combination of the two.   The
efficiency is found to cover the range between 0.53 and 0.88 with most
fields having an efficiency of $\sim80\%$.

\begin{table*}
\caption{Summary of spectroscopic coverage for each target cluster.}
\label{tab:spec_compl}
\begin{center}
\begin{tabular}{lrrrr}
\hline\hline
Field & \#targets & \#redshifts & Compl. & Efficiency\\
\hline
EISJ0946-2133 & 48 & 33 & 0.50 & 0.68 \\
EISJ0949-2145 & 45 & 27 & 0.65 & 0.60 \\
EISJ0949-2046 & 49 & 37 & 0.19 & 0.75 \\
EISJ0950-2133 & 50 & 31 & 0.61 & 0.61 \\
EISJ0951-2052 & 41 & 32 & 0.76 & 0.78 \\
EISJ0951-2026 & 48 & 40 & 0.66 & 0.83 \\
EISJ0951-2145 & 47 & 28 & 0.17 & 0.59 \\
EISJ0952-2150 & 54 & 40 & 0.62 & 0.74 \\
EISJ0952-2103 & 50 & 42 & 0.14 & 0.84 \\
EISJ0952-2144 & 50 & 44 & 0.61 & 0.88 \\
EISJ0952-2018 & 48 & 39 & 0.52 & 0.81 \\
EISJ0953-2053 & 52 & 34 & 0.55 & 0.67 \\
EISJ0953-2156 & 52 & 41 & 0.62 & 0.80 \\
EISJ0953-2017 & 47 & 38 & 0.60 & 0.81 \\
EISJ0955-2123 & 53 & 41 & 0.67 & 0.76 \\
EISJ0955-2151 & 48 & 39 & 0.53 & 0.80 \\
EISJ0955-2037 & 49 & 26 & 0.69 & 0.53 \\
EISJ0955-2020 & 34 & 29 & 0.66 & 0.85 \\
EISJ0956-2054 & 42 & 32 & 0.62 & 0.76 \\
EISJ0957-2051 & 39 & 29 & 0.64 & 0.74 \\
EISJ0957-2143 & 44 & 36 & 0.68 & 0.81 \\
\hline
\end{tabular}
\end{center}
\end{table*}

\pagebreak

 Fig.~\ref{fig:completeness} shows the completeness and efficiency
as function of magnitude for the field of EISJ0953-2017, for which
completeness and efficiency correpond to the typical values as found
from Fig.~\ref{fig:avrg_compl}.  It can be seen that the completeness
is very high at the brightest magnitudes but decreases to $\sim25\%$
at about $I=19.5$. Regarding extraction of the redshifts it can be
seen that the efficiency is quite high, reaching $\sim50\%$ at
$I\sim19.5$.

In order to estimate the uncertainty of the measured redshifts we have
observed several galaxies in two different masks. In total we have
observed 106 galaxies twice. We use the corresponding redshift pairs
to estimate the uncertainty of the individual redshift
measurements. We separate the pairs in three groups: those for which
we did not succeed in measuring the redshift at all, those for which the
redshift could be determined in only one case and those with two
redshift measurements. The first two groups cannot be used for
estimating the uncertainty, but it is interesting to see that the
galaxies in the first group are all fainter than $I\sim18.8$ and in
the second group they are fainter than $I\sim17.7$. The last group
consists of 49 pairs of redshifts for which the magnitudes lie in the
interval $I\sim16.1-20.0$, thus covering the entire magnitude range
investigated here. For these 49 pairs we find that the standard
deviation of the redshift difference between the two independent
measurements is $\Delta z=0.0006$ corresponding to an uncertainty of
the individual measurements of $\sigma_z=0.0004$. This is in good
agreement with the uncertainty of the individual redshift measurements
that was estimated in Paper~I from the width of the peaks of the
correlation function to be $\delta z=0.0005$.


\section{Identification of groups in redshift space}
\label{sec:z_identification}

\begin{table}
\begin{center}
\caption{Redshifts measured for the individual galaxies. This table is
only available in electronic form at the CDS,
http://cdsweb.u-strasbg.fr/cgi-bin/qcat?J/A+A/ .}
\label{tab:galredshifts}
\begin{tabular}{l}
\hline\hline 
This table is only available in electronic form.\\
\hline
\end{tabular}
\end{center}
\end{table}

\begin{table*}
\begin{center}
\caption{Identified groups with a significance of at least 99\% as
obtained by at least one of the methods considered. Those in bold
face are the ones we associate to the cluster detection as discussed
in the text. When $\sigma_v=0$ it indicates that we measured a
velocity dispersion that was smaller than the estimated error.}
\label{tab:EISgroups}
\begin{tabular}{lrcccrrr}
\hline\hline
Cluster Field & Members & $\alpha$ (J2000) & $\delta$ (J2000) & z & $\sigma_v \mathrm{[km/s]}$ & $\sigma_1$ [\%] \\
\hline
{\bf EISJ0946-2133} &   {\bf 7} &  {\bf 09 46 38.5} & {\bf -21 34 57.7} & {\bf 0.141} &   {\bf 183} &  {\bf 99.9} \\
EISJ0946-2133 &   3 &  09 46 38.6 & -21 33 53.3 & 0.153 &     0 &  99.9 \\
EISJ0946-2133 &   3 &  09 46 25.0 & -21 31 58.9 & 0.191 &   302 &  99.6 \\
EISJ0946-2133 &   4 &  09 46 20.3 & -21 34 56.2 & 0.351 &   562 &  99.7 \\
\hline
EISJ0949-2145 &   4 &  09 49 56.2 & -21 42 06.3 & 0.159 &     0 &  99.3 \\
EISJ0949-2145 &   4 &  09 49 53.9 & -21 44 09.3 & 0.184 &  1105 &  99.5 \\
\hline
{\bf EISJ0949-2046} &  {\bf 16} &  {\bf 09 49 50.3} & {\bf -20 46 26.8} & {\bf 0.143} &   {\bf 286} &  {\bf 99.9} \\
EISJ0949-2046 &   7 &  09 49 57.3 & -20 46 08.5 & 0.266 &   229 &  99.9 \\
\hline
{\bf EISJ0950-2133} &  {\bf 6} &  {\bf 09 50 43.0} & {\bf -21 34 20.0} & {\bf 0.131} &   {\bf 126} &  {\bf 99.9} \\
EISJ0950-2133 &   3 &  09 50 49.8 & -21 36 06.6 & 0.184 &     0 &  99.9 \\
EISJ0950-2133 &   7 & 09 50 40.5 & -21 35 19.7 & 0.235 &   902 &  99.4 \\
\hline
EISJ0951-2052 &   5 &  09 51 00.2 & -20 53 39.2 & 0.205 &   285 &  99.9 \\
{\bf EISJ0951-2052} &  {\bf 15} &  {\bf 09 51 09.6} & {\bf -20 51 56.3} & {\bf 0.243} &   {\bf 833} &  {\bf 99.9} \\
\hline
EISJ0951-2026 &   4 &  09 51 26.7 & -20 26 35.0 & 0.183 &     0 &  99.9 \\
{\bf EISJ0951-2026} &  {\bf 25} &  {\bf 09 51 32.9} & {\bf -20 27 01.1} & {\bf 0.242} &   {\bf 544} &  {\bf 99.9} \\
\hline
{\bf EISJ0951-2145} &  {\bf 16} &  {\bf 09 51 48.1} & {\bf -21 45 27.6} & {\bf 0.185} &   {\bf 555} &  {\bf 99.9} \\
EISJ0951-2145 &   5 &  09 51 46.3 & -21 45 48.1 & 0.233 &   488 &  99.7 \\
\hline
EISJ0952-2150 &   6 &  09 52 52.3 & -21 49 27.9 & 0.149 &   204 &  99.9 \\
{\bf EISJ0952-2150} &  {\bf 12} &  {\bf 09 52 47.5} & {\bf -21 49 27.9} & {\bf 0.183} &   {\bf 613} &  {\bf 99.9} \\
EISJ0952-2150 &   5 &  09 53 07.4 & -21 46 00.3 & 0.215 &   208 &  99.9 \\
\hline
EISJ0952-2103 &   3 &  09 52 56.3 & -21 05 11.1 & 0.108 &   124 &  99.4 \\
EISJ0952-2103 &   3 &  09 53 05.5 & -21 06 34.2 & 0.129 &   133 &  99.9 \\
{\bf EISJ0952-2103} &  {\bf 18} &  {\bf 09 52 54.3} & {\bf -21 03 53.6} & {\bf 0.236} &   {\bf 838} &  {\bf 99.9}\\
\hline
EISJ0952-2144 &   5 &  09 52 54.4 & -21 45 01.8 & 0.149 &    0 &  99.8 \\
{\bf EISJ0952-2144} & {\bf  17} &  {\bf 09 52 51.0} & {\bf -21 44 45.5} & {\bf 0.183} &   {\bf 595} &  {\bf 99.9} \\
EISJ0952-2144 &   8 &  09 53 06.8 & -21 45 09.0 & 0.216 &    86 &  99.9 \\
EISJ0952-2144 &   5 &  09 52 31.9 & -21 42 26.7 & 0.234 &    0 &  99.9 \\
EISJ0952-2144 &   3 &  09 52 31.1 & -21 41 23.2 & 0.267 &    74 &  99.9 \\
\hline
EISJ0952-2018 &   6 &  09 52 57.9 & -20 21 34.0 & 0.163 &    0 &  99.9 \\
{\bf EISJ0952-2018} &  {\bf 14} &  {\bf 09 53 01.7} & {\bf -20 21 20.5} & {\bf 0.252} &   {\bf 444} &  {\bf 99.9} \\
\hline
EISJ0953-2053 &   3 &  09 52 54.6 & -20 52 31.3 & 0.204 &  0 &  99.9 \\
{\bf EISJ0953-2053} &  {\bf 12} &  {\bf 09 53 06.7} & {\bf -20 52 47.6} & {\bf 0.235} &   {\bf 437} &  {\bf 99.9} \\
\hline
EISJ0953-2156 &   5 &  09 53 43.5 & -21 56 20.9 & 0.162 &  1009 &  99.9 \\
{\bf EISJ0953-2156} &   {\bf 6} &  {\bf 09 53 40.0} & {\bf -21 54 52.9} & {\bf 0.181} &     {\bf 0} &  {\bf 99.9} \\
EISJ0953-2156 &   5 &  09 53 25.3 & -21 55 33.7 & 0.233 &   246 &  99.3 \\
EISJ0953-2156 &   4 &  09 53 43.2 & -21 54 18.2 & 0.330 &    86 &  99.9 \\
\hline

EISJ0953-2017 &   3 &  09 54 06.9 & -20 14 42.9 & 0.064 &     0 &  99.9 \\
{\bf EISJ0953-2017} &   {\bf 9} &  {\bf 09 53 56.2} & {\bf -20 17 26.2} & {\bf 0.095} &   {\bf 195} &  {\bf 99.9} \\
EISJ0953-2017 &   3 &  09 54 03.8 & -20 16 24.0 & 0.173 &    0 &  99.6 \\
EISJ0953-2017 &   7 &  09 53 59.6 & -20 16 38.7 & 0.282 &   469 &  99.9 \\
\hline
EISJ0955-2123 &   6 &  09 55 12.6 & -21 21 45.0 & 0.111 &   880 &  99.9 \\
{\bf EISJ0955-2123} &  {\bf 16} &  {\bf 09 54 59.0} & {\bf -21 22 19.1} & {\bf 0.203} &   {\bf 739} &  {\bf 99.9} \\
EISJ0955-2123 &   3 &  09 54 49.2 & -21 22 17.6 & 0.269 &   179 &  99.6 \\
EISJ0955-2123 &   3 &  09 54 44.0 & -21 21 50.9 & 0.415 &     0 &  99.9 \\
\hline
\end{tabular}
\end{center}
\end{table*}

\addtocounter{table}{-1}

\begin{table*}
\begin{center}
\caption{\it  -- Continued}
\begin{tabular}{lrcccrrr}
\hline\hline
Cluster Field & Members & $\alpha$ (J2000) & $\delta$ (J2000) & z & $\sigma_v \mathrm{[km/s]}$ & $\sigma_1$ [\%] \\
\hline
EISJ0955-2151 &   5 &  09 55 26.4 & -21 52 19.9 & 0.105 &   272 &  99.9 \\
{\bf EISJ0955-2151} &   {\bf 9} &  {\bf 09 55 00.5} & {\bf -21 52 13.7} & {\bf 0.114} &   {\bf 465} &  {\bf 99.9} \\
EISJ0955-2151 &   3 &  09 54 57.1 & -21 52 55.8 & 0.203 &   752 &  99.2 \\
EISJ0955-2151 &   8 &  09 55 00.8 & -21 52 54.2 & 0.217 &   348 &  99.9 \\
\hline
EISJ0955-2037 &   5 &  09 55 18.8 & -20 34 18.2 & 0.234 &   369 &  99.9 \\
EISJ0955-2037 &   3 &  09 54 57.5 & -20 37 08.9 & 0.248 &    0 &  99.9 \\
{\bf EISJ0955-2037} &   {\bf 6} &  {\bf 09 55 08.4} & {\bf -20 35 30.6} & {\bf 0.283} &   {\bf 221} &  {\bf 99.9} \\
\hline
{\bf EISJ0955-2020} &   {\bf 8} &  {\bf 09 55 16.0} & {\bf -20 19 44.3} & {\bf 0.064} &   {\bf 389} &  {\bf 99.9} \\
EISJ0955-2020 &   6 &  09 55 23.1 & -20 19 58.1 & 0.104 &   872 &  99.9 \\
EISJ0955-2020 &   3 &  09 55 32.6 & -20 19 11.1 & 0.285 &   257 &  99.6 \\
\hline
EISJ0956-2054 &   5 &  09 55 56.4 & -20 54 23.7 & 0.245 &   200 &  99.9 \\
{\bf EISJ0956-2054} &  {\bf 17} &  {\bf 09 56  4.4} & {\bf -20 55 25.7} & {\bf 0.279} &   {\bf 753} &  {\bf 99.9} \\
\hline
EISJ0957-2051 &   5 &  09 57 22.2 & -20 52 40.5 & 0.148 &   559 &  99.9 \\
EISJ0957-2051 &   3 &  09 57 13.8 & -20 51 20.8 & 0.203 &   523 &  99.4 \\
{\bf EISJ0957-2051} &   {\bf 8} &  {\bf 09 57  4.6} & {\bf -20 52 49.9} & {\bf 0.241} &   {\bf 387} &  {\bf 99.9} \\
\hline
EISJ0957-2143 &   3 &  09 57 23.9 & -21 42 40.0 & 0.104 &     0 &  99.9 \\
{\bf EISJ0957-2143} &  {\bf 16} &  {\bf 09 57  9.9} & {\bf -21 44  3.5} & {\bf 0.202} &   {\bf 264} &  {\bf 99.9} \\
EISJ0957-2143 &   5 &  09 57 19.9 & -21 43  5.6 & 0.228 &   301 &  99.9 \\
\hline
\end{tabular}
\end{center}
\end{table*}

\begin{figure*}
\begin{center}
\resizebox{0.23\textwidth}{!}{\includegraphics{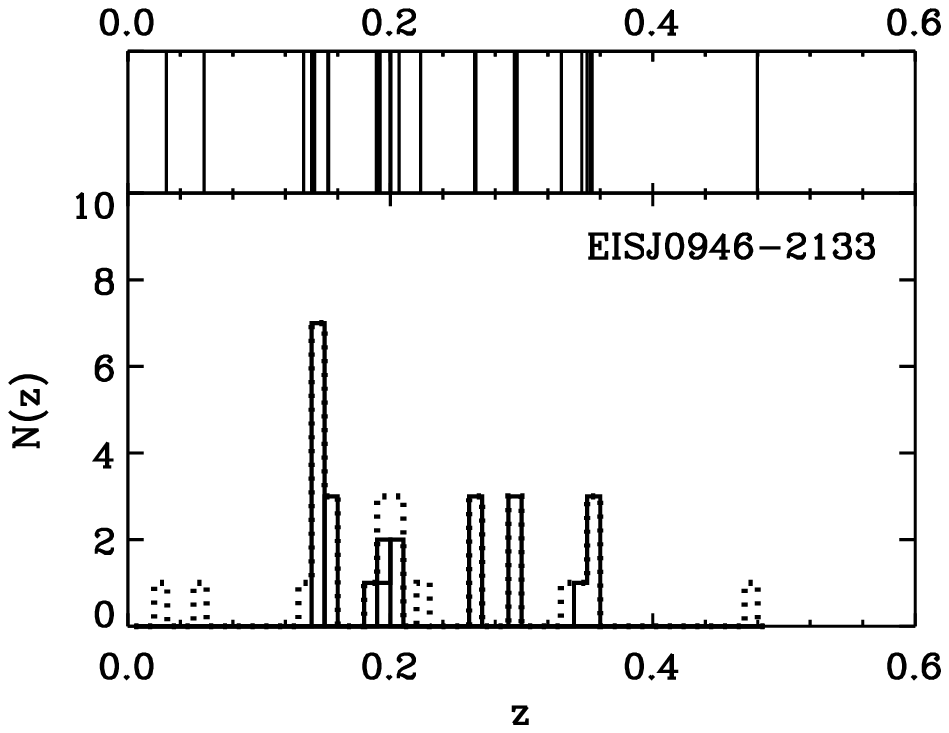}}
\resizebox{0.23\textwidth}{!}{\includegraphics{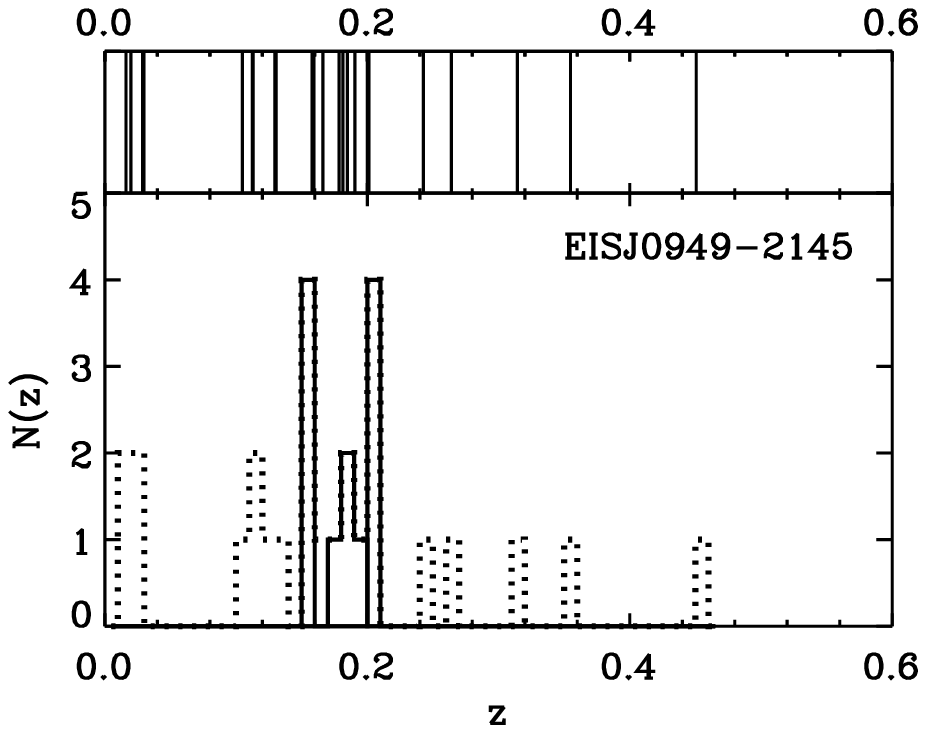}}
\resizebox{0.23\textwidth}{!}{\includegraphics{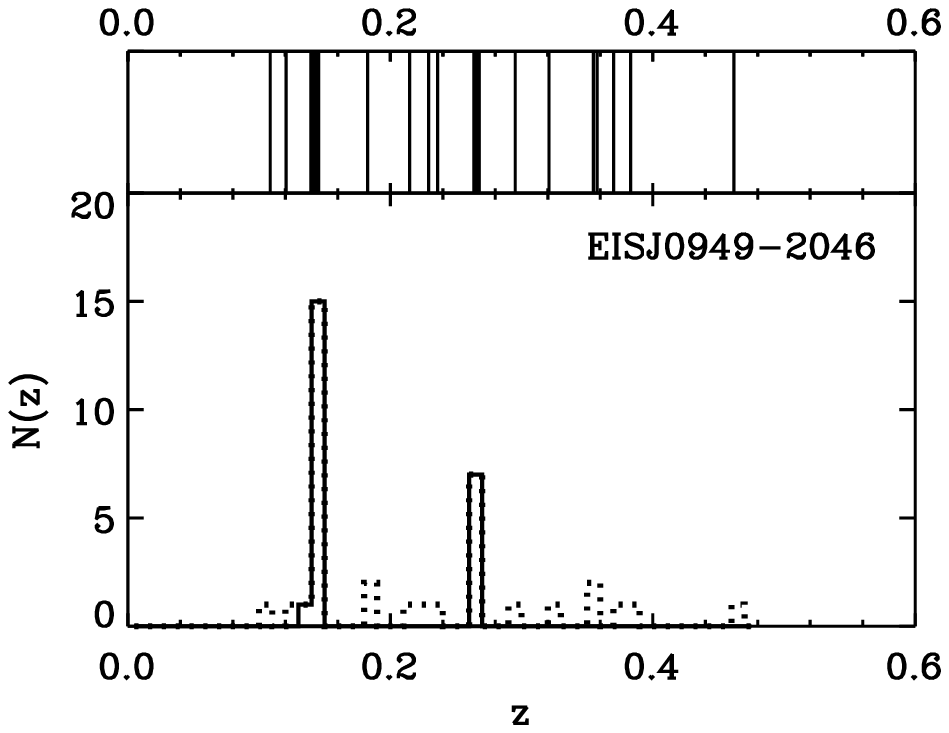}}
\resizebox{0.23\textwidth}{!}{\includegraphics{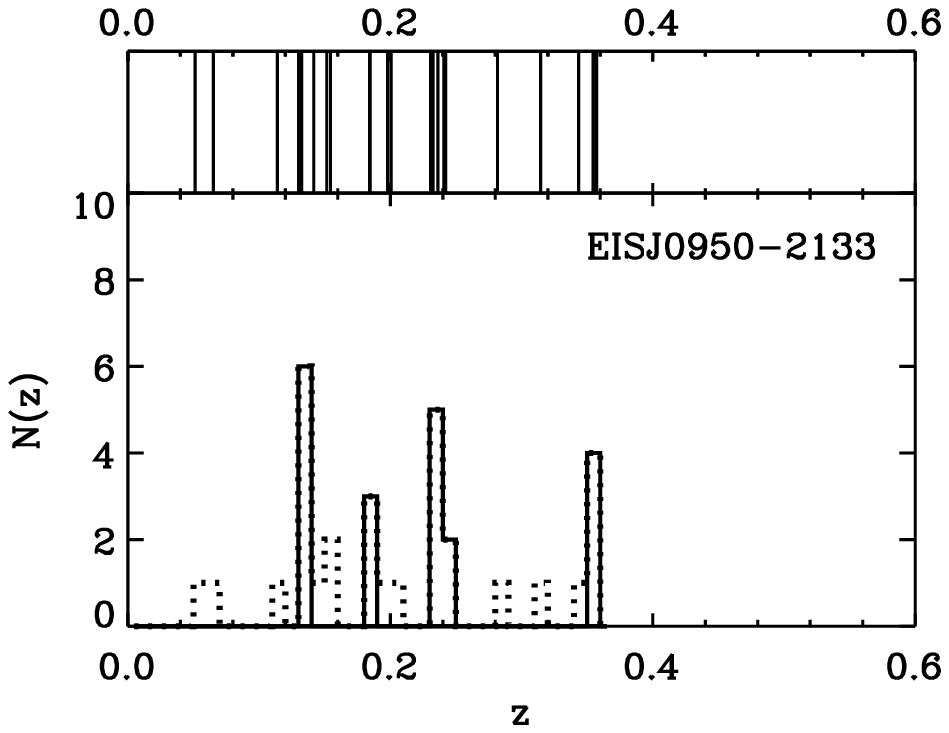}}
\resizebox{0.23\textwidth}{!}{\includegraphics{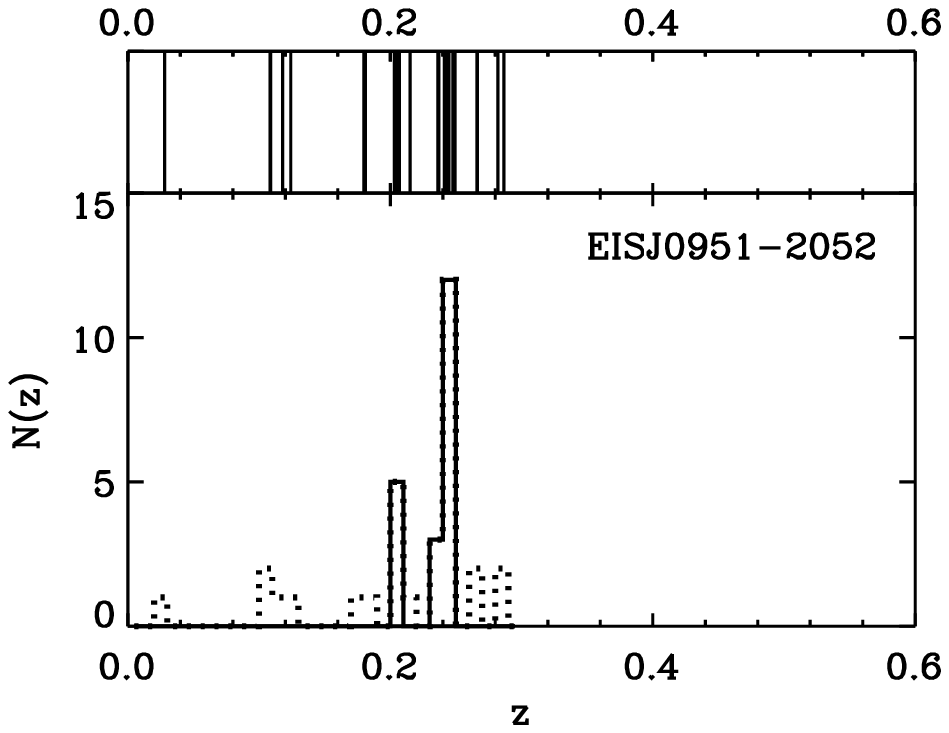}}
\resizebox{0.23\textwidth}{!}{\includegraphics{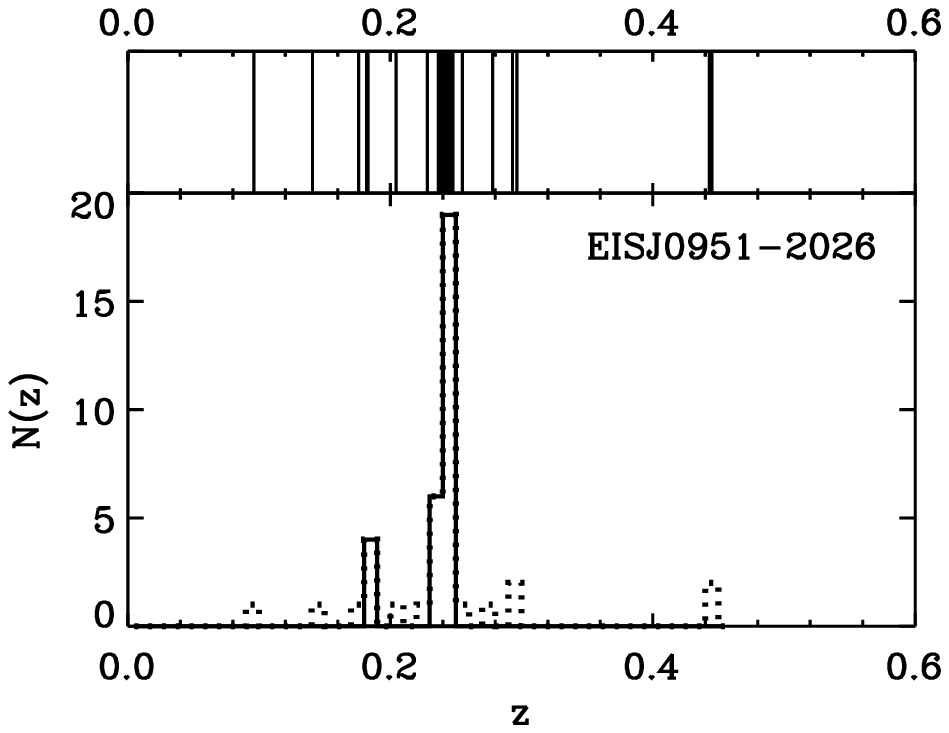}}
\resizebox{0.23\textwidth}{!}{\includegraphics{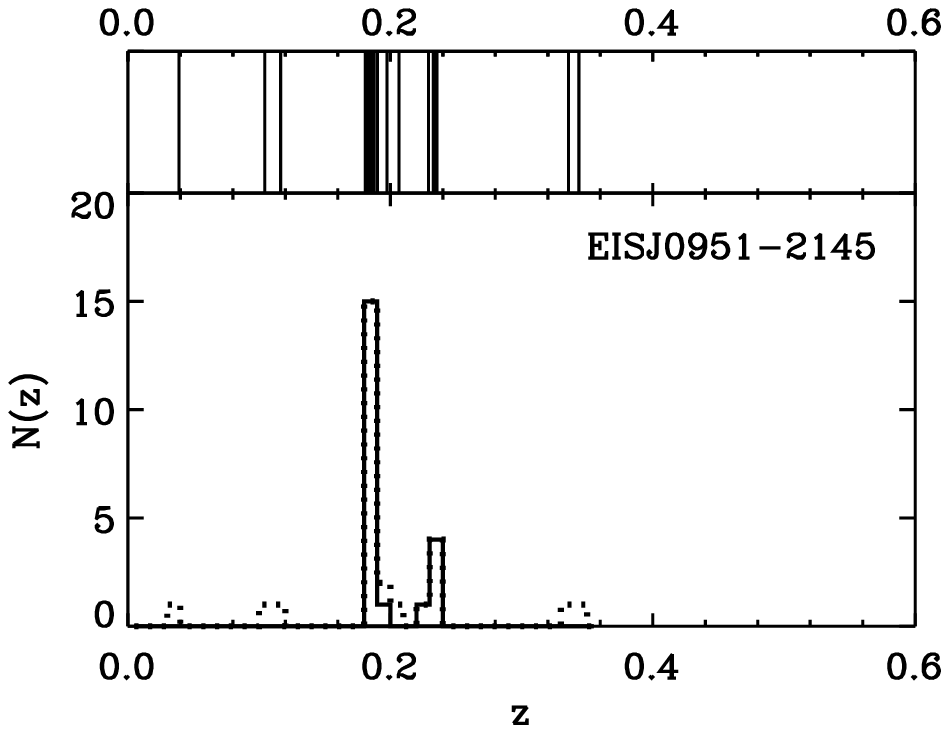}}
\resizebox{0.23\textwidth}{!}{\includegraphics{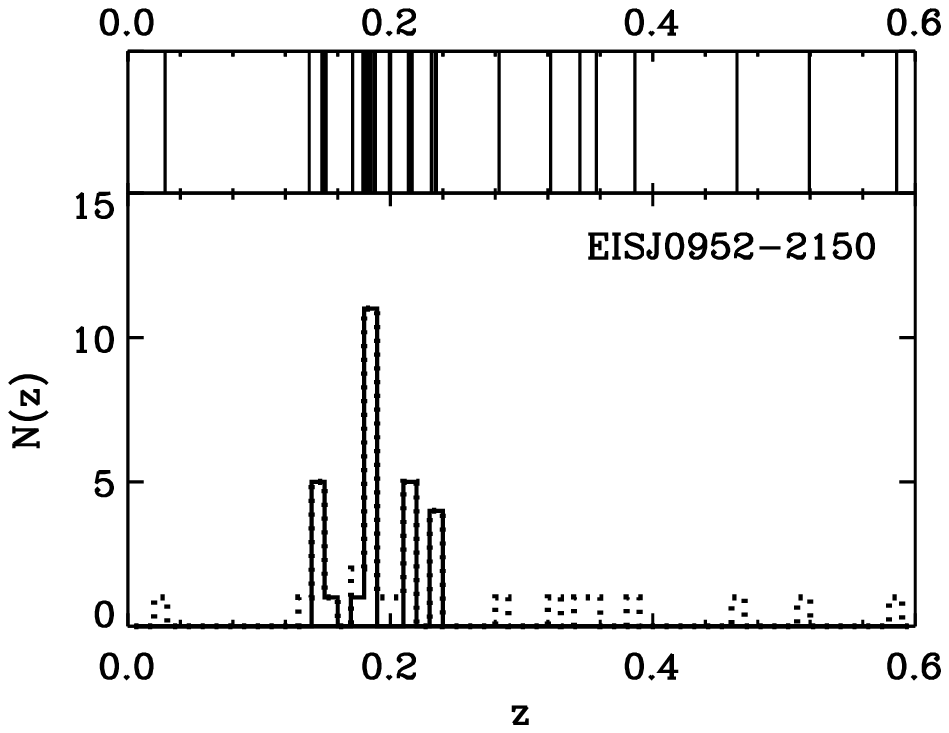}}
\resizebox{0.23\textwidth}{!}{\includegraphics{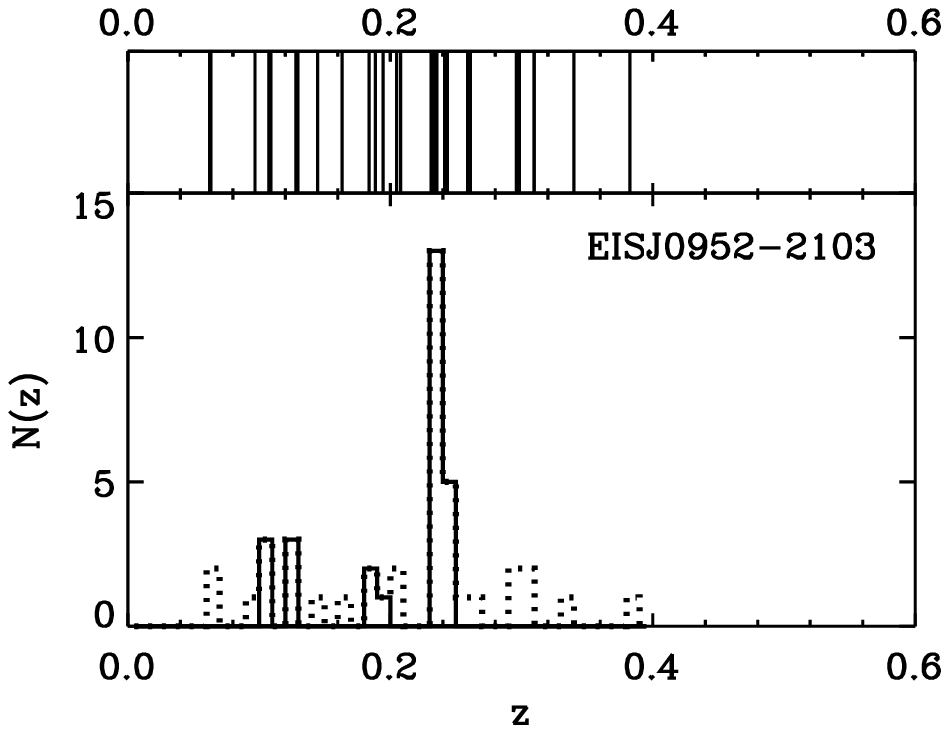}}
\resizebox{0.23\textwidth}{!}{\includegraphics{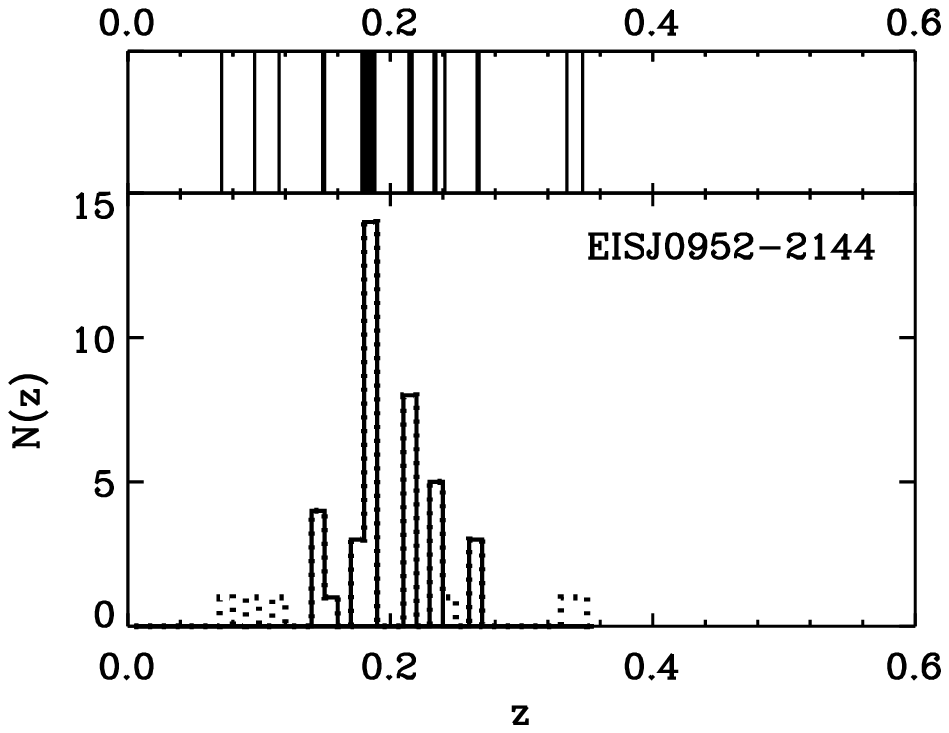}}
\resizebox{0.23\textwidth}{!}{\includegraphics{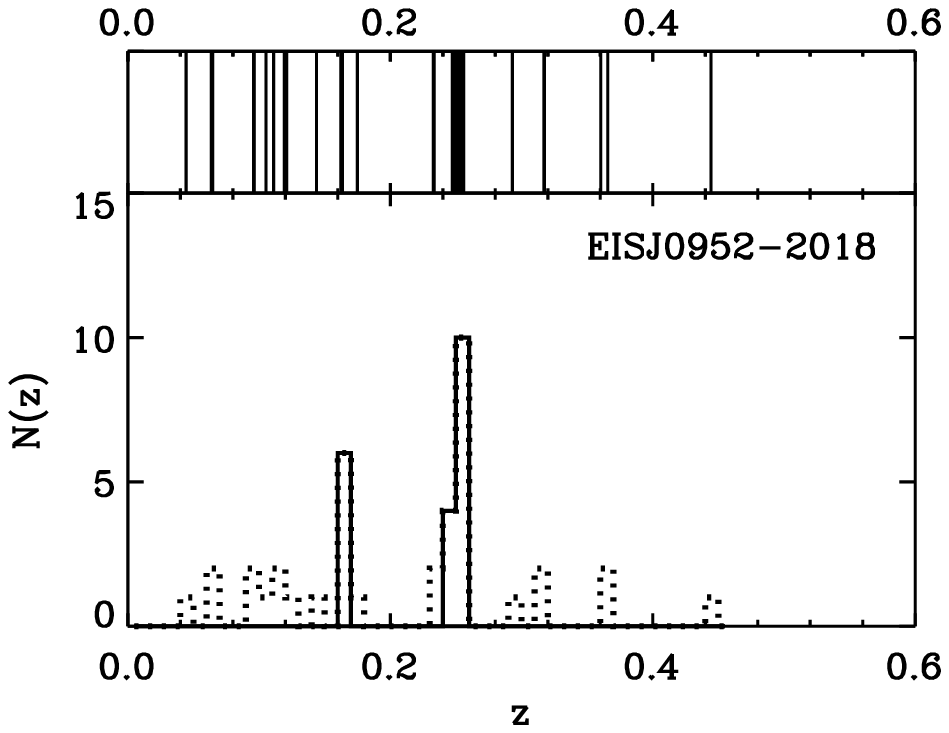}}
\resizebox{0.23\textwidth}{!}{\includegraphics{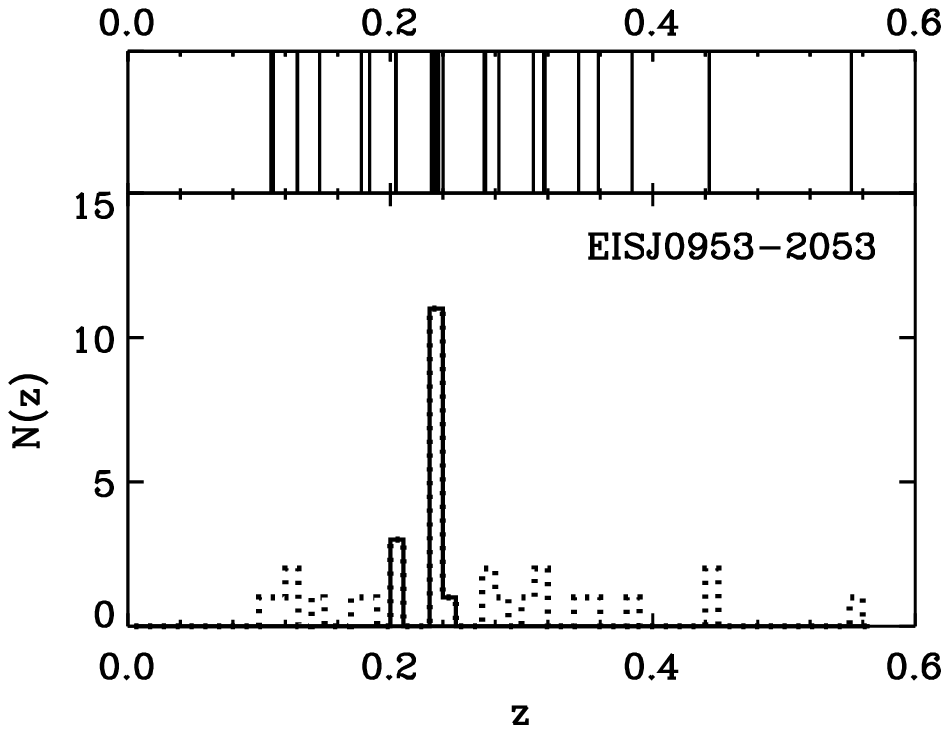}}
\resizebox{0.23\textwidth}{!}{\includegraphics{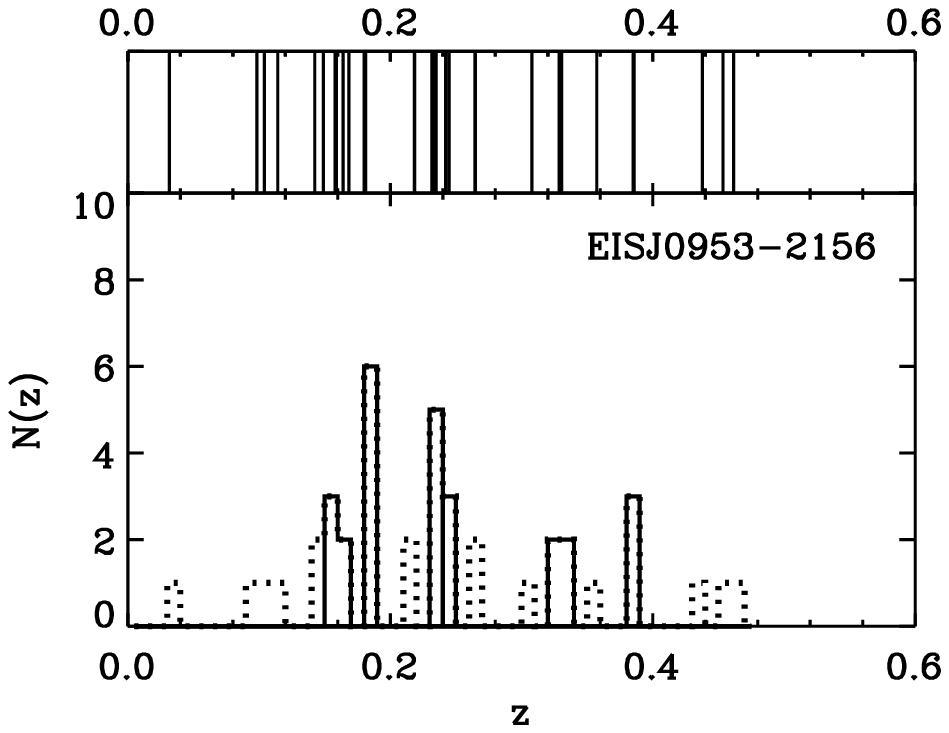}}
\resizebox{0.23\textwidth}{!}{\includegraphics{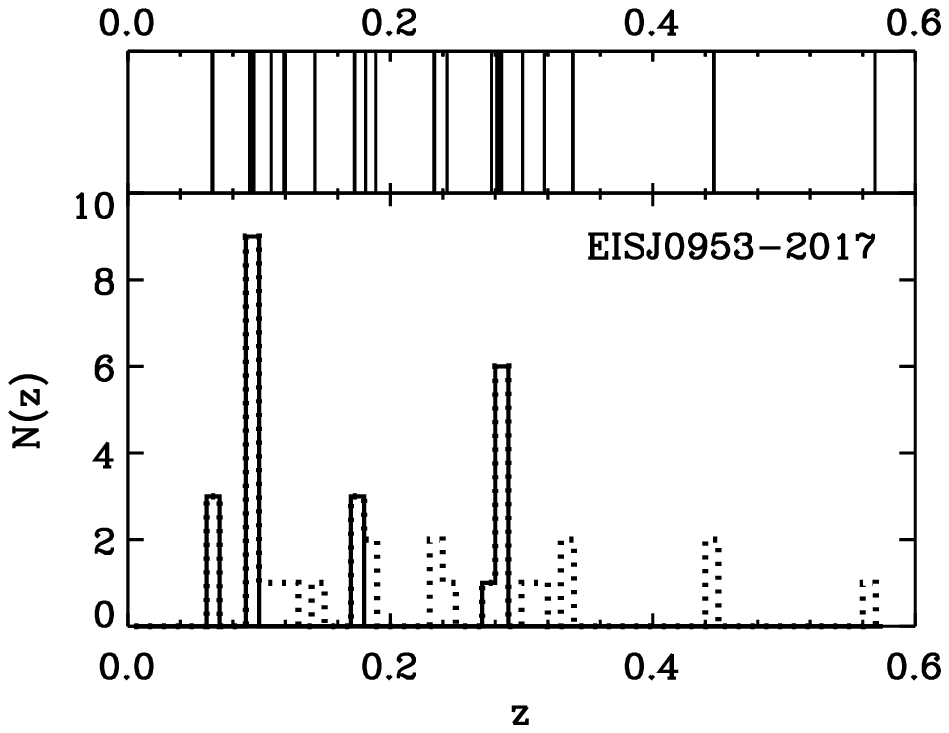}}
\resizebox{0.23\textwidth}{!}{\includegraphics{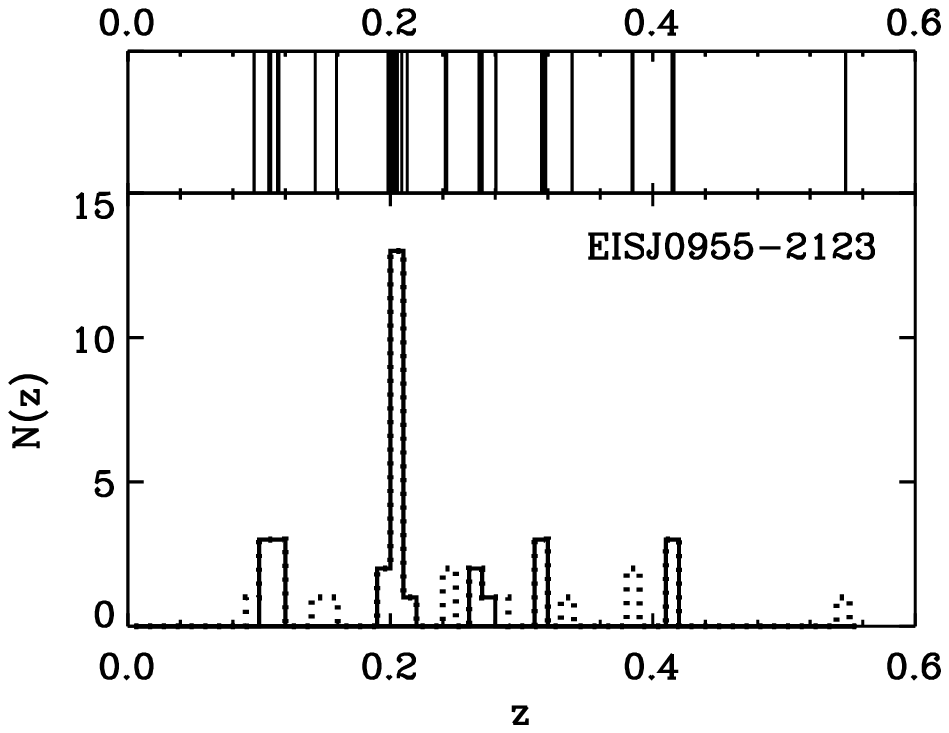}}
\resizebox{0.23\textwidth}{!}{\includegraphics{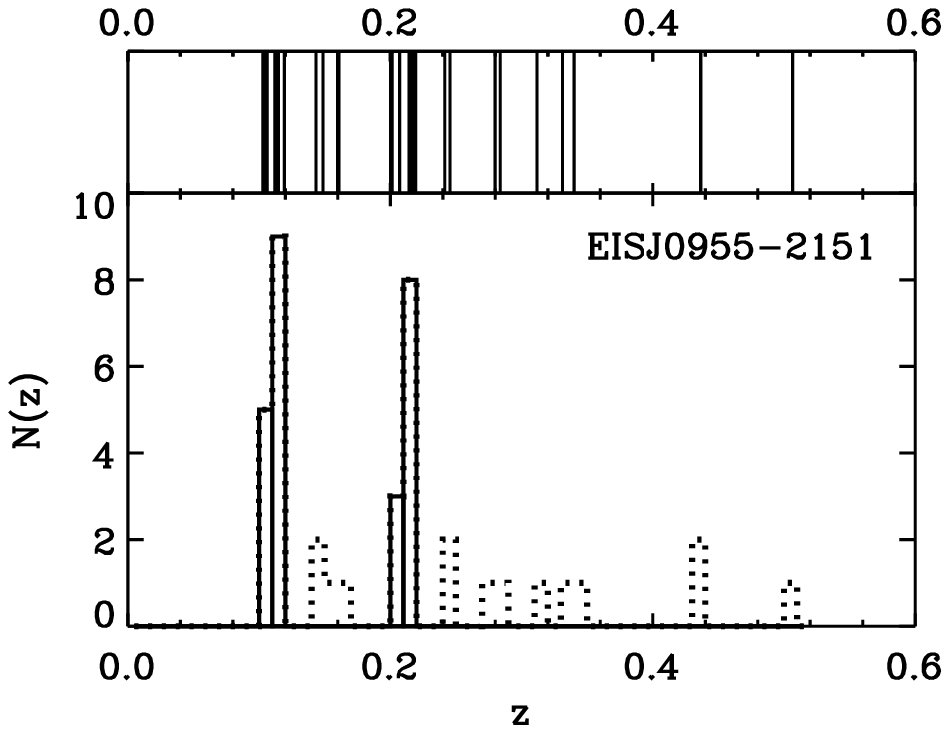}}
\resizebox{0.23\textwidth}{!}{\includegraphics{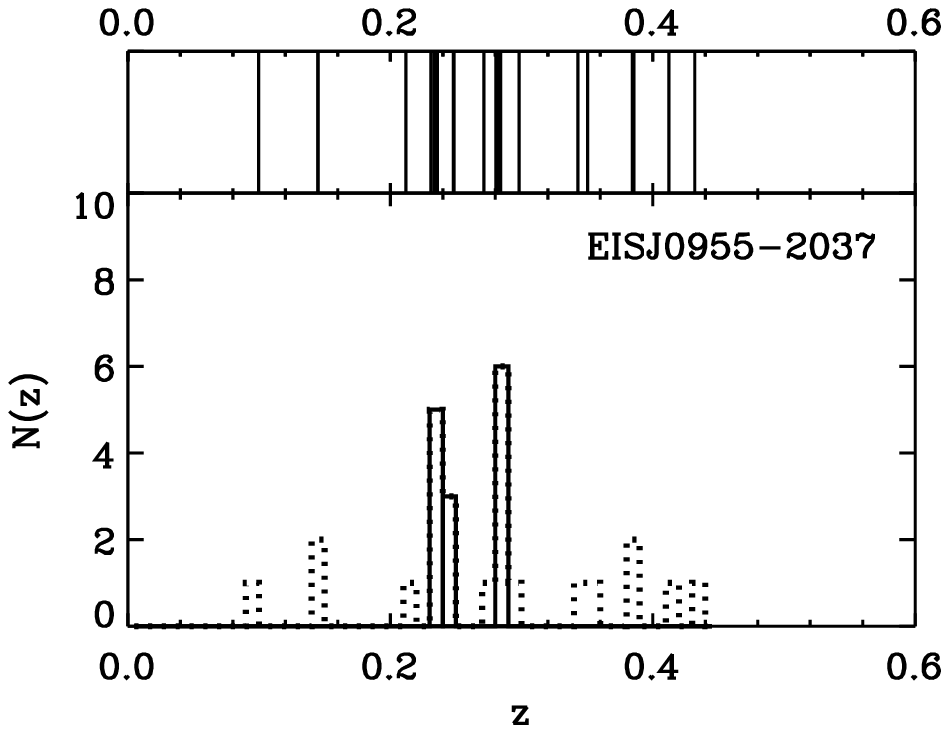}}
\resizebox{0.23\textwidth}{!}{\includegraphics{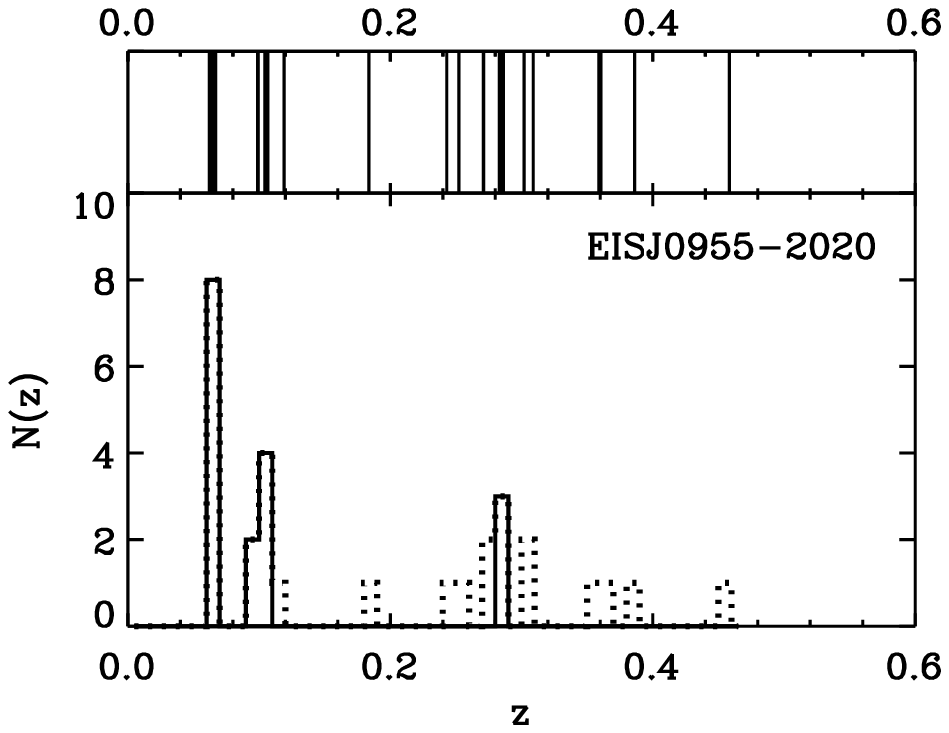}}
\resizebox{0.23\textwidth}{!}{\includegraphics{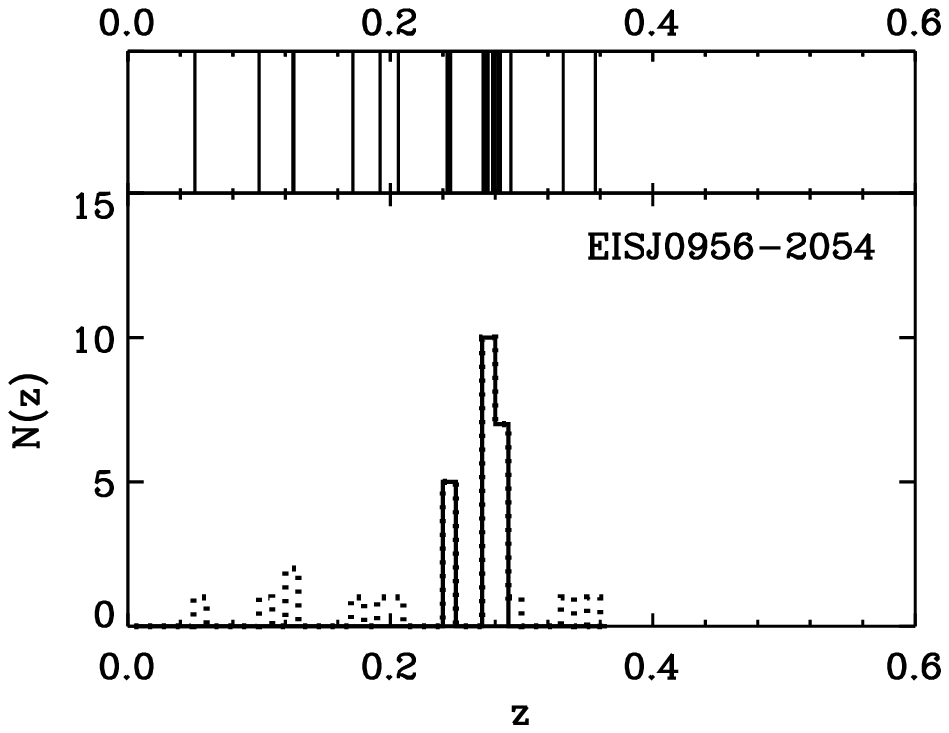}}
\resizebox{0.23\textwidth}{!}{\includegraphics{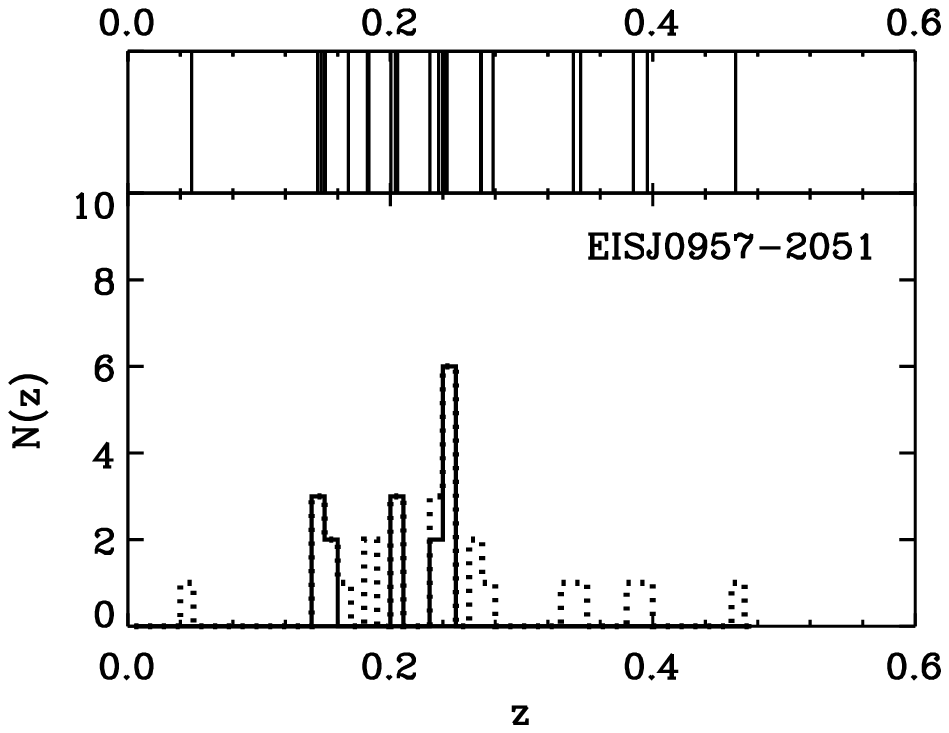}}
\resizebox{0.23\textwidth}{!}{\includegraphics{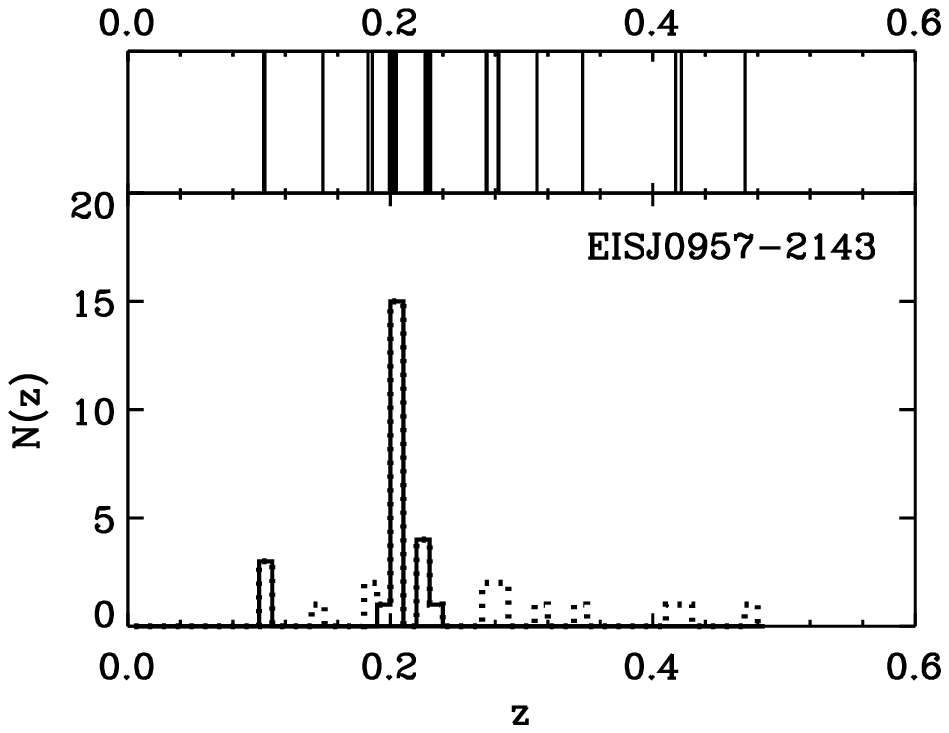}}
\caption{Redshift distributions for the 21 observed cluster
fields. Note that the scale of the y-axis differs between the
panels. The upper panels show bar diagrams of the measured redshifts,
while the lower panels give the corresponding histograms of the
redshift distributions  (dotted line). The solid lines mark the
detected groups.}
\label{fig:redshift_dists}
\end{center}
\end{figure*}

\begin{figure*}
\begin{center}
\resizebox{\textwidth}{!}{\includegraphics{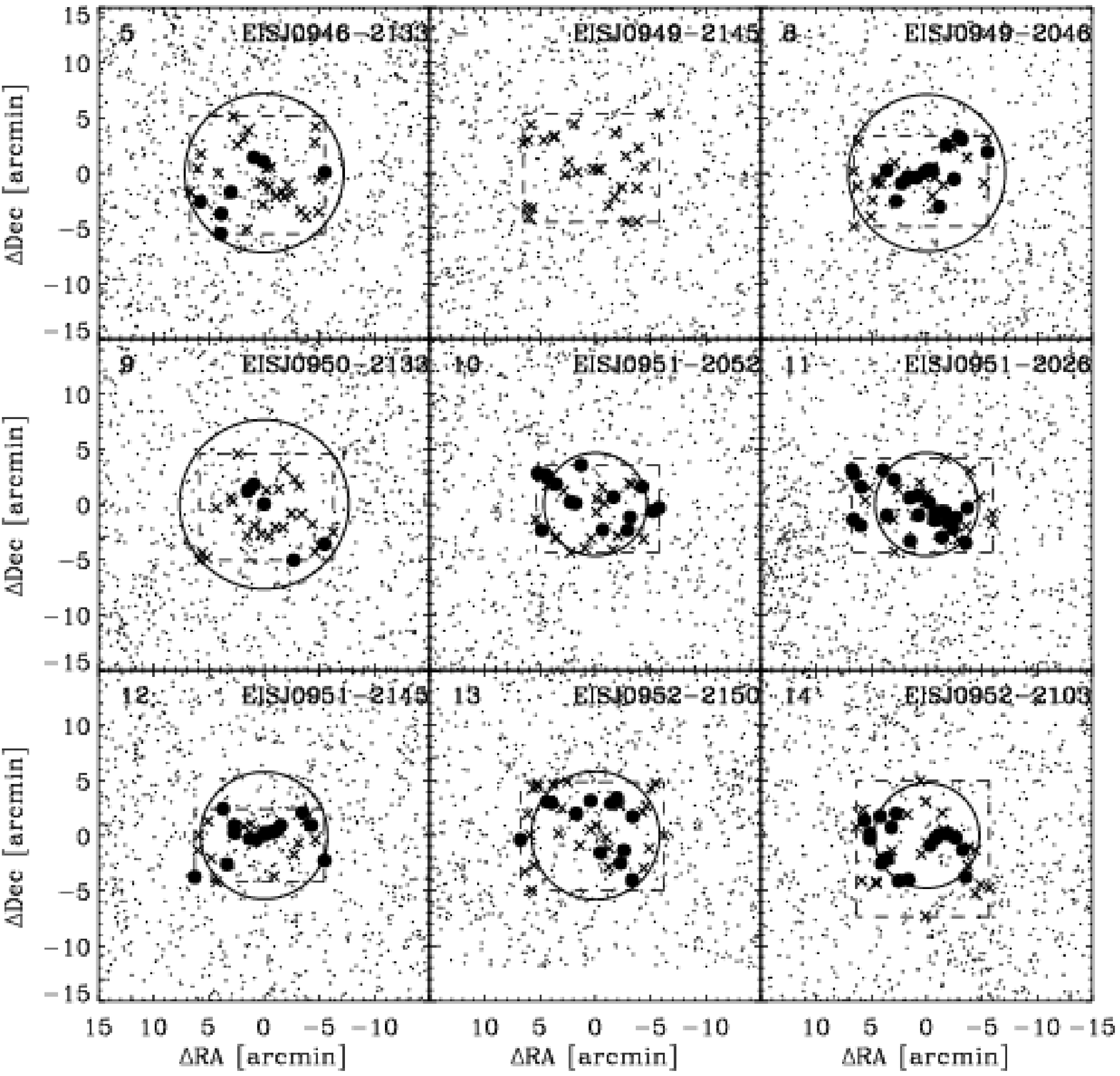}}
\end{center}
\caption{ Projected distributions of $I\leq19.5$ galaxies (small
symbols) in each of the cluster fields. The number on the left-hand
side refers to Table~\protect\ref{tab:colour}.  The matched filter
center of the cluster is in the center of the plots. The dashed line
marks the region covered by the MOS-masks. In some cases the MOS-masks
are not centered on the cluster center due to the distribution of the
bright galaxies. The filled circles mark spectroscopic members of the
confirmed systems, and the crosses galaxies with redshifts not
belonging to the group. The large circles indicate a region of
$0.5h_{75}^{-1}\mathrm{Mpc}$ at the redshift of the confirmed group. }
\label{fig:spatial_dist}
\end{figure*}

\addtocounter{figure}{-1}

\begin{figure*}
\begin{center}
\resizebox{\textwidth}{!}{\includegraphics{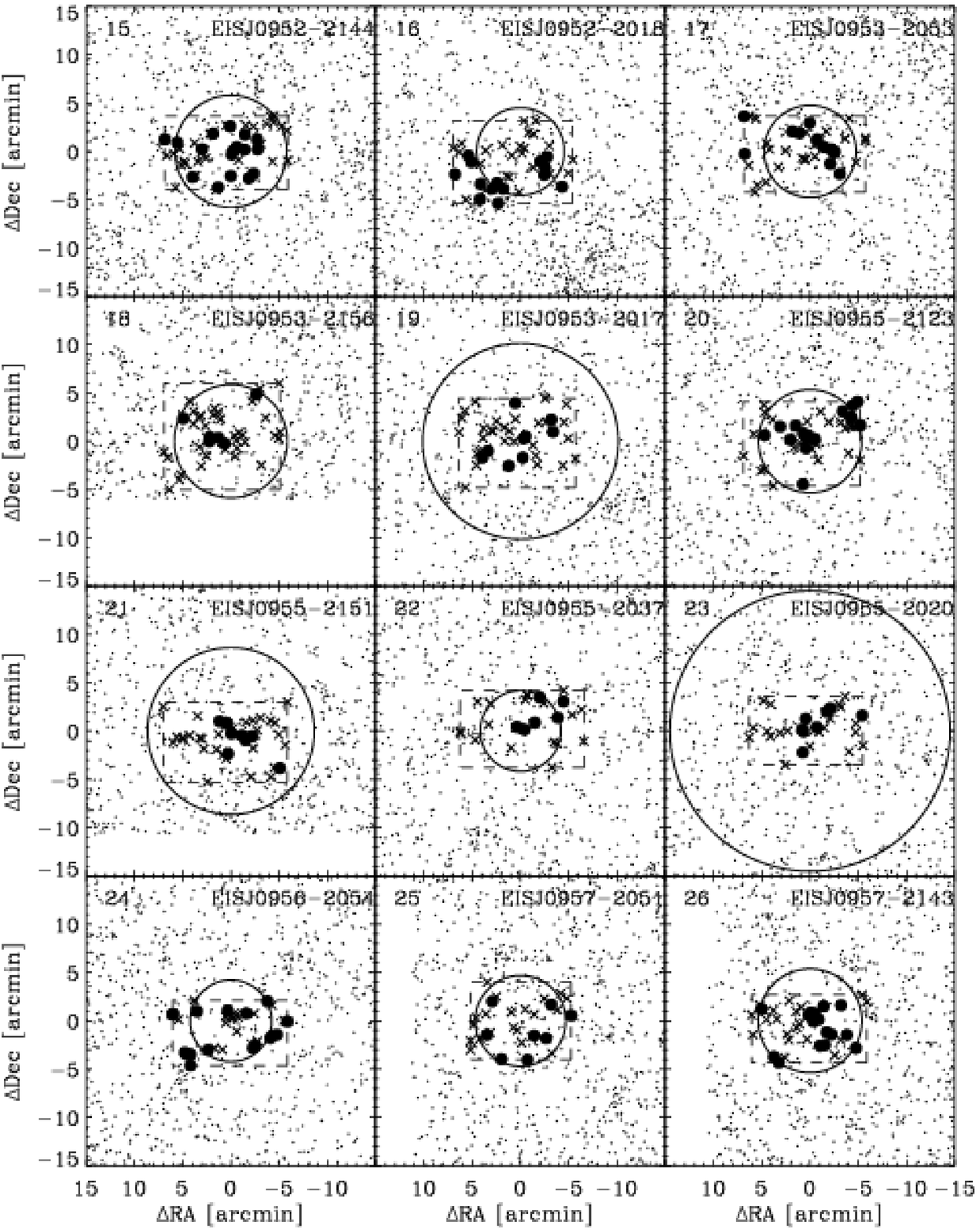}}
\end{center}
\caption{\it -- Continued }
\end{figure*}

We have obtained 738 redshifts for galaxies in 21 EIS cluster
candidate fields.  Table~\ref{tab:galredshifts}, available at the CDS\footnote{http://cdsweb.u-strasbg.fr/cgi-bin/qcat?J/A+A/},
lists: in Col.~1 a running identifier for each galaxy; in Cols.~2 and 3
the right ascension and declination in J2000 for the galaxy; Col.~4
the I-magnitude from the EIS object catalogues \citep{benoist99}; and
in Col.~5 the measured redshift. A colon (``:'') marks a measurement
for which we could not identify any features to confirm the redshift,
and an ``e'' indicates that the spectrum had emission lines.

The number of derived redshifts ranges between 26 and 44 per field. In
Fig.~\ref{fig:redshift_dists} we show the redshifts for each
field. The upper parts show the bar diagram of the redshifts while the
lower part gives the redshift histogram with a bin size of $\Delta
z=0.01$.  As described in Paper II we use the ``gap''-technique of
\cite{katgert96} to identify groups in redshift
space. Fig.~\ref{fig:redshift_dists} shows the identified groups as
the solid histograms. We have selected a gap-size of $\Delta
z=0.005(1+z)$ corresponding to $1500\mathrm{km/s}$ in the
restframe. For assessing the significance of the identified groups we
use the CNOC2 0223+00 catalogue \citep{yee00}. The significance is
determined from the probability of finding a group with the same
number of objects or more at the same redshift \citep[see ][ for more
details]{olsen03}. This significance is referred to as $\sigma_1$.

In Table~\ref{tab:EISgroups} we list all groups with significance
larger than $99\%$ identified in each cluster field.  The Table lists:
in Col~1 the cluster field name; in Col.~2 the number of spectroscopic
members of the group; in Cols.~3 and 4 the mean position in J2000; in
Col.~5 the mean redshift of the group members; in Col.~6 the velocity
dispersion corrected for our measurement accuracy. In cases where the
measured velocity dispersion is smaller than the measurement error we
list the value of $\sigma_v=0$; and in Cols.~7 the significance as
defined above.

The table lists 62 significant groups, ranging between two and five
groups per cluster field, having from 3 to 25 members. In many cases a
single group dominates the field, with many more members than the
others. In these cases, there is little ambiguity in associating this
group to the matched filter detection.  This is the case for 12 of 21
($\sim57\%$) fields.  In the remaining cases the redshift distribution
is more complex and consequently, the association of a group to the
matched filter detection is more difficult.  In other words, this is
 the result of the projection effects that plague the
identification of clusters from a projected distribution of galaxies
using a single passband.

The group associated with the matched filter detection is chosen as
follows: 1) The richest group in the field, if it has a significantly
larger number of members than the other groups; 2) The one closest to
the EIS position, if two groups have roughly the same number of
members; 3) The most concentrated group, if two groups are close to
the EIS position and have almost the same number of members.  Note
that in all but one case (EISJ0950-2133) we associated the richest
significant group with the EIS detection.

In Fig.~\ref{fig:spatial_dist} we show the projected distribution of
all galaxies with $I\leq19.5$ in the cluster regions. The solid
circles mark galaxies belonging to the group associated with the EIS
cluster candidate, and the crosses mark galaxies with redshifts
outside the group. The large circles mark the area within
$0.5h_{75}^{-1}\mathrm{Mpc}$ from the cluster center. From this
analysis we find that 20 out of 21 ($\sim95\%$) cluster candidates are
confirmed as overdensities in redshift space.  In one case
(EISJ0949-2145), we do not consider any of the groups as representing
the matched filter detection, since the number of members of the
groups is small and they are spread over most of the surveyed area.

\subsection{Projection effects}

One of the main problems in detecting clusters from the projected galaxy
distribution is the contamination along the line of sight. This
effect may have two origins: one is the superposition of galaxy
systems and the other the contamination by field galaxies.

As noted above, all the surveyed fields studied in the present paper
contain more than one significant group in redshift space indicating
 that superposition effects cannot be neglected. Following
\cite{katgert96}, we consider systems to be significantly affected by
superposition if the ratio of the number of member galaxies in
the confirmed system to the number of members in the second largest
system is  smaller than two.  For each confirmed system we compute this
ratio and find that 8 (40\%) out of the 20 systems (EISJ0946-2133,
EISJ0950-2133, EISJ0953-2156, EISJ0953-2017, EISJ0955-2151,
EISJ0955-2037, EISJ0955-2020, EISJ0957-2051) are likely to be affected
by superposition and may have overestimated richnesses. In addition,
it should be noted that these systems are also rather poor with less
than 10 members and may thus also be affected by field contamination.

In the following sections we will combine the present sample with
those of Paper~I (3 confirmed systems out of 3 observed) and Paper~II
(9 confirmed systems out of 10 observed). Therefore, we have also
investigated the projection effects for those systems. We find that
one case (EISJ2241-3949) is likely to be affected by superposition
even though it has 18 member galaxies and thus the field contamination
is relatively low. Furthermore, we find four cases with less than ten
members (EISJ0046-2925, EISJ2243-4013, EISJ2244-3955, EISJ2246-4012A)
where only one group is identified and these systems are thus
likely to be significantly contaminated by field galaxies.

In summary, we estimate that 13 out of 32 systems ($\sim40\%$) are
likely to be contaminated and care should be taken when interpreting
results based on these systems.

\subsection{Summary}

Combining the results of the present paper with those of Papers~I and
II we find an overall confirmation rate of $\sim94\%$ (32
clusters of 34 candidates) covering a region of about
$\sim9~\mathrm{square~degrees}$. This confirmation rate is consistent
with the expected rate of false detecions of $\sim1$ at $z_{MF}=0.2$
within the area considered here estimated by \cite{olsen00}.  The
results are also in good agreement with those of \cite{holden99},
\cite{holden00} and \cite{postman02}, who carried out spectroscopic
follow-up of cluster candidates detected using a similar matched
filter technique.  \cite{holden99} and \cite{holden00} studied 9
candidates with $z_{MF}\leq0.3$ of which they confirmed 8 and
\cite{postman02} confirmed 13 out of 15 clusters with $z_{MF}=0.3$.

\section{Properties of the detected systems}
\label{sec:properties}

\begin{table*}
\caption{Properties of the confirmed EIS clusters and groups. This table is also available in electronic form at the CDS, http://cdsweb.u-strasbg.fr/cgi-bin/qcat?J/A+A/ .}
\label{tab:colour}
\begin{center}
\begin{tabular}{rlrrrlrrrrrrr}
\hline\hline
Id & Cluster &  \#mem  & $z_{spec}$ & \multicolumn{2}{c}{$\sigma_v \mathrm[km/s]$ } & $\Lambda_{cl, new}$ & $C$ & $(V-I)_{ph}$ & $\sigma_{S/N}$ & $(V-I)_{sp}$ & $\sigma_{spec}$ & Scatter\\
\hline

\vspace{1mm}

1 & EISJ0045-2923 & 25 & 0.257 &  $ 674$&$^{+  78}_{- 139}$ & 36.4 & 0.54 & 1.800 &  99.9  & 1.800 & $>99.9$  &0.228\\ 

\vspace{1mm}

2 & EISJ0046-2925 & 7  & 0.167 &  $ 970$&$^{+  88}_{- 842}$ & 15.4 & 0.30 & $-$   &  $-$   &$-$   &  $-$  & $-$\\

\vspace{1mm}

3 & EISJ0052-2923 & 13 & 0.114 &  $ 615$&$^{+  72}_{- 124}$ & 10.8 & 0.95 & 1.350 &  97.3 & 1.350 & $>99.9$  & 0.064\\

\vspace{1mm}

4 & EISJ0946-2029 & 28 & 0.111 &  $ 460$&$^{+  55}_{-  93}$ & 32.0 & 0.35 & 1.200 &  99.2 &  1.200 & $>99.9$ & 0.031\\

\vspace{1mm}

5 & EISJ0946-2133 & 7  & 0.141 &  $ 289$&$^{+   0}_{- 158}$ & 23.5 & 0.78 & $-$   &  $-$  & $-$   &  $-$  & $-$\\

\vspace{1mm}

6 & EISJ0947-2120 & 12 & 0.191 &  $ 233$&$^{+  55}_{-  82}$ & 43.0 & 0.90 & 1.500 &  97.4 & 1.500 & $>99.9$ &0.065\\

\vspace{1mm}

7 & EISJ0948-2044 & 27 & 0.182 &  $ 472$&$^{+  68}_{-  87}$ & 34.9 & 1.04 & 1.500 &  98.0 &  1.425 & $>99.9$ & 0.060\\

\vspace{1mm}

8 & EISJ0949-2046 & 16 & 0.143 &  $ 311$&$^{+  84}_{- 119}$ & 18.0 & 0.74 & $-$   &  $-$  &  $-$ & $-$ & $-$\\

\vspace{1mm}

9 & EISJ0950-2133 & 6  & 0.131 &  $27$&$^{+  178}_{-27}$    & 18.2 & 0.48 & $-$   &  $-$  &  $-$   &  $-$ &$-$\\

\vspace{1mm}

10 & EISJ0951-2052 & 15 & 0.243 &  $1139$&$^{+  27}_{- 293}$ & 15.4 & 0.90 & $-$   &  $-$  &  1.575 & $>99.9$  & 0.25\\

\vspace{1mm}

11 & EISJ0951-2026 & 25 & 0.242 &  $ 703$&$^{+  72}_{- 120}$ & 31.9 & 0.48 & $-$   &  $-$  &   1.500 & $>99.9$& 0.093\\

\vspace{1mm}

12 & EISJ0951-2145 & 16 & 0.185 &  $ 662$&$^{+ 101}_{- 127}$ & 50.0 & 0.70 & 1.650 &  98.6 & $-$  & $-$ & $-$\\

\vspace{1mm}

13 & EISJ0952-2150 & 12 & 0.183 &  $ 717$&$^{+ 103}_{- 208}$ & 30.3 & 0.51 & $-$   &  $-$  & 1.650 & $>99.9$ & 0.132\\

\vspace{1mm}

14 & EISJ0952-2103 & 18 & 0.236 &  $ 161$&$^{+ 915}_{-  40}$ & 20.0 & 0.30 & $-$   &  $-$  &   1.725 & 99.5 & 0.353\\

\vspace{1mm}

15 & EISJ0952-2144 & 17 & 0.183 &   $ 709$&$^{+  99}_{- 577}$& 25.0 & 0.35 & $-$ &  $-$ &   1.650 & $>99.9$& 0.043\\

\vspace{1mm}

16 & EISJ0952-2018 & 14 & 0.252 &  $ 570$&$^{+  78}_{- 166}$ & 20.0 & 0.37 & $-$   &  $-$ &  1.500 & $>99.9$ & 0.097\\

\vspace{1mm}

17 & EISJ0953-2053 & 12 & 0.235 &  $ 488$&$^{+ 125}_{- 212}$ & 27.2 & 0.52 & $-$ &  $-$ &  1.575 & $>99.9$ & 0.070\\

\vspace{1mm}

18 & EISJ0953-2156 & 6  & 0.181 &  $0$&$^{+ 127}_{- 0}$      & 30.2 & 0.64 & $-$   &  $-$ &  $-$   &  $-$ & $-$\\

\vspace{1mm}

19 & EISJ0953-2017 & 9  & 0.095 &  $ 191$&$^{+  85}_{-  54}$ & 10.7 & 0.35 & $-$   &  $-$ &  $-$   &  $-$ & $-$\\

\vspace{1mm}

20 & EISJ0955-2123 & 16 & 0.203 &  $ 774$&$^{+ 202}_{- 279}$ & 41.8 & 0.74 & 1.575   &  91.4 & 1.575 & $>99.9$ & 0.114 \\

\vspace{1mm}

21 & EISJ0955-2151 & 9  & 0.114 &   $ 163$&$^{+ 238}_{-  28}$& 22.4 & 0.26 & $-$   &  $-$ &  $-$ & $-$ & $-$\\

\vspace{1mm}

22 & EISJ0955-2037 & 6  & 0.283 &  $ 330$&$^{+   7}_{- 213}$ & 39.9 & 0.43 & $-$   &  $-$ &  $-$   &  $-$ & $-$\\

\vspace{1mm}

23 & EISJ0955-2020 & 8  & 0.064 &  $ 406$&$^{+  66}_{- 148}$ & 20.5 & 0.34 & $-$   &  $-$ &  1.125 & $99.7$ &0.161\\

\vspace{1mm}

24 & EISJ0956-2054 & 17 & 0.279 &  $ 962$&$^{+ 114}_{- 282}$ & 42.7 & 0.74 & 1.650   &  90.4 & 1.800 & 99.6 & 0.338\\

\vspace{1mm}

25 & EISJ0957-2051 & 8  & 0.241 &  $ 483$&$^{+  65}_{- 311}$ & 17.0 & 0.37 & $-$   &  $-$ &  $-$   &  $-$ & $-$\\

\vspace{1mm}

26 & EISJ0957-2143 & 16 & 0.202 &  $ 325$&$^{+  61}_{- 107}$ & 32.6 & 0.51 & 1.650 &  93.6 & 1.725 & $>99.9$ & 0.062\\

\vspace{1mm}

27 & EISJ2237-3932 & 35 & 0.244 & $1210$&$^{+ 108}_{- 106}$  & 33.0 & 0.51 & N.A. & N.A. &  N.A. & N.A.  & N.A.\\

\vspace{1mm}

28 & EISJ2241-3949 & 18 & 0.185 & $ 205$&$^{+  59}_{- 107}$  & 44.8 & 0.78 & N.A. & N.A. &  N.A.  & N.A.  & N.A.\\

\vspace{1mm}

29 & EISJ2243-4013 & 4  & 0.183 &   $ 714$&$^{+  48}_{- 345}$& 26.8 & 0.45 & $-$   &  $-$ & $-$   &  $-$ & $-$\\

\vspace{1mm}

30 & EISJ2243-4025 & 18 & 0.246 &  $ 280$&$^{+  51}_{-  69}$ & 32.2 & 0.54 & 1.725 &  97.2 & 1.725 & $>99.9$& 0.037\\

\vspace{1mm}

31 & EISJ2244-3955 & 4 & 0.097 &  $ 416$&$^{+  89}_{- 220}$ &  11.2 & 0.30 & N.A. & N.A. &  N.A. & N.A. & N.A.\\
32 & EISJ2246-4012A& 6  & 0.150 &  $ 860$&$^{+ 140}_{- 395}$ & 21.5 & 0.60 & $-$ &  $-$ & $-$   &  $-$ & $-$\\
\hline
\end{tabular}
\end{center}
\end{table*}

In the previous section we established the existence of systems in
redshift space that we consider confirmations of the EIS clusters. In
this section we will further characterize the 32 confirmed galaxy
overdensities (3 from Paper~I, 9 from paper~II and 20 from the present
work) by establishing the reliability of the redshifts and velocity
dispersions. To characterize the systems in more detail we also
describe their richness and concentration parameters as well as
determine the colours of their galaxy populations. In
Table~\ref{tab:colour} we summarize the properties of the confirmed
systems. The table gives: in Col.~1 a running number identifying the
system; in Col.~2 the name of the cluster field; in Col.~3 the number
of spectroscopic members; in Col.~4 the spectroscopic redshift; in
Col.~5 the velocity dispersion with 68\% bootstrap errors; in Col. 6
updated $\Lambda_{cl, new}$-richnesses as described below; in Col. 7
the concentration index as computed in Sect.~\ref{sec:conc}; in Col.~8
and 9 the colour of the identified photometric red sequence and the
confidence level as described in Sect.~\ref{sec:colours}; in Col.~10
and 11 the colour of the red sequence of the spectroscopic members and
its significance as also described in Sect.~\ref{sec:colours}; in
Col.~12 the measured colour scatter for the spectroscopic members.
For three systems we do not have colour information available, so we
mark the relevant entries by N.A. in the table.

\subsection{Redshifts}

To investigate the reliability of the measured spectroscopic redshifts
for the confirmed clusters, we have compared the results from three
different redshift estimators: the traditional mean redshift, the
median and the biweight location \citep{beers90} of the redshift of
the identified systems. We find that the three estimators give
consistent results  with small deviations of the order $\delta
z\sim0.002$. Hence, we continue to report  mean values in order to
be compatible with our previously published redshifts.

\begin{figure}
\resizebox{0.5\textwidth}{!}{\includegraphics{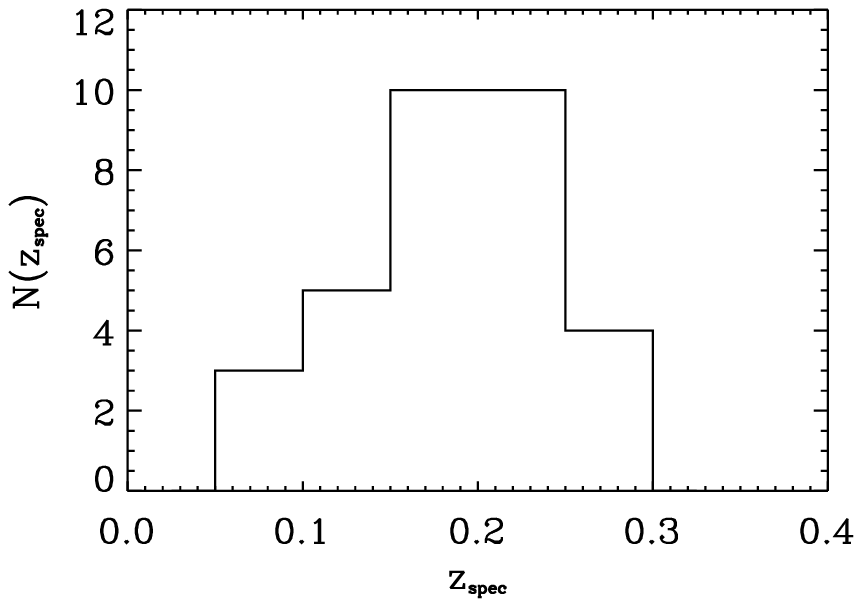}}
\caption{ The distribution of spectroscopic redshifts of all the
confirmed systems.}
\label{fig:redshifts}
\end{figure}

In Fig.~\ref{fig:redshifts} we show the distribution of
redshifts for all confirmed systems.  We find that the average
redshift of the spectroscopically confirmed systems is $\langle z
\rangle = 0.186$ with a standard deviation of
$\sigma_{zspec}=0.058$. This is in good agreement with the matched
filter estimated redshift of $z_{MF}=0.2$ given that the uncertainty
in the estimated redshifts is $\Delta z=0.1$, mainly due to the
spacing between redshift shells in the matched filter detection
procedure.

\subsection{Velocity dispersions}

\begin{figure*}
\begin{center}
\resizebox{0.23\textwidth}{!}{\includegraphics{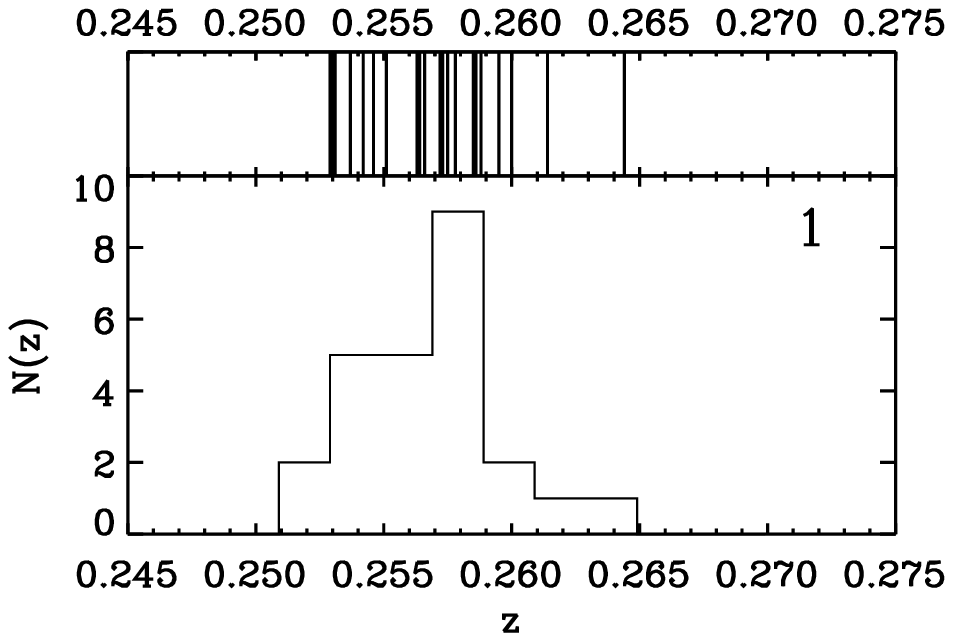}}
\resizebox{0.23\textwidth}{!}{\includegraphics{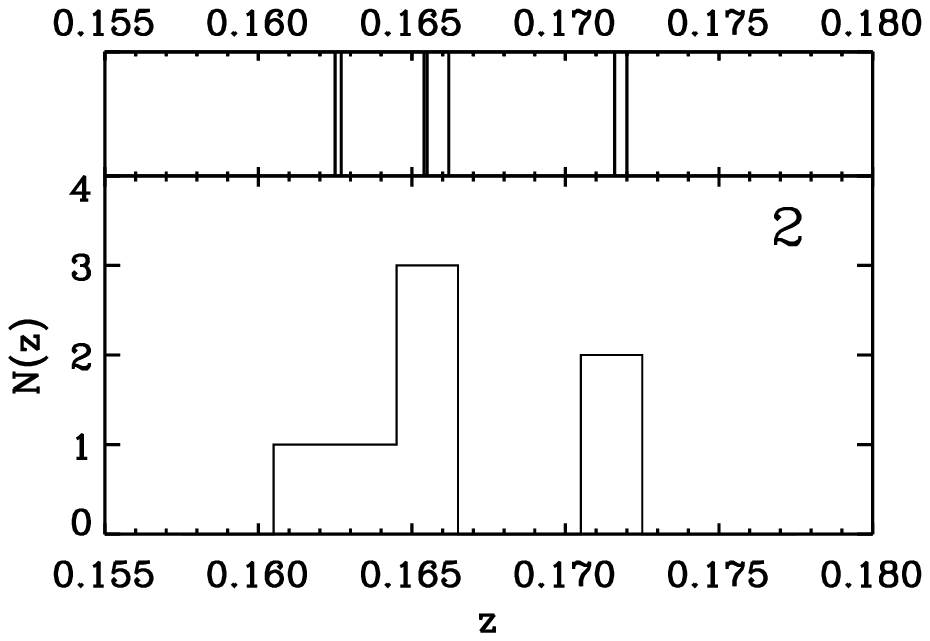}}
\resizebox{0.23\textwidth}{!}{\includegraphics{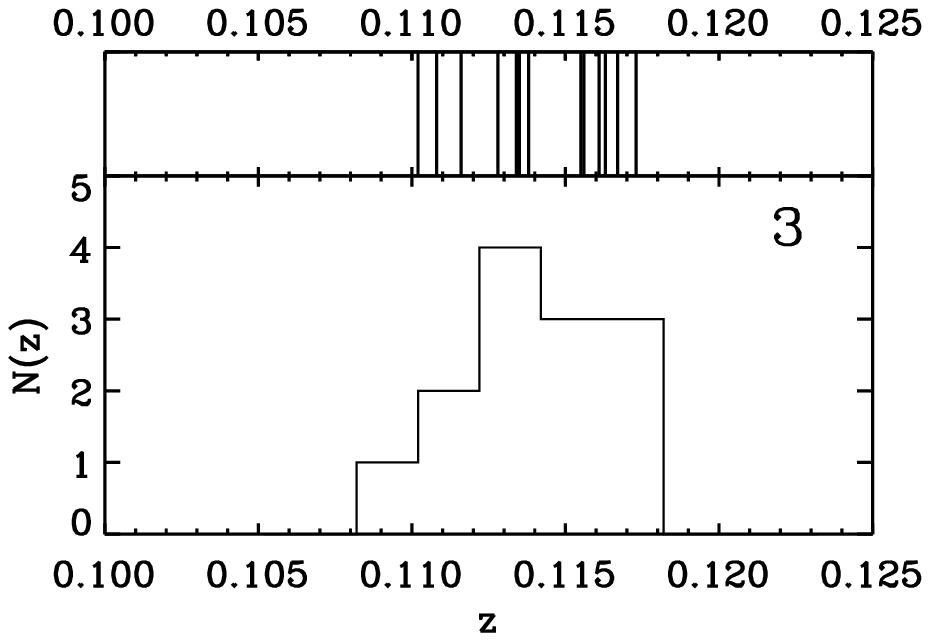}}
\resizebox{0.23\textwidth}{!}{\includegraphics{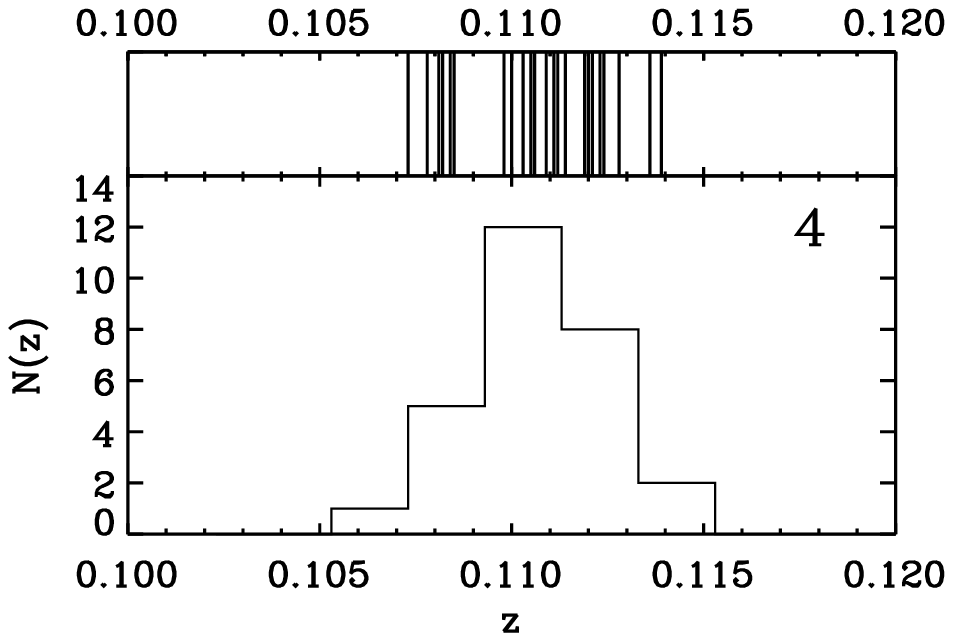}}

\vspace{-1mm}

\resizebox{0.23\textwidth}{!}{\includegraphics{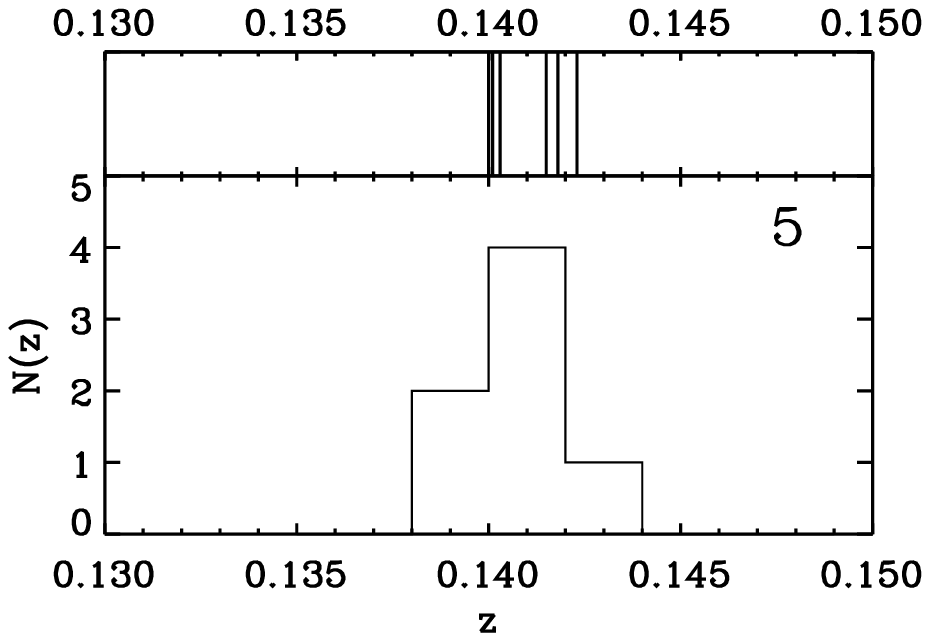}}
\resizebox{0.23\textwidth}{!}{\includegraphics{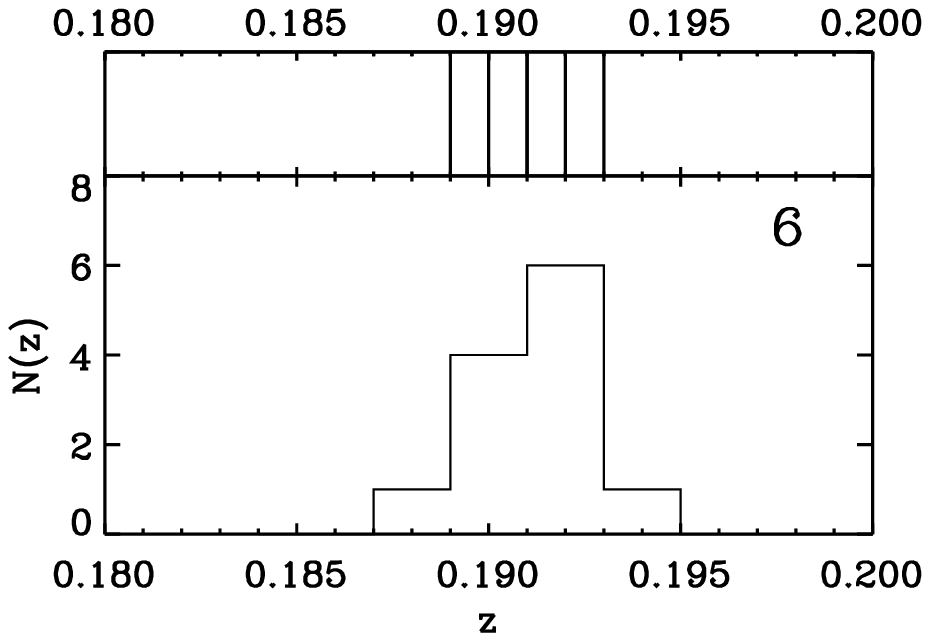}}
\resizebox{0.23\textwidth}{!}{\includegraphics{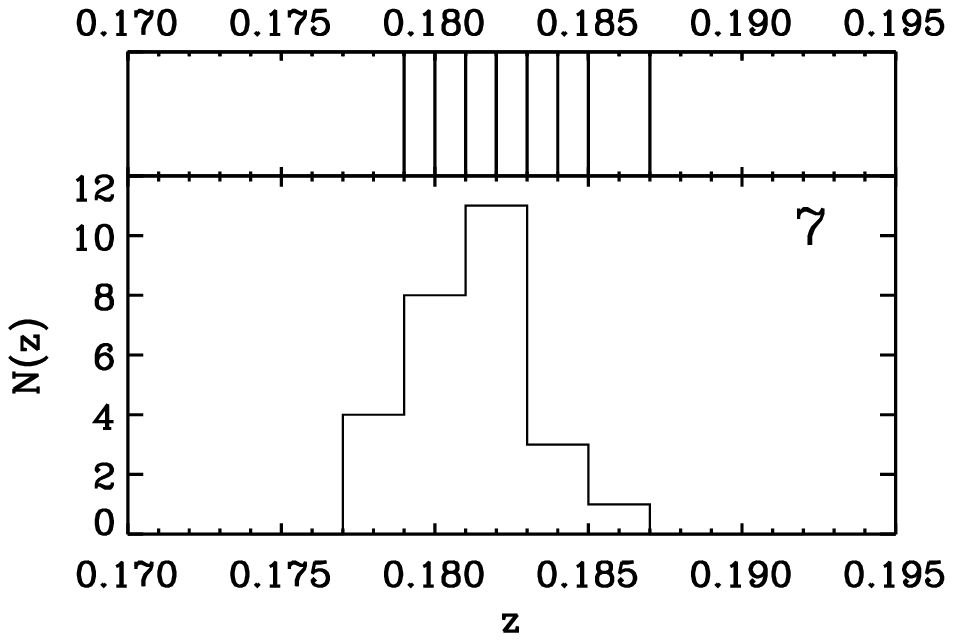}}
\resizebox{0.23\textwidth}{!}{\includegraphics{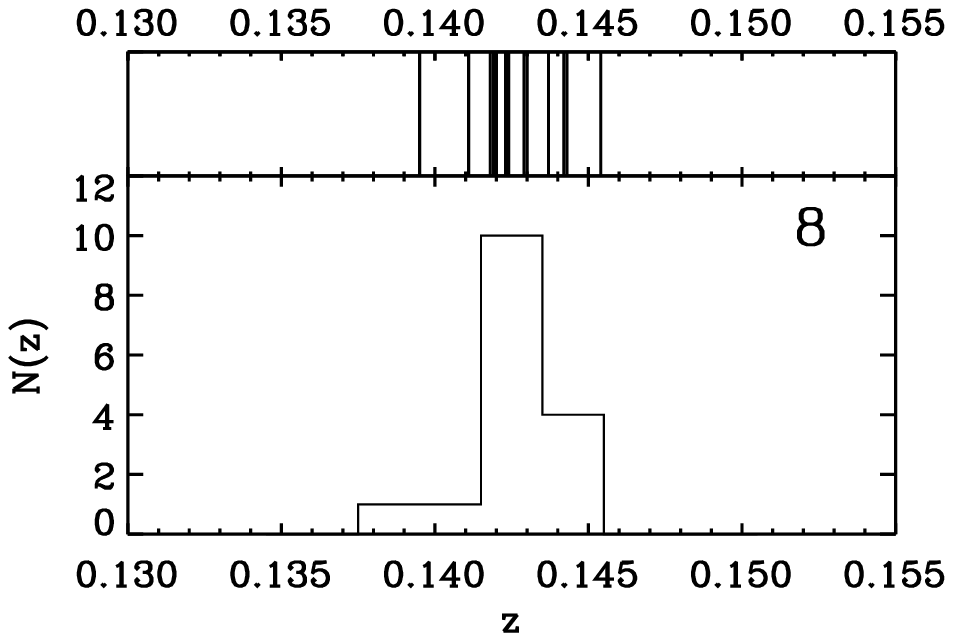}}

\vspace{-1mm}

\resizebox{0.23\textwidth}{!}{\includegraphics{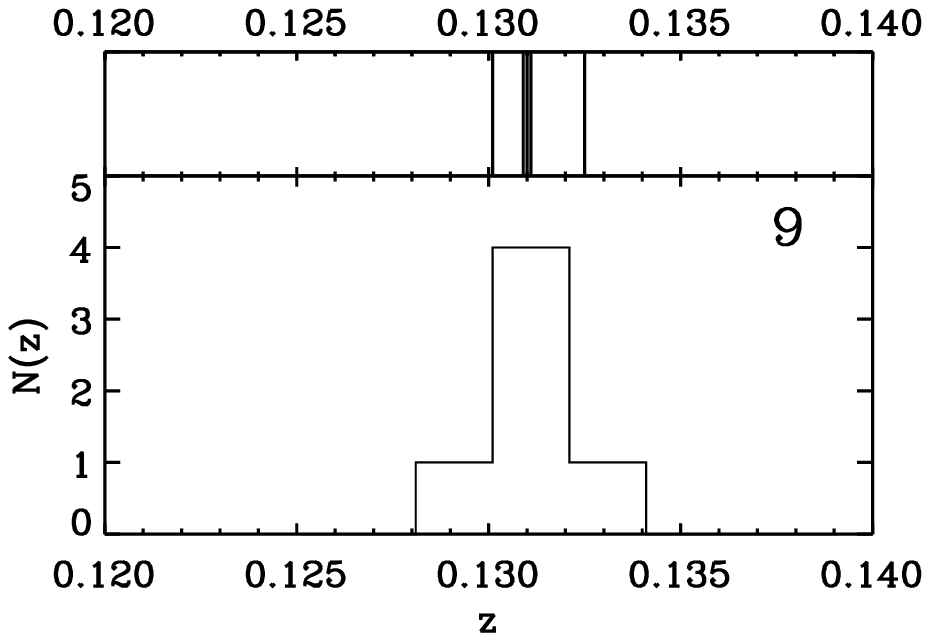}}
\resizebox{0.23\textwidth}{!}{\includegraphics{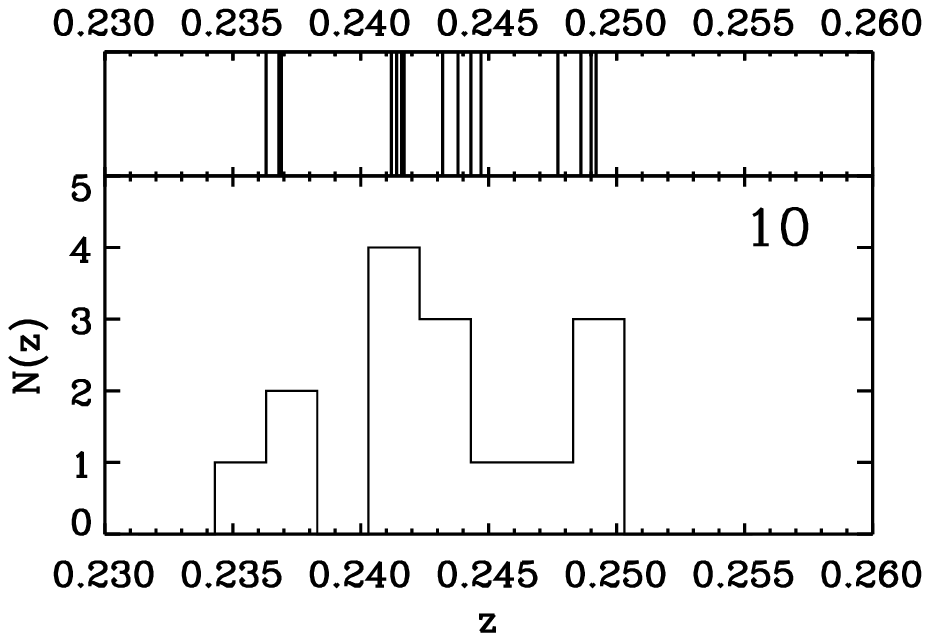}}
\resizebox{0.23\textwidth}{!}{\includegraphics{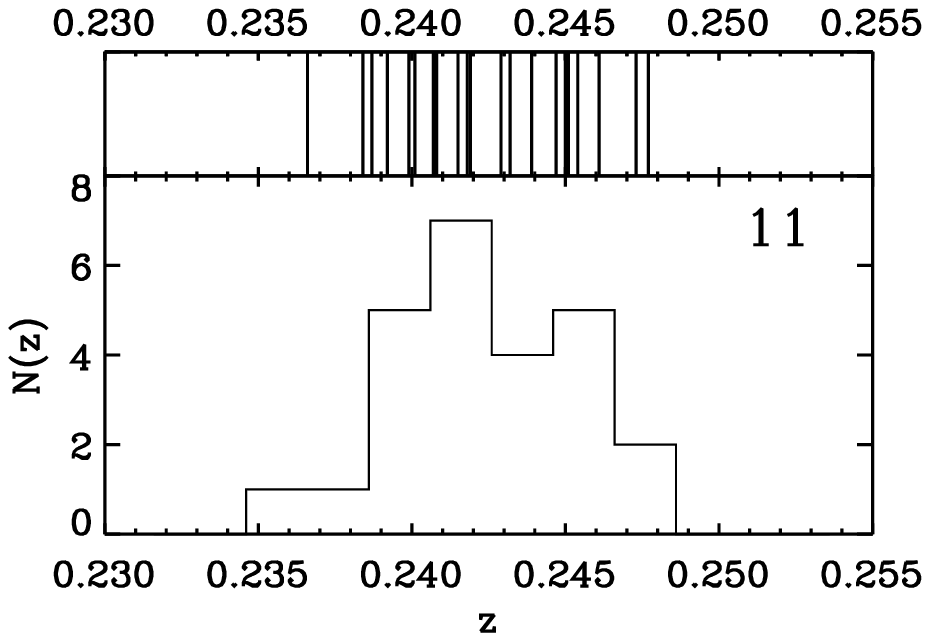}}
\resizebox{0.23\textwidth}{!}{\includegraphics{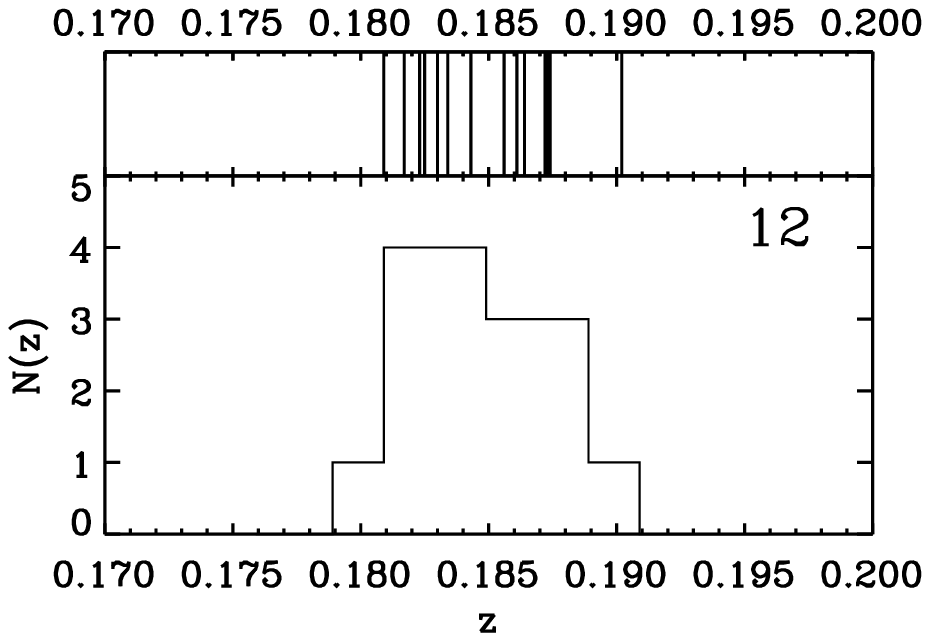}}

\vspace{-1mm}

\resizebox{0.23\textwidth}{!}{\includegraphics{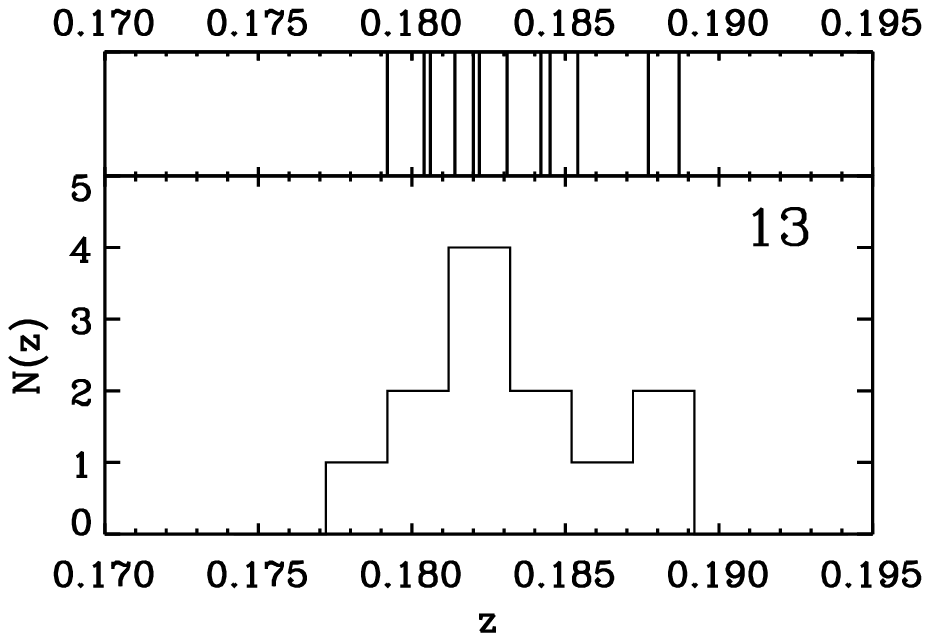}}
\resizebox{0.23\textwidth}{!}{\includegraphics{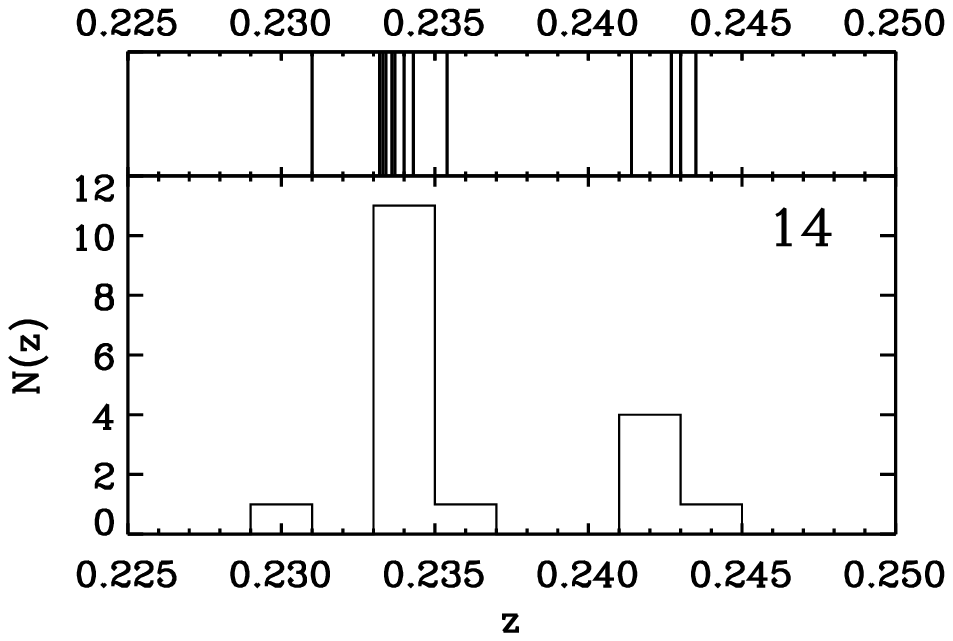}}
\resizebox{0.23\textwidth}{!}{\includegraphics{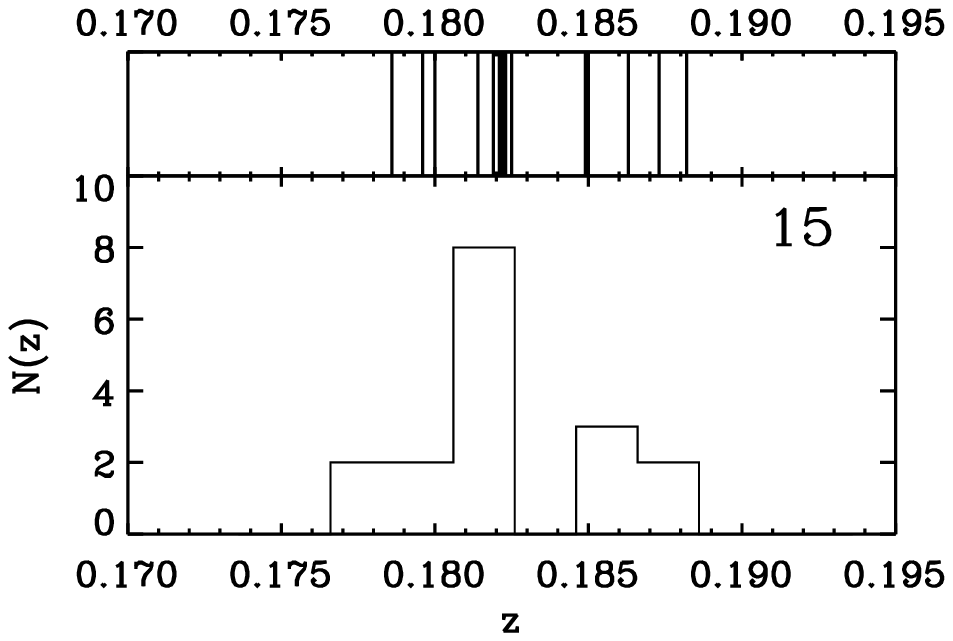}}
\resizebox{0.23\textwidth}{!}{\includegraphics{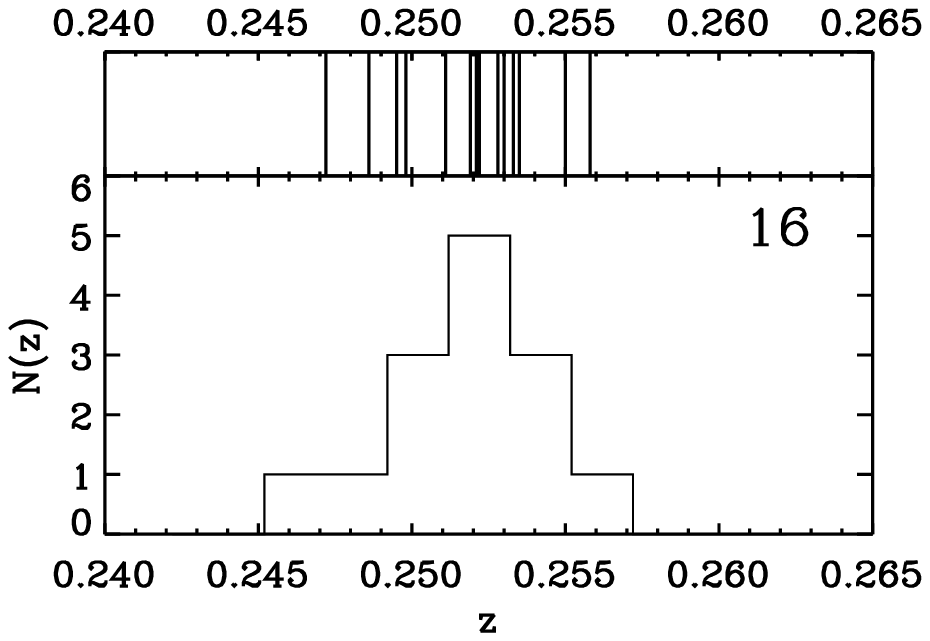}}

\vspace{-1mm}

\resizebox{0.23\textwidth}{!}{\includegraphics{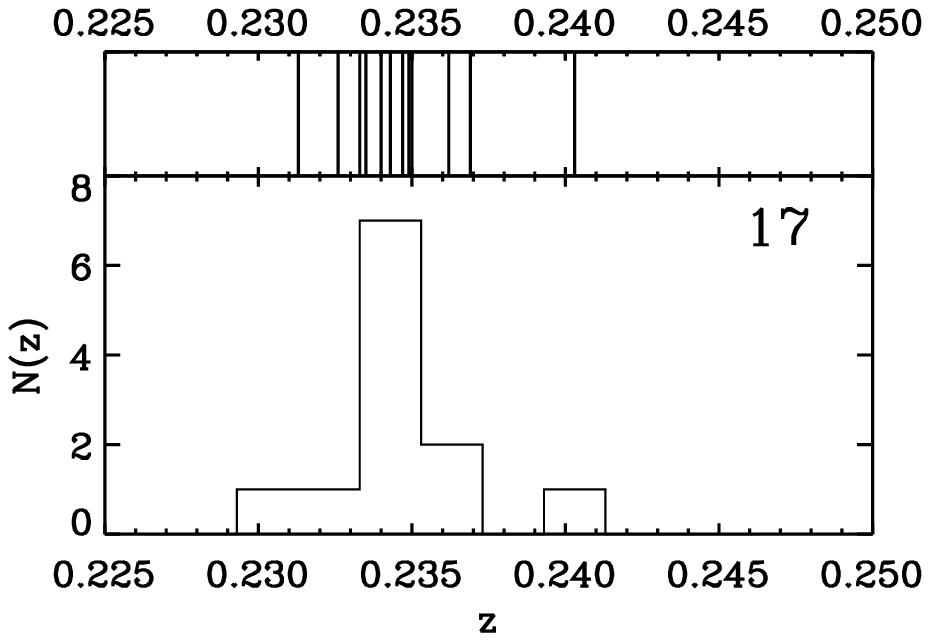}}
\resizebox{0.23\textwidth}{!}{\includegraphics{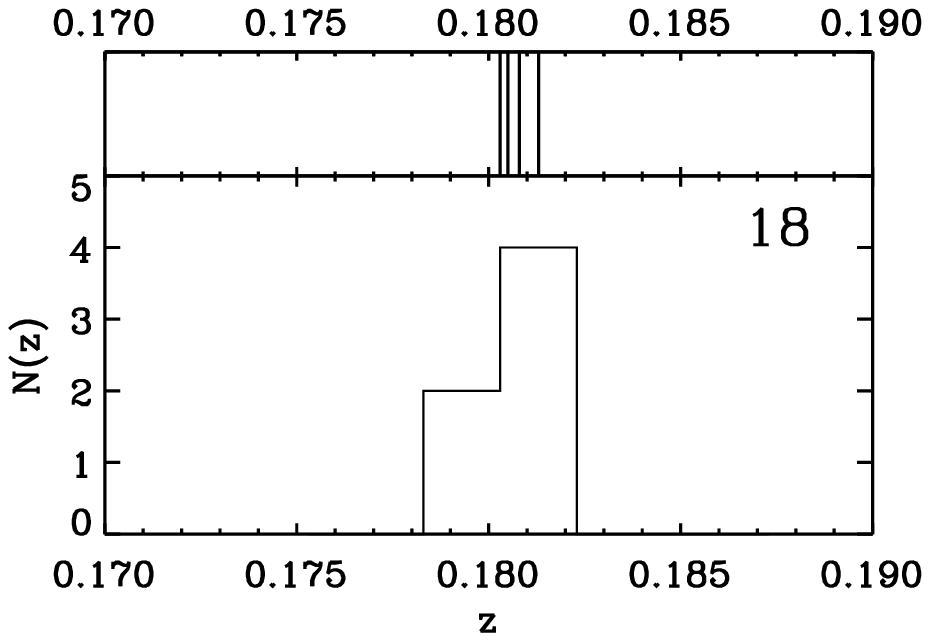}}
\resizebox{0.23\textwidth}{!}{\includegraphics{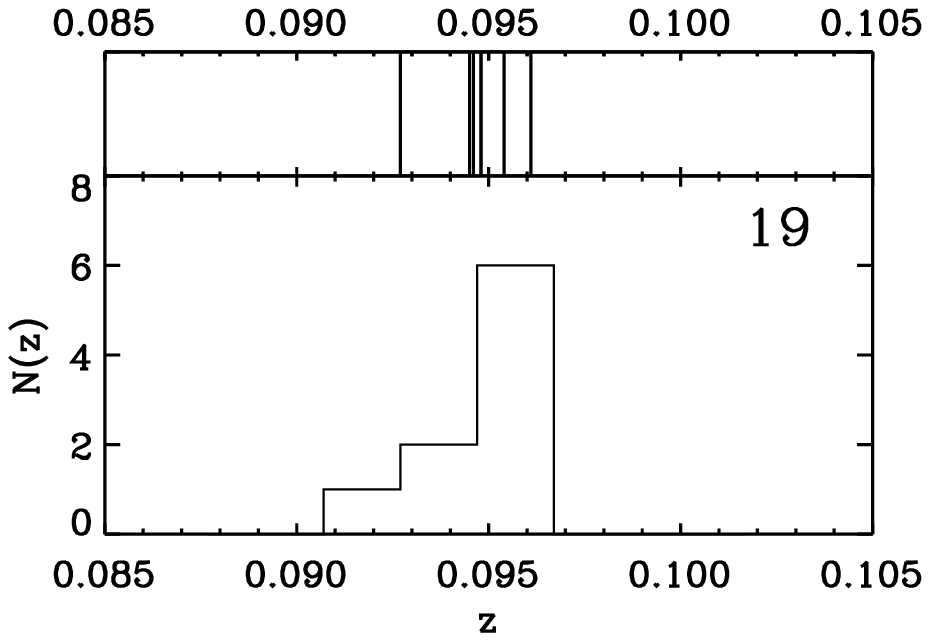}}
\resizebox{0.23\textwidth}{!}{\includegraphics{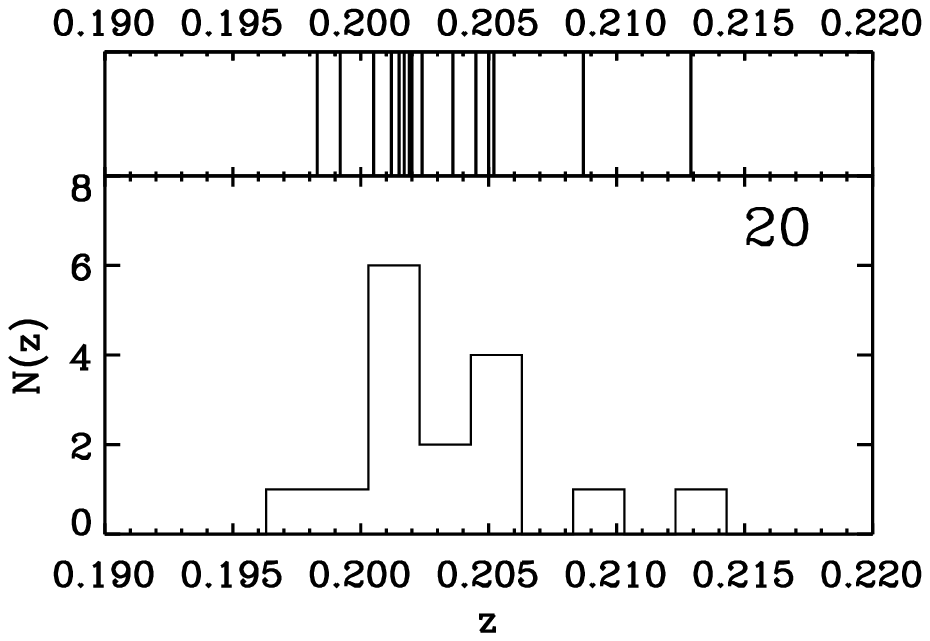}}

\vspace{-1mm}

\resizebox{0.23\textwidth}{!}{\includegraphics{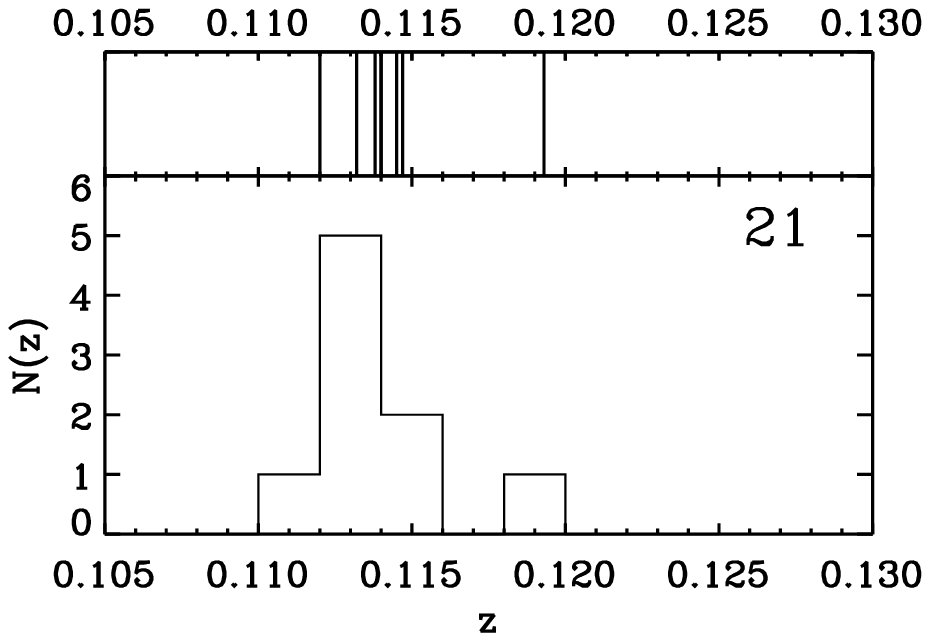}}
\resizebox{0.23\textwidth}{!}{\includegraphics{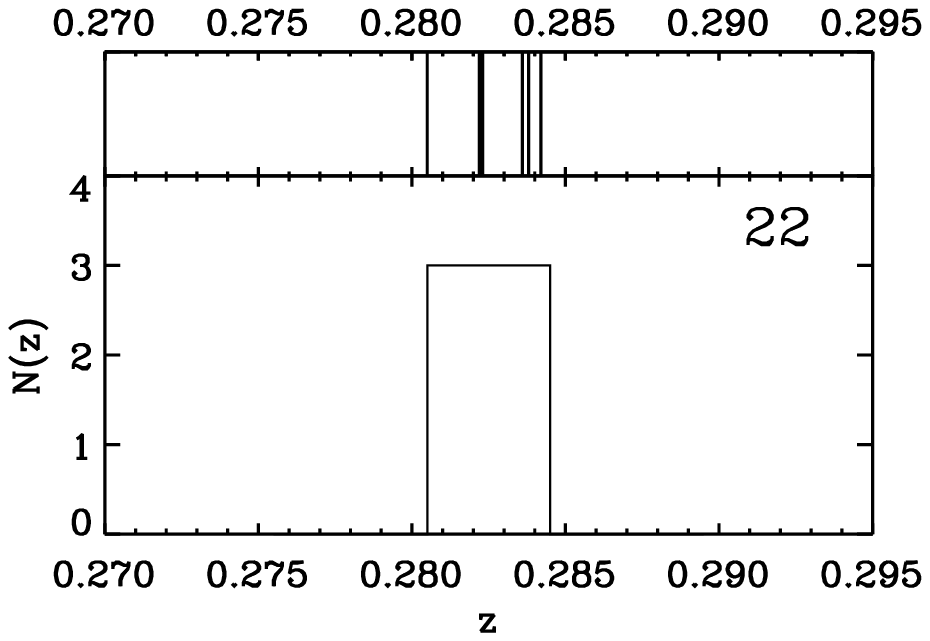}}
\resizebox{0.23\textwidth}{!}{\includegraphics{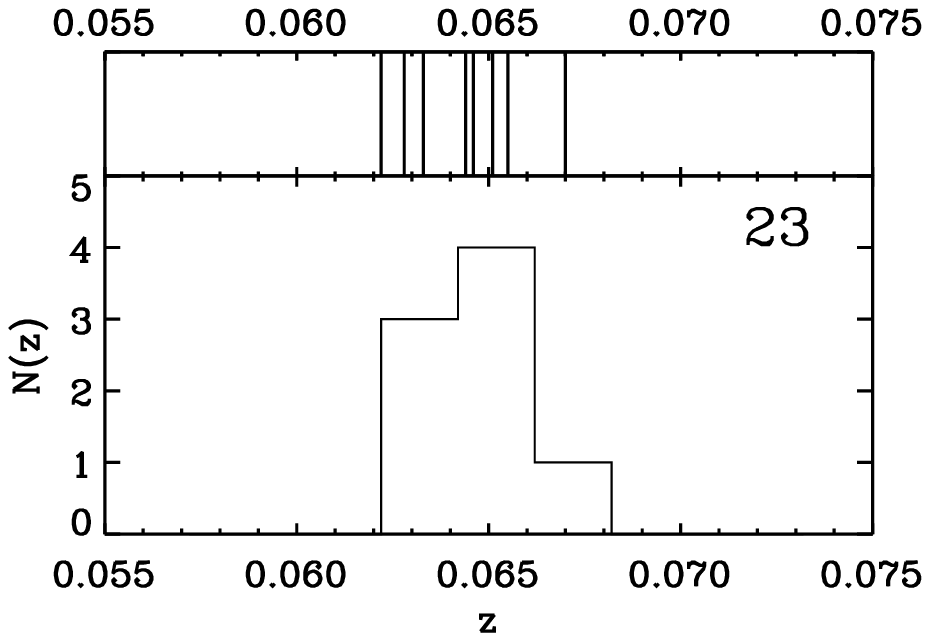}}
\resizebox{0.23\textwidth}{!}{\includegraphics{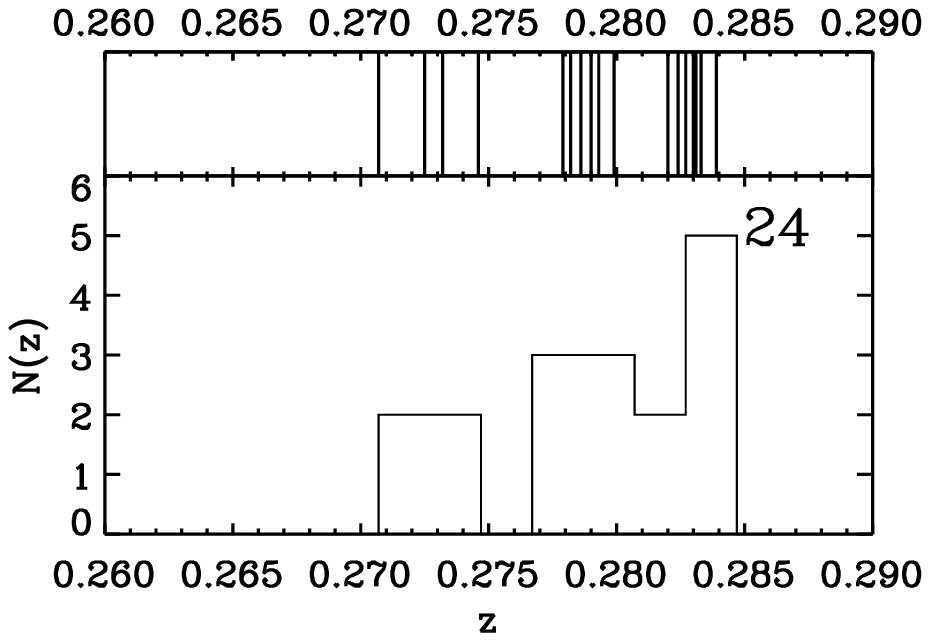}}

\vspace{-1mm}

\resizebox{0.23\textwidth}{!}{\includegraphics{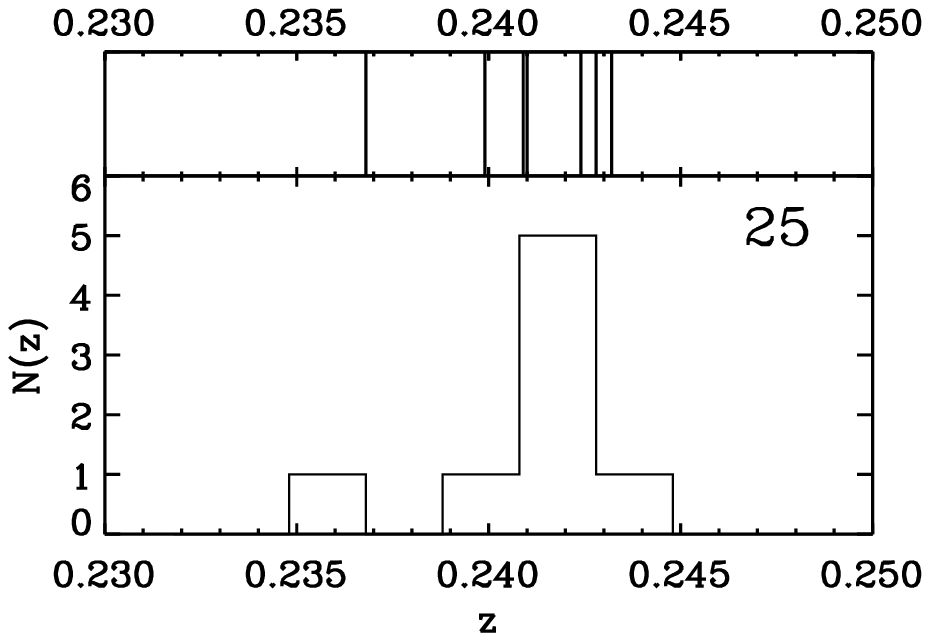}}
\resizebox{0.23\textwidth}{!}{\includegraphics{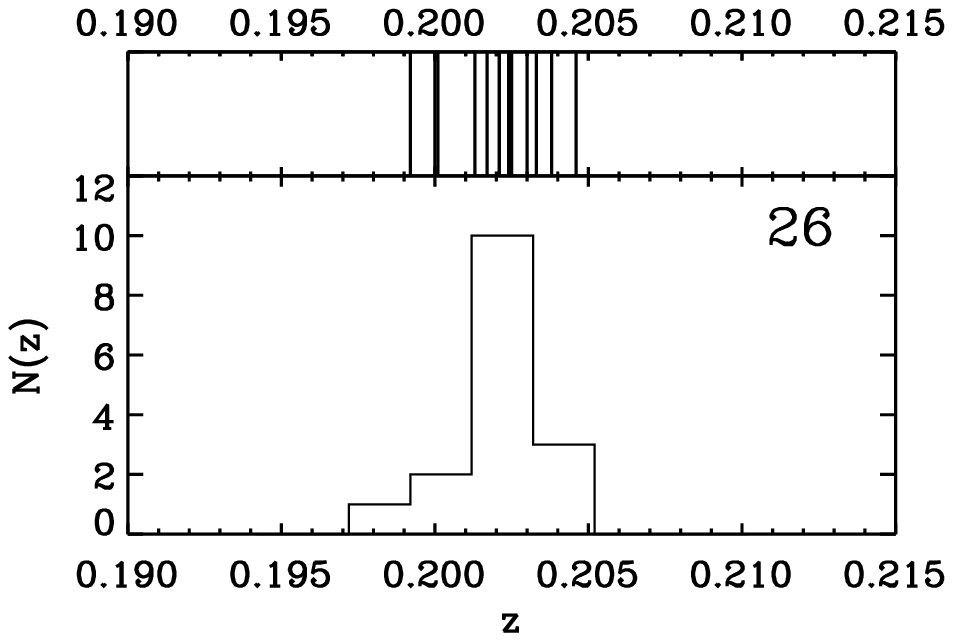}}
\resizebox{0.23\textwidth}{!}{\includegraphics{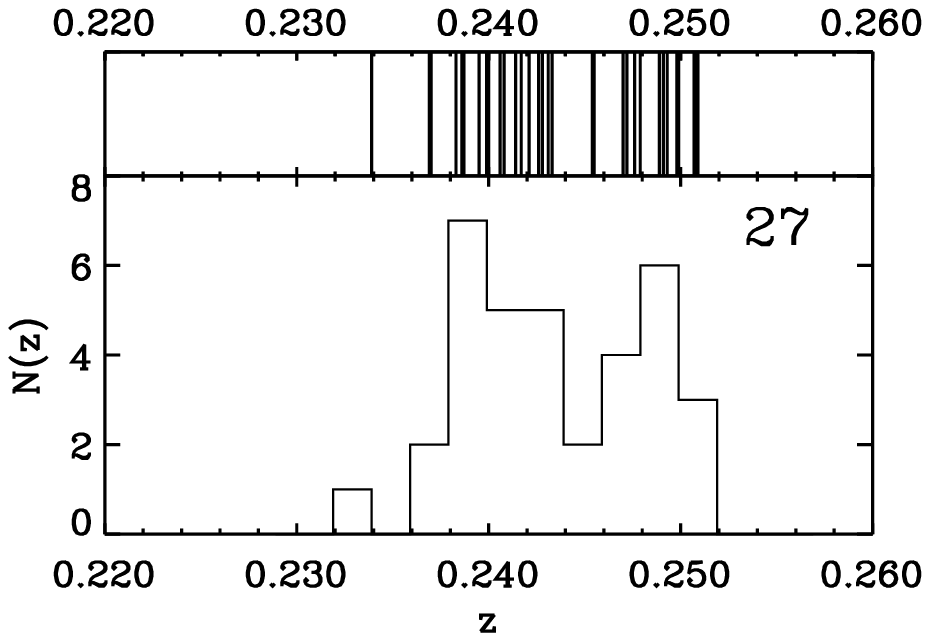}}
\resizebox{0.23\textwidth}{!}{\includegraphics{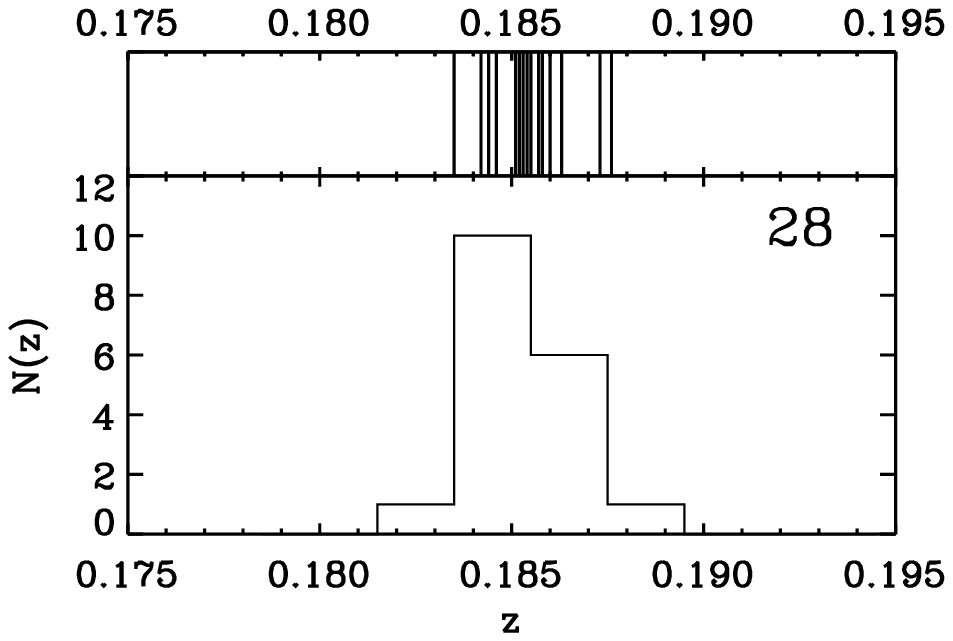}}

\vspace{-1mm}

\resizebox{0.23\textwidth}{!}{\includegraphics{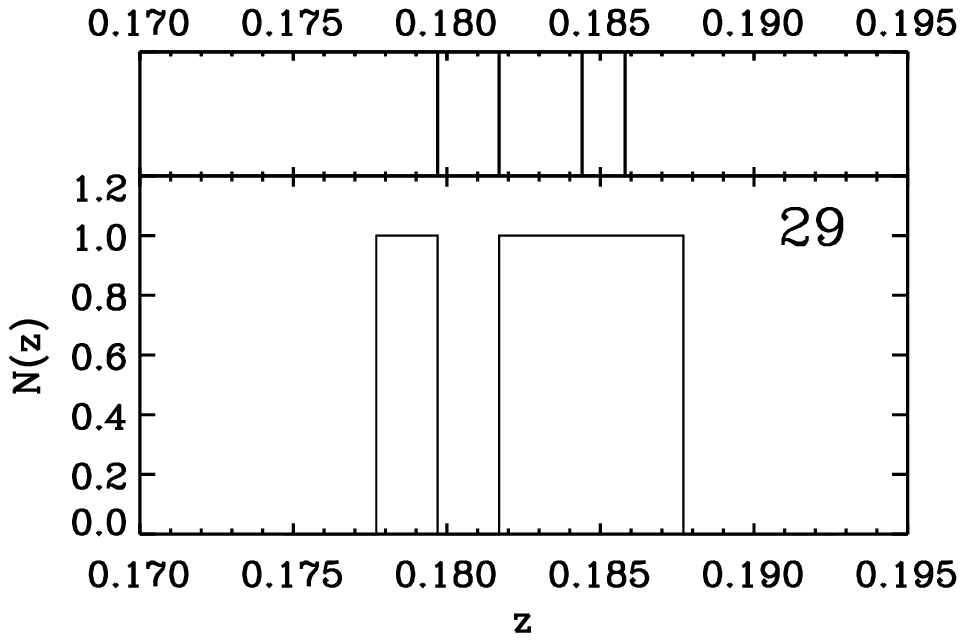}}
\resizebox{0.23\textwidth}{!}{\includegraphics{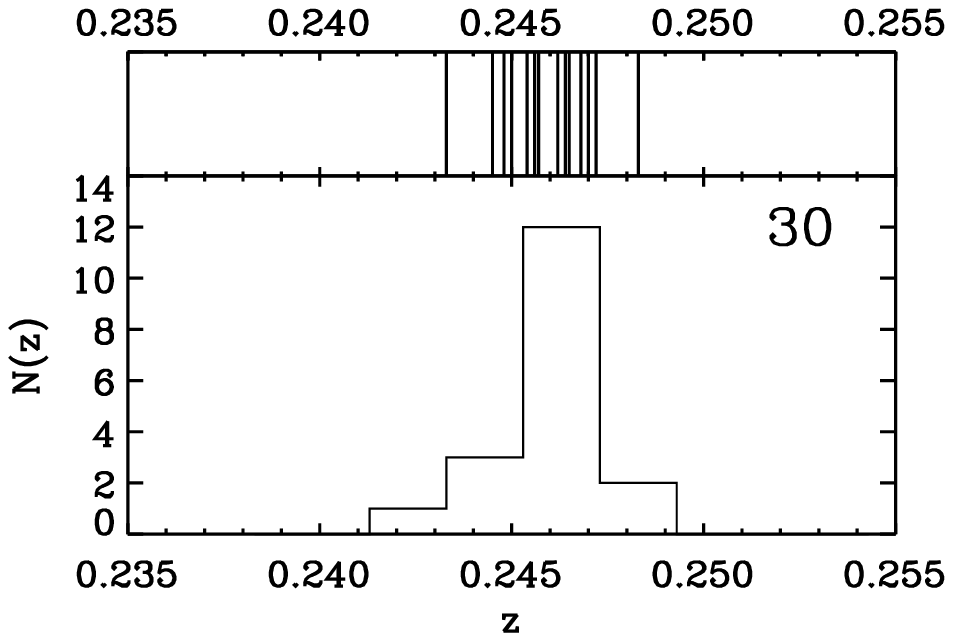}}
\resizebox{0.23\textwidth}{!}{\includegraphics{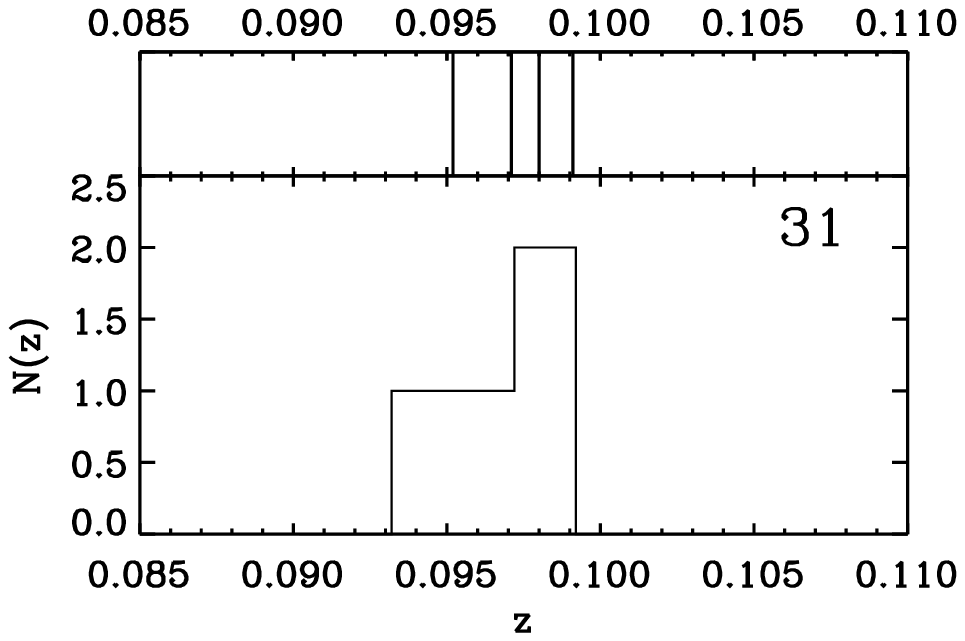}}
\resizebox{0.23\textwidth}{!}{\includegraphics{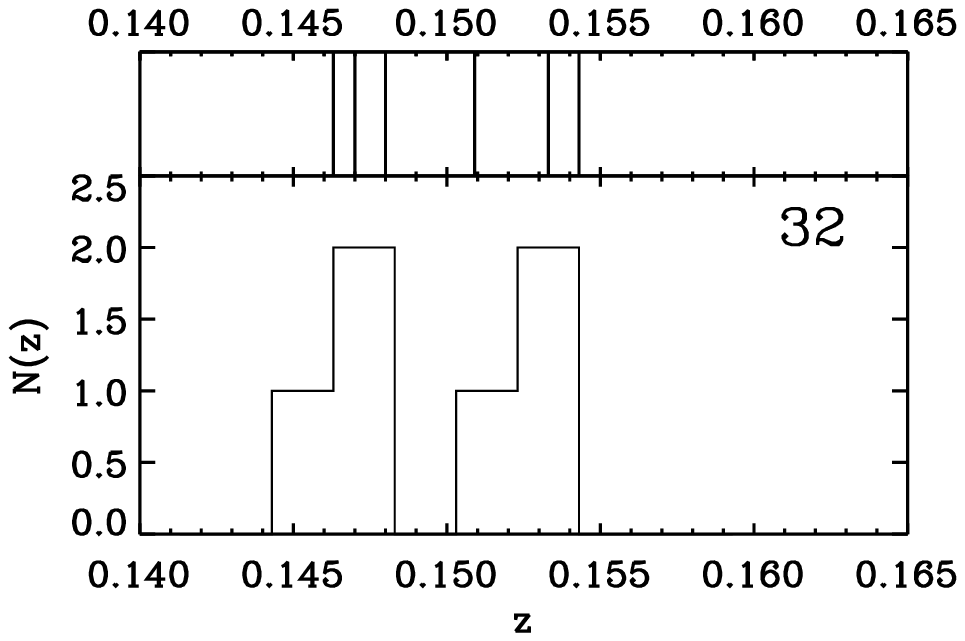}}

\vspace{-1mm}

\caption{Detailed redshift diagrams for all spectroscopically
confirmed systems. The identification numbers in each panel refer to
Table~\protect\ref{tab:colour}. Upper panels show a bar diagram of the
redshifts of cluster members. The lower part shows the redshift
histogram with a bin size of $\Delta z=0.002$.}
\label{fig:redshift_detail}
\end{center}
\end{figure*}

The velocity dispersion of galaxy systems is an important indicator of
their dynamical state. In Fig.~\ref{fig:redshift_detail} we show the
redshift distributions for each confirmed system.  From these
distributions it is clear that some systems have a small number of
members, while others show signs of substructure.  Therefore, it is
important to determine the velocity dispersion using an estimator that
is robust both for small samples and with respect to outliers.
\cite{beers90} investigated the problem in detail  and suggested
the use of the biweight scale or gapper estimators. 

We have compared the estimates of the velocity dispersion for our
systems using these two estimators as well as the traditional standard
deviation.  In general, the velocity dispersions estimated by the
different methods are consistent with the differences of the
order $\lesssim50\mathrm{km/s}$, much smaller than the
bootstrap-estimated errors.  However, in two cases the biweight
estimator gives significantly smaller values than the other two
(though with large errors and still the values agree within the
errors). These two cases are EISJ0952-2103 (panel 14 of
Fig.~\ref{fig:redshift_detail}) and EISJ0955-2151 (panel 21) with the
traditional standard deviation giving estimates that are significantly
higher.

Inspecting the redshift distribution of EISJ0952-2103 we find that the
18 members split into three groups containing one, five and twelve
members, respectively.  We find that the biweight estimator reflects
the velocity dispersion of the largest, and most concentrated group,
while the other methods are more sensitive to the outlying members.
The large error given in Table~\ref{tab:colour} probably reflects the
presence of the outliers.


The other case is EISJ0955-2151 (panel 21) with only 9 members, one
being an outlier, which explains the large discrepancy reported above
and the large error in Table~\ref{tab:colour}. It is worth
emphasizing, that this system is one of those we, in
Sect.~\ref{sec:z_identification}, found to be affected by projection
effects.

Inspecting Fig.~\ref{fig:redshift_detail} we also find outliers in
panels~15, 20 and 25. However, in these cases the different velocity
dispersion estimates are not as discrepant as in the aforementioned
cases, but large errors, comparable to those discussed above, reflect
the presence of outliers.

\begin{figure}
\resizebox{\columnwidth}{!}{\includegraphics{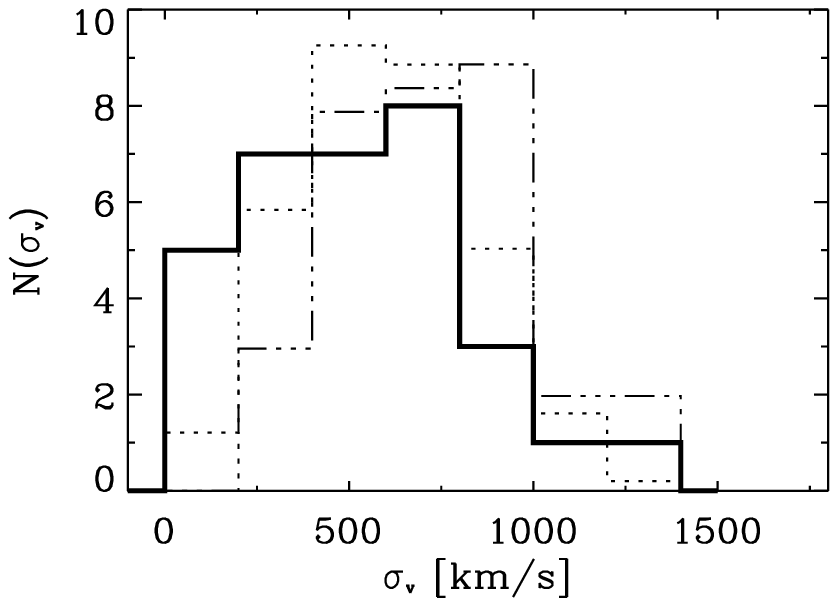}}
\caption{Distribution of velocity dispersions (solid line) compared
with those  of \cite{fadda96} -- dotted histogram -- and those of
\cite{zabludoff90} -- triple dot-dashed histogram.}
\label{fig:veldisp_dist}
\end{figure}

Hereafter, we adopt the biweight method to compute the velocity
dispersions listed in Col.~5 of Table~\ref{tab:colour}  with 68\%
bootstrap errors.  The distribution of these velocity dispersions
(solid line) is shown in Fig.~\ref{fig:veldisp_dist}. This
distribution is compared to those of \cite{fadda96} and
\cite{zabludoff90}, which will be discussed in
Sect.~\ref{sec:discussion}. As seen from this figure and
Table~\ref{tab:colour} we find systems with very low velocity
dispersions. In particular, the clusters EISJ0950-2133 (panel~9) and
EISJ0953-2156 (panel~18) have implausibly low velocity
dispersions,  possibly indicating severe undersampling.


\subsection{Richness}

\begin{figure}
\resizebox{\columnwidth}{!}{\includegraphics{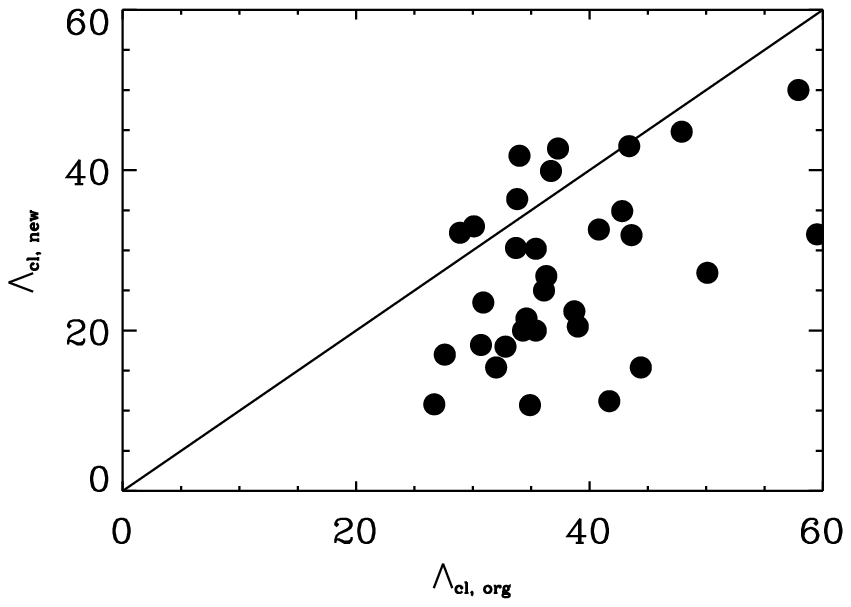}}
\caption{The relation between the original matched filter estimated
$\Lambda_{cl, org}$ richnesses and the new ones, $\Lambda_{cl, new}$.
The solid line marks $\Lambda_{cl, org}=\Lambda_{cl, new}$.}
\label{fig:rich_org_new}
\end{figure}

The matched filter algorithm also provides a measure of the richness
($\Lambda_{cl}$) for the cluster candidates based on the estimated
redshifts, where the $\Lambda_{cl}$-richness is equivalent to the
number of $L^*$-galaxies {contributing to the matched filter signal of
a particular detection.}  This computation depends on the apparent
Schechter magnitude and angular extent of the cluster, which at these
redshifts vary rapidly. Therefore, even though the spectroscopic and
estimated redshifts are in good agreement we have to recompute the
cluster richness using the assigned spectroscopic redshift.  The new
richness values ($\Lambda_{cl, new}$) are listed in
Table~\ref{tab:colour} and compared with the original estimates
($\Lambda_{cl, org}$) in Fig.~\ref{fig:rich_org_new}.  In general the
new richnesses are smaller than the original ones.  This is in good
agreement with the, on average, lower  redshift relative to the
$z_{MF}$ estimate since the expected apparent Schechter magnitude is
now brighter and thus the observed cluster luminosity corresponds to
fewer $L^*$-galaxies. The lower redshift also causes a larger region
to be used for the richness measurement, but the added luminosity from
those large cluster-centric distances is small.

\begin{figure}
\resizebox{\columnwidth}{!}{\includegraphics{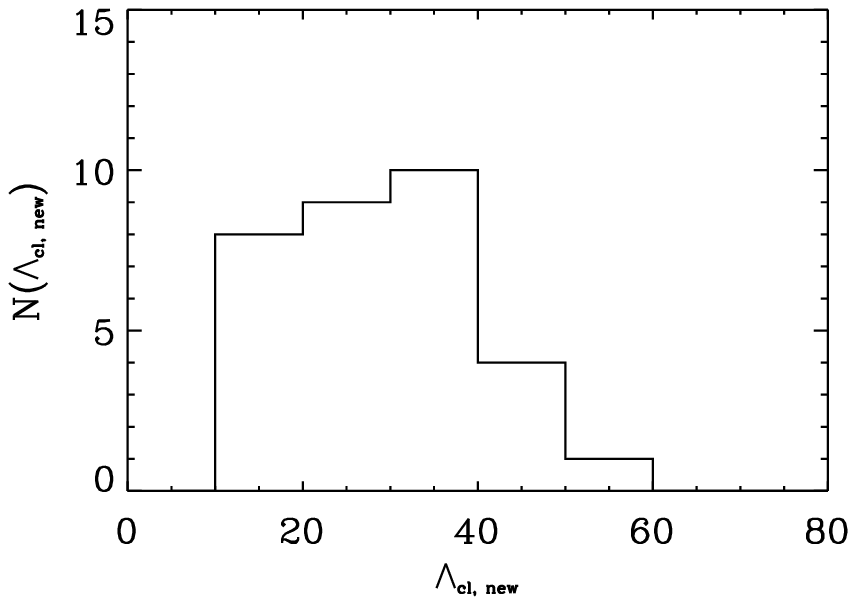}}
\caption{The distribution of $\Lambda_{cl, new}$-richnesses.}
\label{fig:lambda_dist}
\end{figure}

Finally, in Fig.~\ref{fig:lambda_dist} we show the distribution of
$\Lambda_{cl,new}$-richnesses that covers the range $10\lesssim
\Lambda_{cl, new} \lesssim 50$.  This range corresponds to Abell
richness class $\leq1$, typical of poor galaxy clusters.  This is not
surprising considering the relatively small volume surveyed for
clusters at the low redshift range considered here.

\subsection{Concentration}
\label{sec:conc}

Using the definition by \cite{butcher84} we have computed
concentration indices for all the clusters. The concentration index is
defined as $C=\log\left(R_{60}/R_{20}\right)$, where $R_n$ is the
radius encircling n\% of the galaxies. Butcher \& Oemler found that,
in the local Universe, clusters with $C\gtrsim0.4$ were centrally
concentrated and dominated by ellipticals, while those with $C\sim0.3$
are closer to uniform-density spheres and would be dominated by
spirals.

\begin{figure}
\resizebox{\columnwidth}{!}{\includegraphics{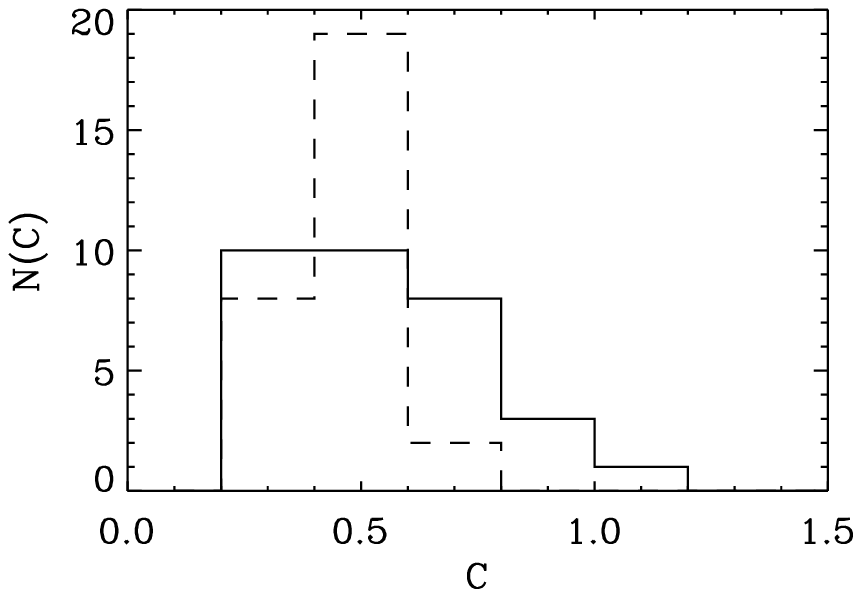}}
\caption{The distribution of concentration indices for the confirmed
systems (solid line) compared with the distribution from
\cite{butcher84} given by the dashed histogram.}
\label{fig:concentration_bo}
\end{figure}

We compute the concentration index based on the background corrected
galaxy counts within $2h_{75}^{-1}\mathrm{Mpc}$ from the cluster
center with $I\leq I^*(z)+2$, where $I^*(z)$ is the expected I-band
Schechter magnitude at the redshift of the cluster (see
Sect.~\ref{sec:compl}). The background correction is based on
galaxies with magnitudes in the same range and clustercentric distance
between $2h_{75}^{-1}\mathrm{Mpc}$ and $3h_{75}^{-1}\mathrm{Mpc}$. The
computed concentration indices are given in Table~\ref{tab:colour} and
their distribution (solid line) is shown in
Fig.~\ref{fig:concentration_bo}. For comparison, the distribution of
concentration indices found by Butcher \& Oemler is also shown and
will be discussed in more detail in Sect.~\ref{sec:discussion}.  The
EIS systems cover a broad range of concentration indices from
`` uniform'' to highly concentrated systems.

\subsection{Colour of the galaxy population}
\label{sec:colours}

\subsubsection{Detection method}

\begin{figure*}
\begin{center}
\resizebox{0.3\textwidth}{!}{\includegraphics{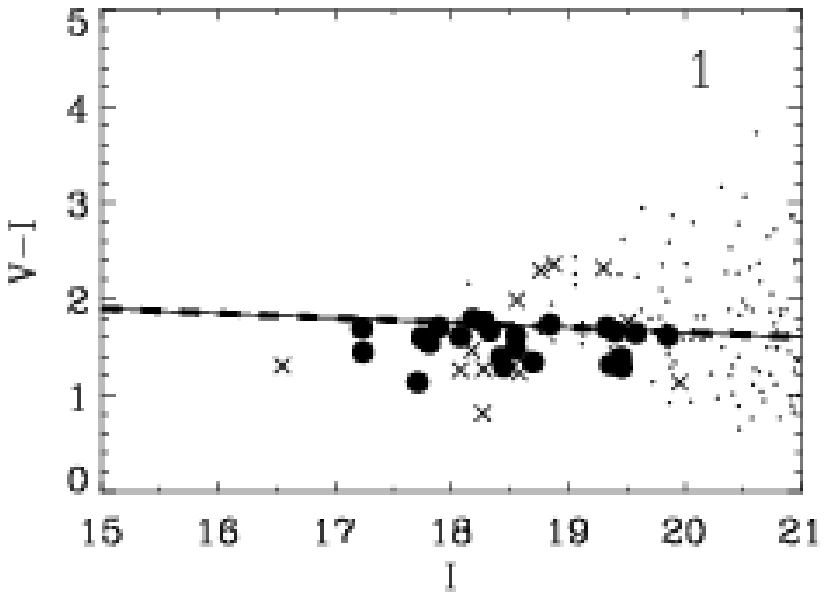}}
\resizebox{0.3\textwidth}{!}{\includegraphics{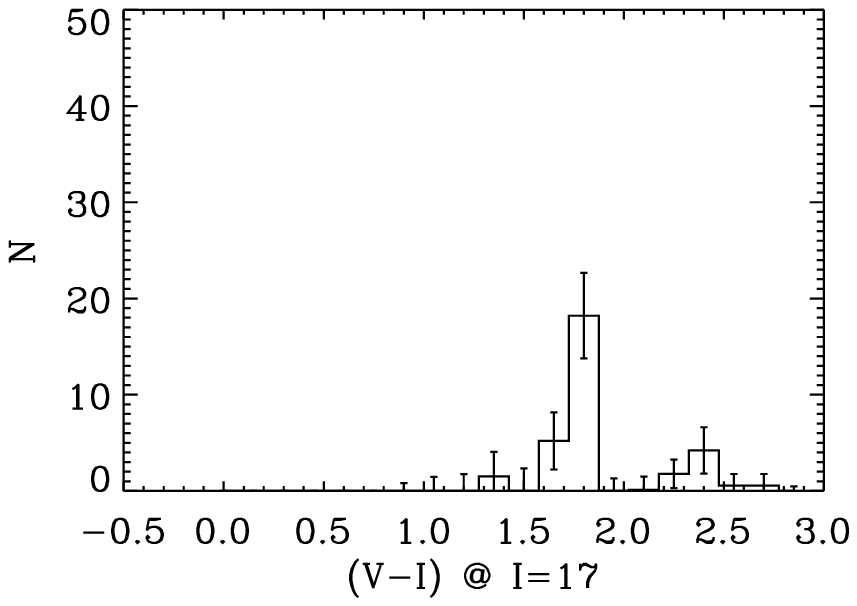}}
\resizebox{0.3\textwidth}{!}{\includegraphics{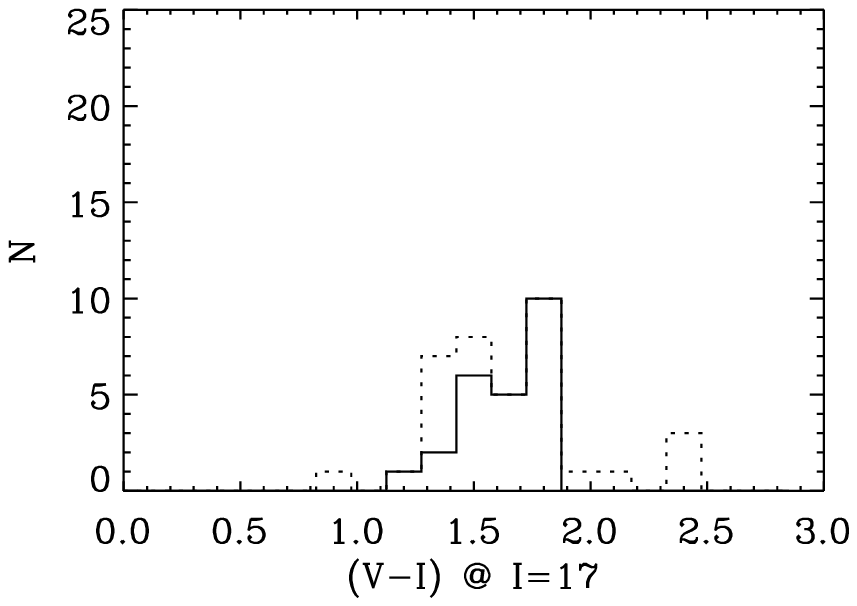}}
\vspace{4mm}

\resizebox{0.3\textwidth}{!}{\includegraphics{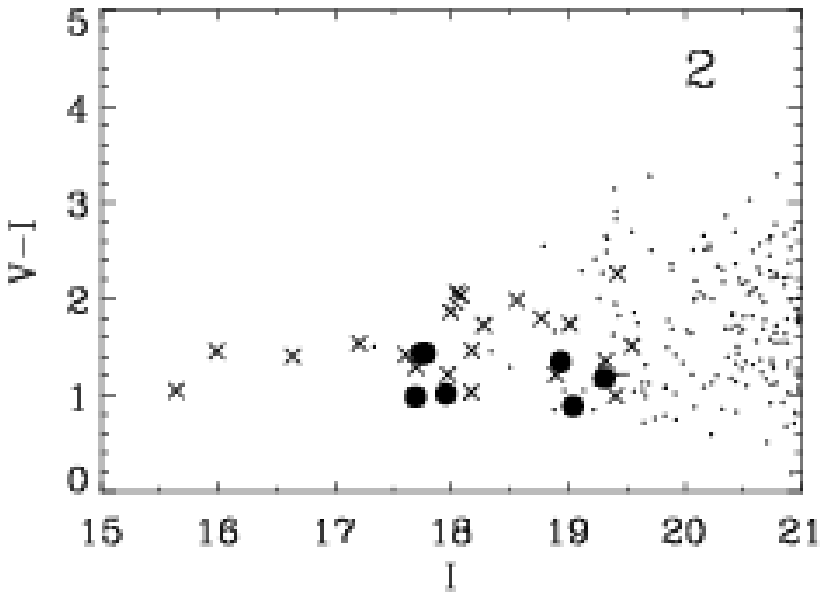}}
\resizebox{0.3\textwidth}{!}{\includegraphics{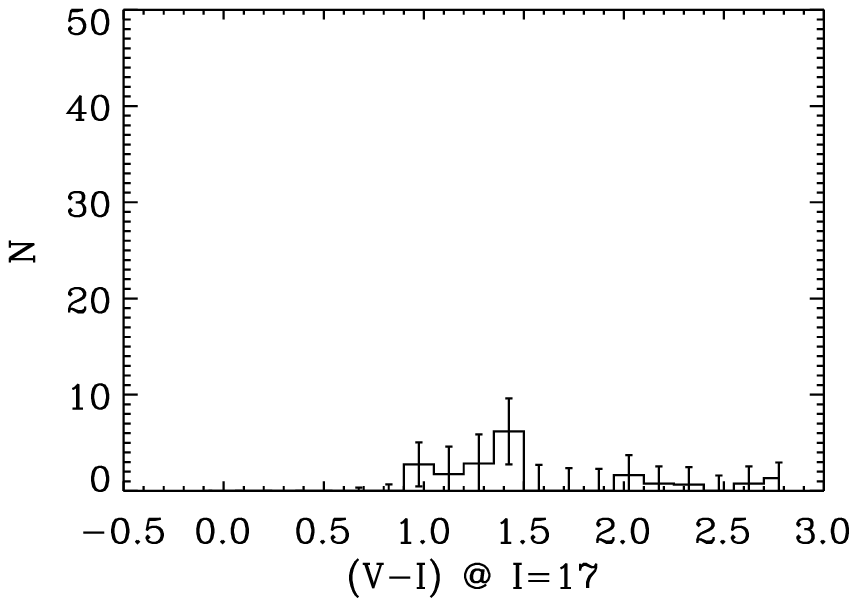}}
\resizebox{0.3\textwidth}{!}{\includegraphics{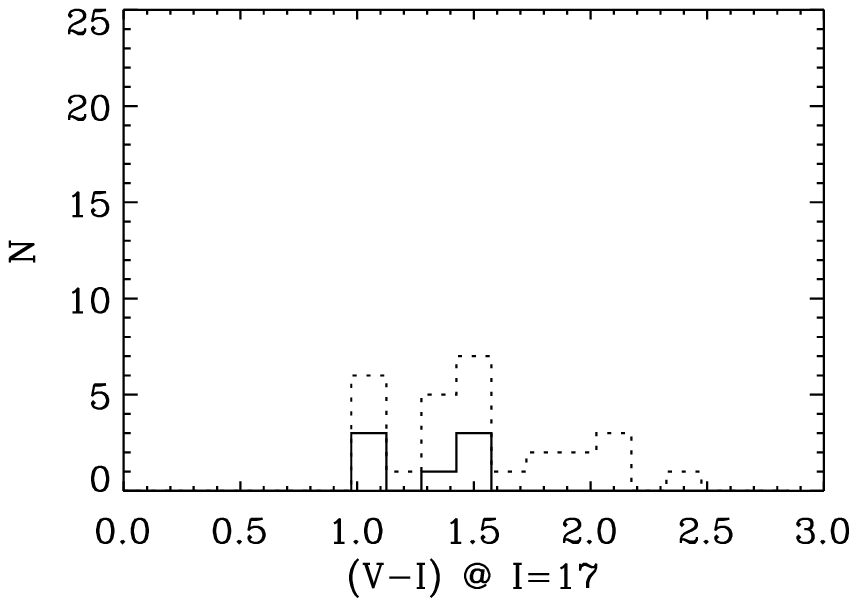}}
\vspace{4mm}

\resizebox{0.3\textwidth}{!}{\includegraphics{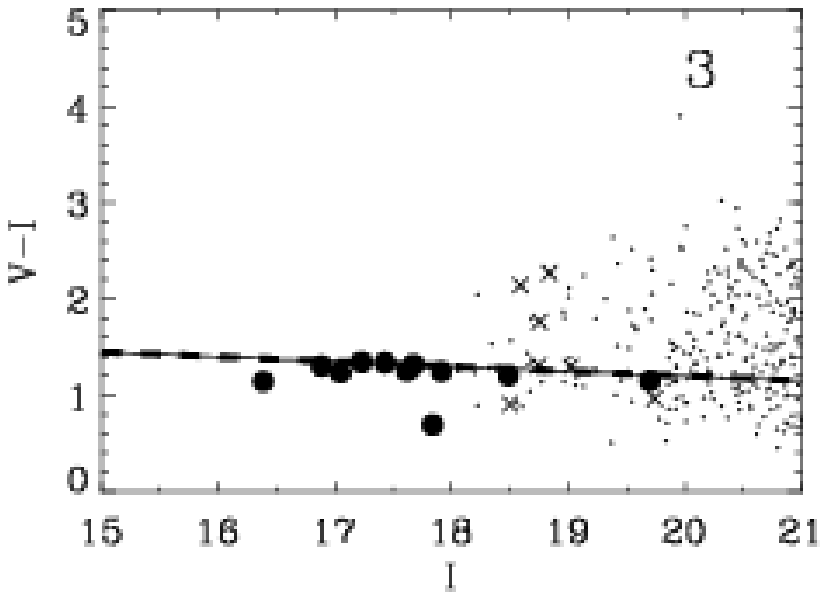}}
\resizebox{0.3\textwidth}{!}{\includegraphics{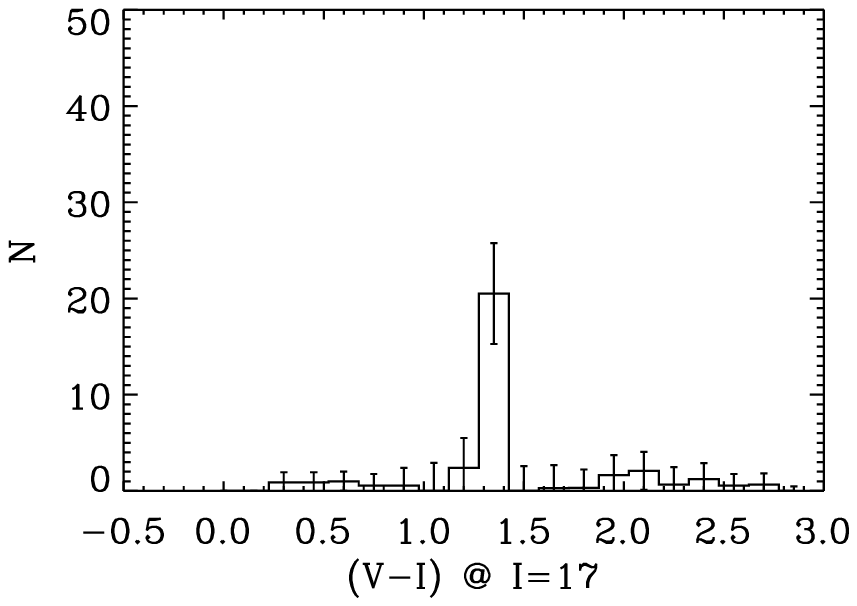}}
\resizebox{0.3\textwidth}{!}{\includegraphics{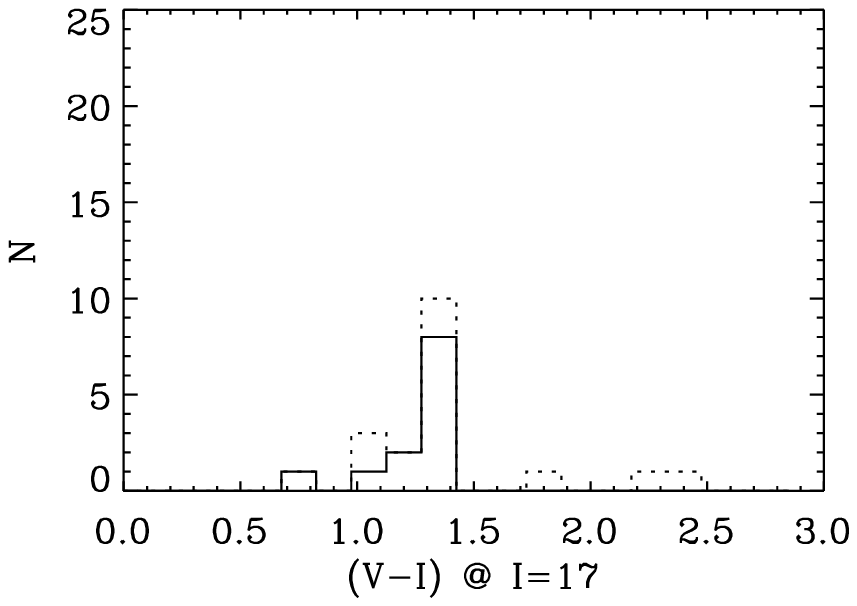}}
\vspace{4mm}

\resizebox{0.3\textwidth}{!}{\includegraphics{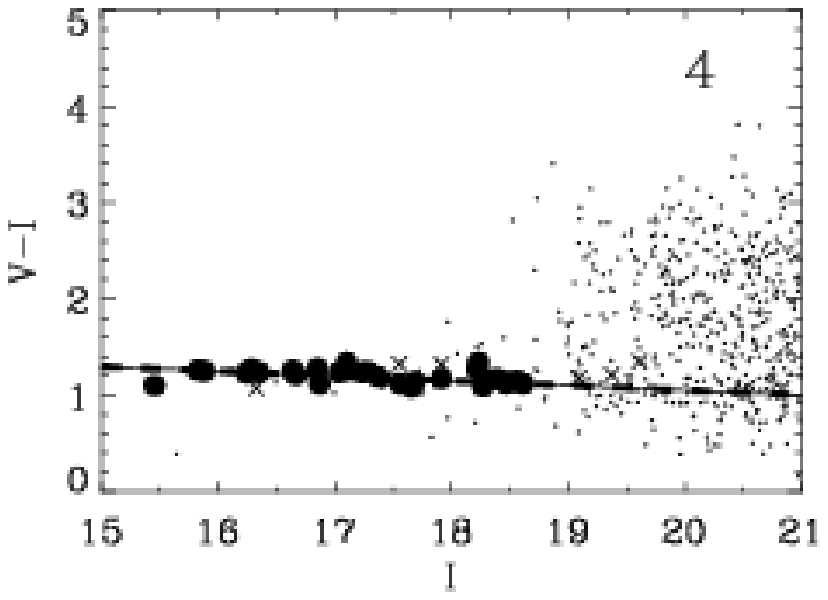}}
\resizebox{0.3\textwidth}{!}{\includegraphics{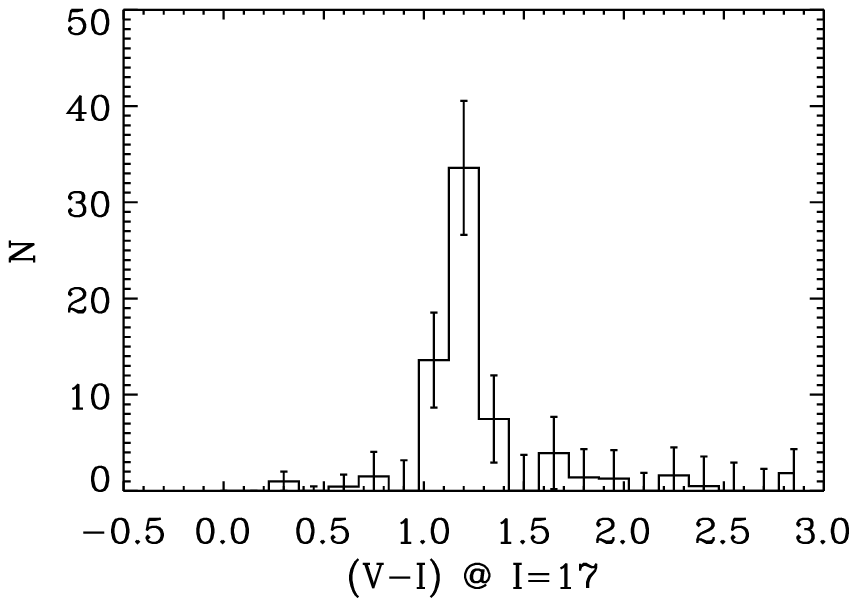}}
\resizebox{0.3\textwidth}{!}{\includegraphics{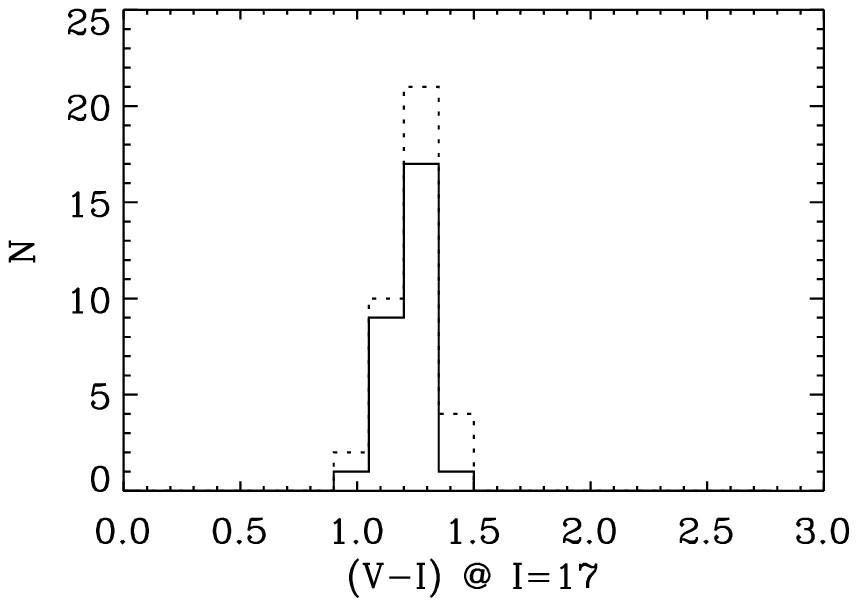}}
\vspace{4mm}

\resizebox{0.3\textwidth}{!}{\includegraphics{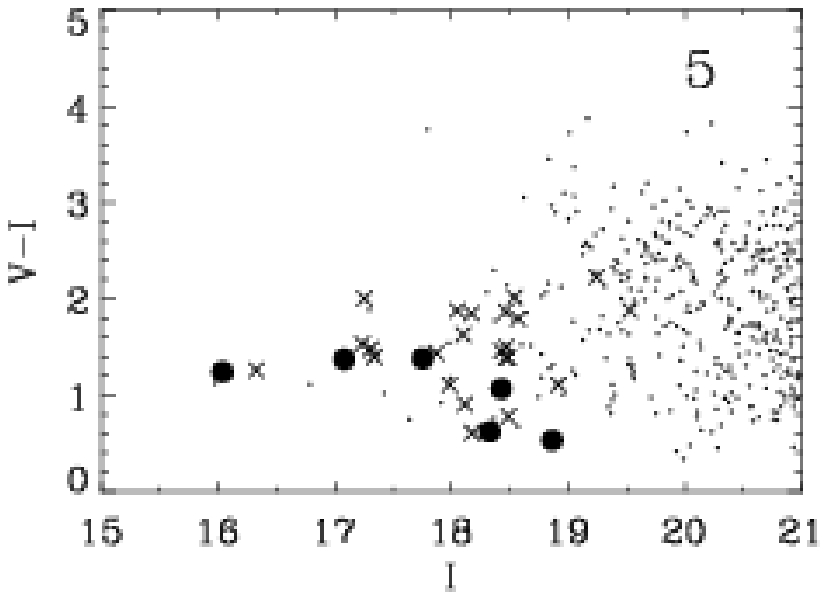}}
\resizebox{0.3\textwidth}{!}{\includegraphics{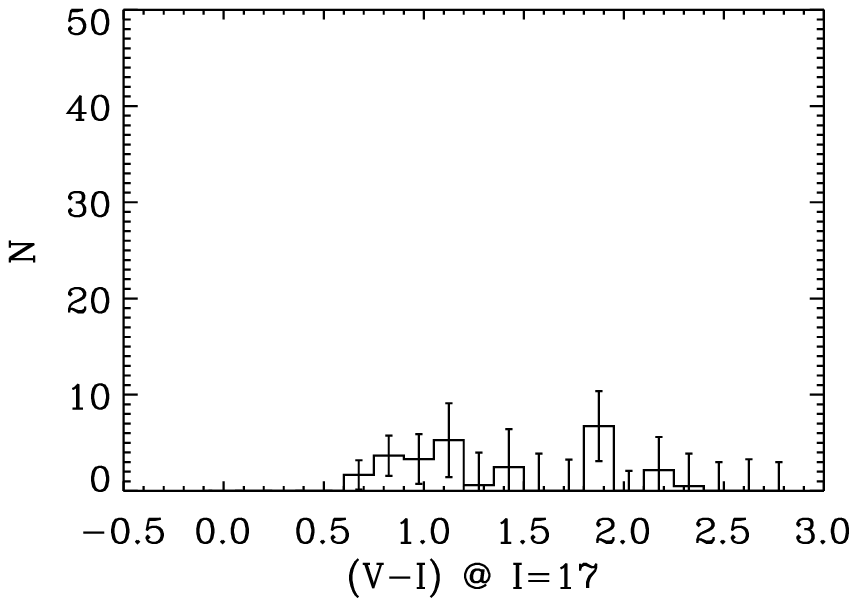}}
\resizebox{0.3\textwidth}{!}{\includegraphics{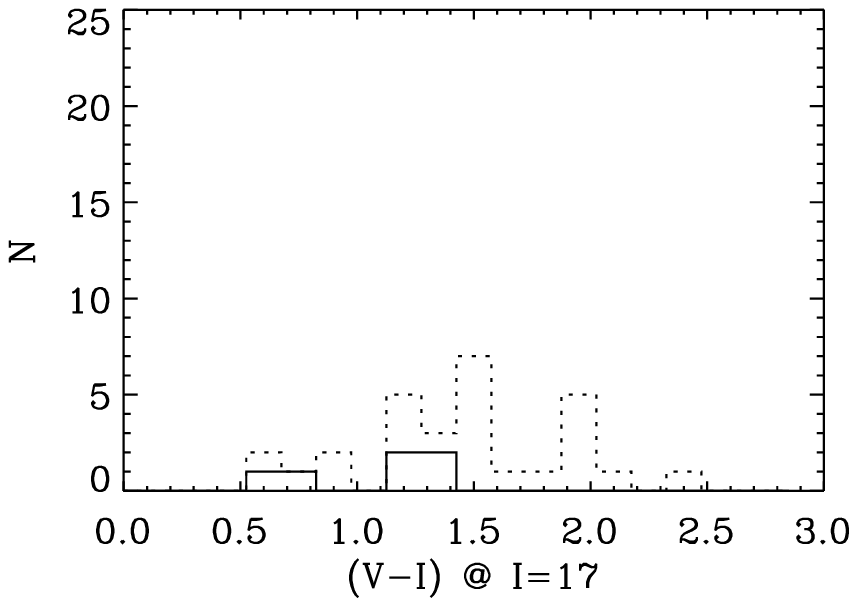}}
\end{center}
\caption{For each cluster we show 3 diagrams with the cluster
identification number indicated in the first one. The first diagram is
the colour-magnitude diagram for all galaxies within
$0.75h_{75}^{-1}\mathrm{Mpc}$ from the cluster center (dots). On top
of that we mark by solid circles the spectroscopic members of the
confirmed group and by crosses the remaining galaxies with
redshifts. The solid line is the locus of the red sequence detected
from the spectroscopic members and the dashed line is the one detected
in the photometric analysis. In both cases we only show the line if we
consider the sequence significant (see the text for details). The
second plot is the ``tilted colour histogram'' for the galaxies
with $I\leq19.5$ in the same region statistically corrected for the
background contribution. The last panel is the ``tilted colour
histogram'' for the spectroscopic members (solid line) and for all
galaxies with a redshift (dotted line).}
\label{fig:cmd}
\end{figure*}

\addtocounter{figure}{-1}

\begin{figure*}
\begin{center}
\resizebox{0.3\textwidth}{!}{\includegraphics{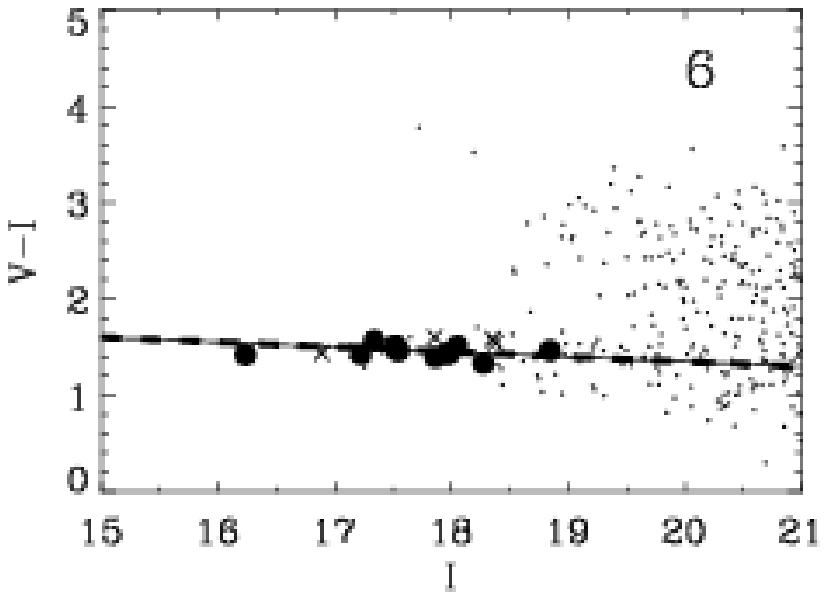}}
\resizebox{0.3\textwidth}{!}{\includegraphics{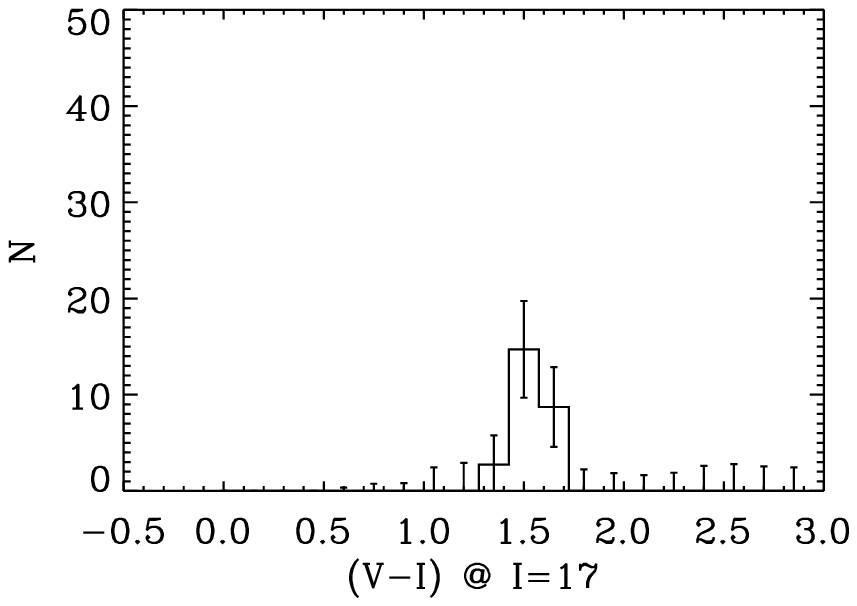}}
\resizebox{0.3\textwidth}{!}{\includegraphics{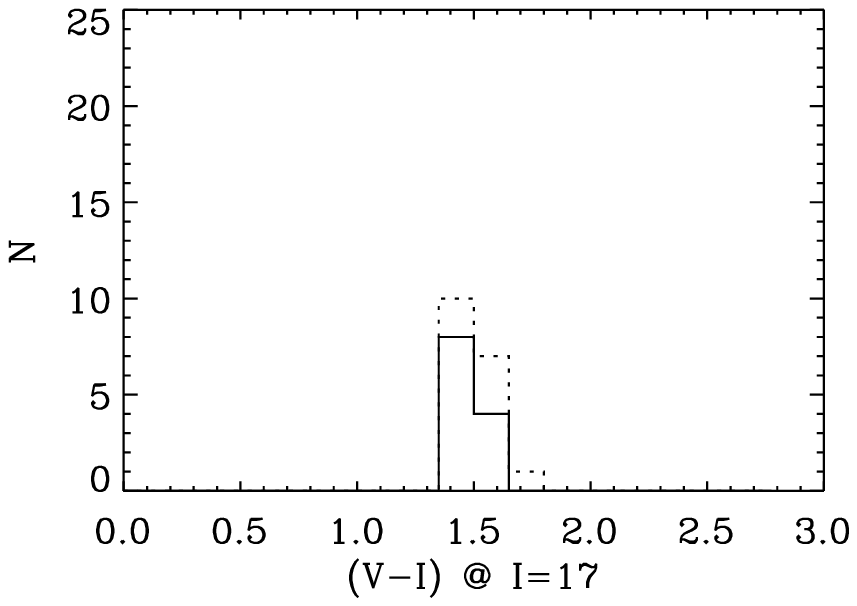}}
\resizebox{0.3\textwidth}{!}{\includegraphics{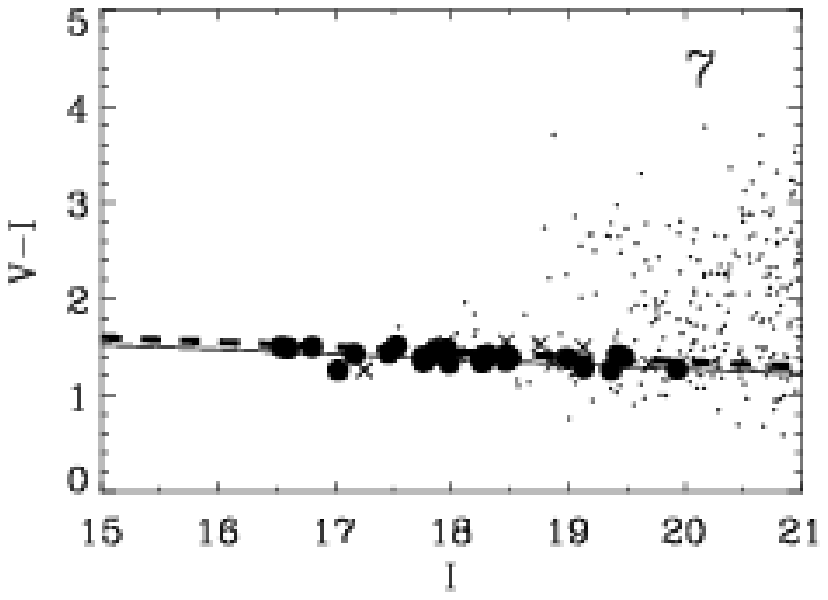}}
\resizebox{0.3\textwidth}{!}{\includegraphics{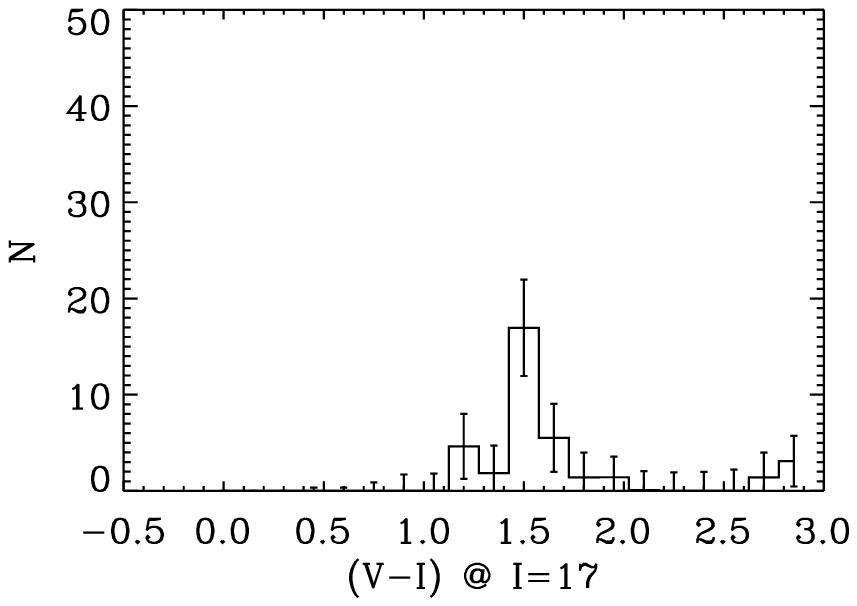}}
\resizebox{0.3\textwidth}{!}{\includegraphics{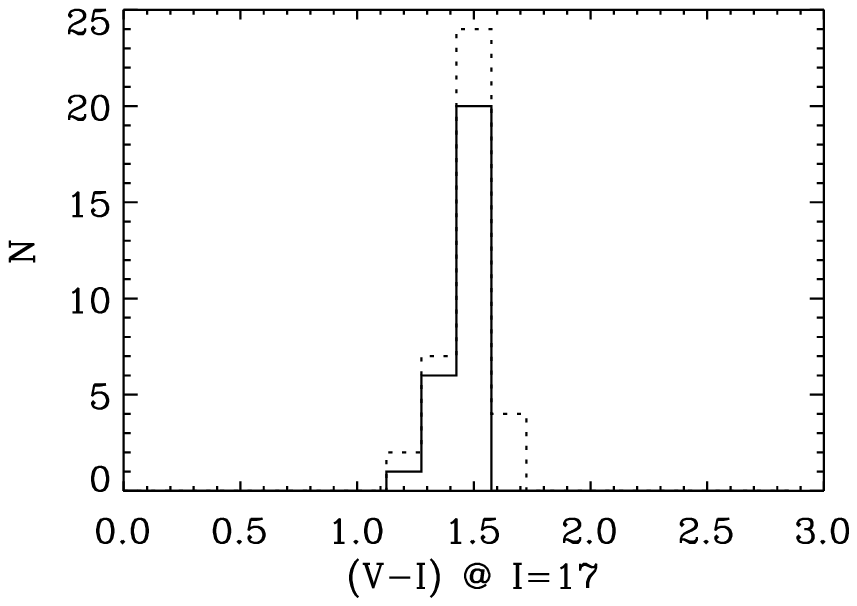}}
\resizebox{0.3\textwidth}{!}{\includegraphics{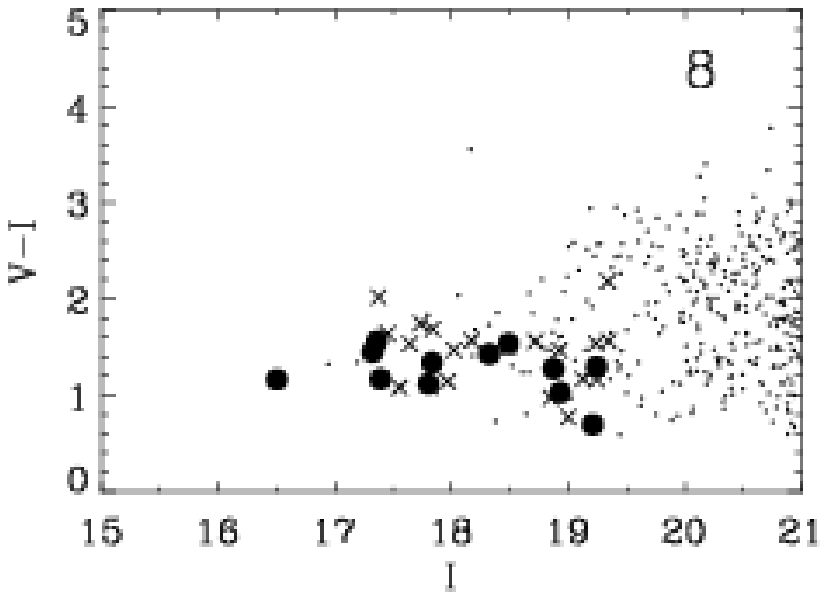}}
\resizebox{0.3\textwidth}{!}{\includegraphics{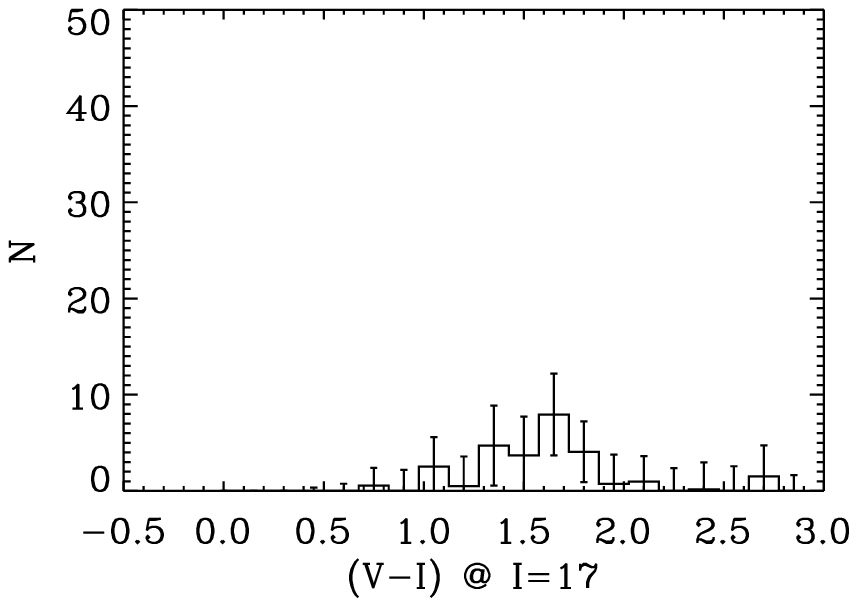}}
\resizebox{0.3\textwidth}{!}{\includegraphics{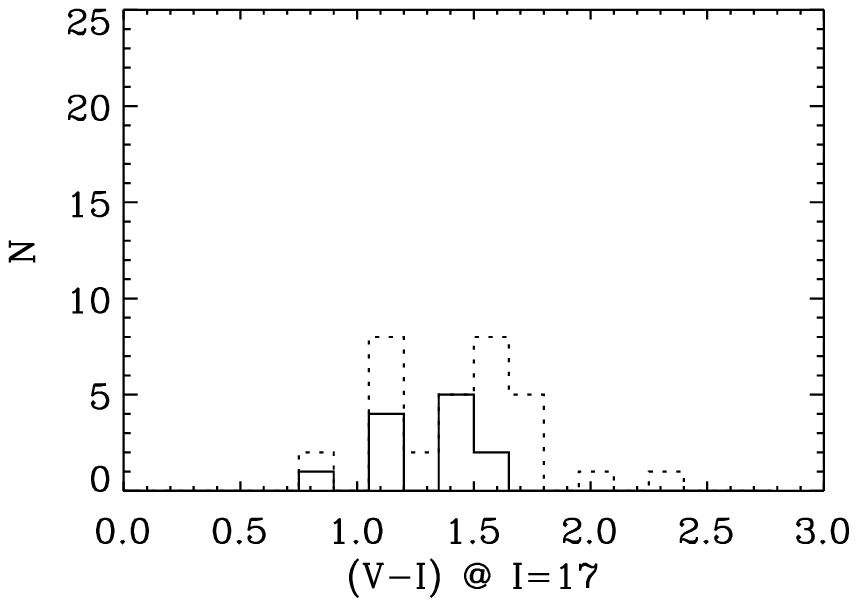}}
\resizebox{0.3\textwidth}{!}{\includegraphics{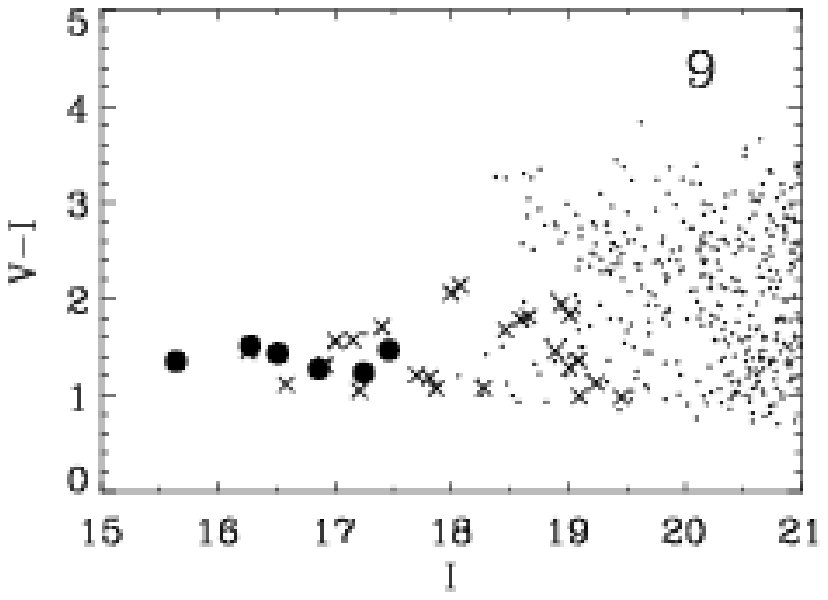}}
\resizebox{0.3\textwidth}{!}{\includegraphics{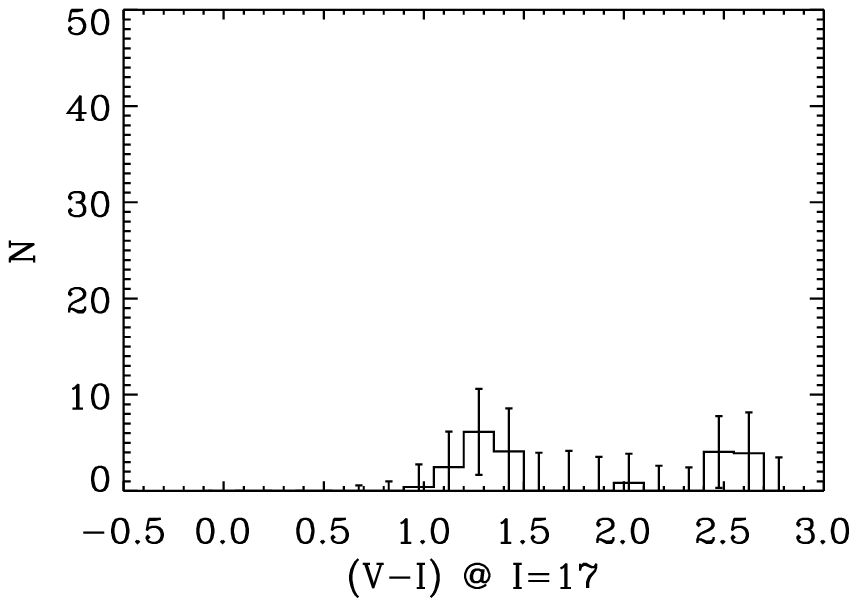}}
\resizebox{0.3\textwidth}{!}{\includegraphics{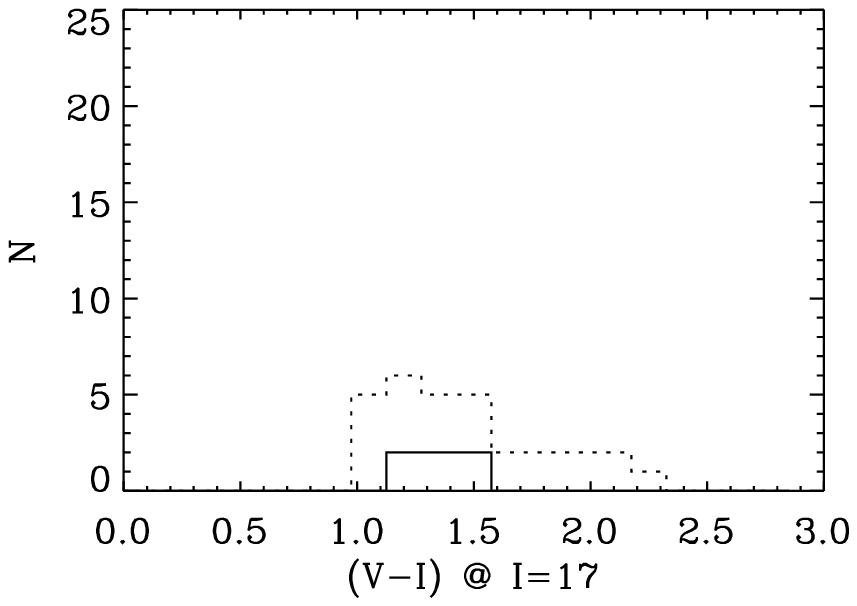}}
\resizebox{0.3\textwidth}{!}{\includegraphics{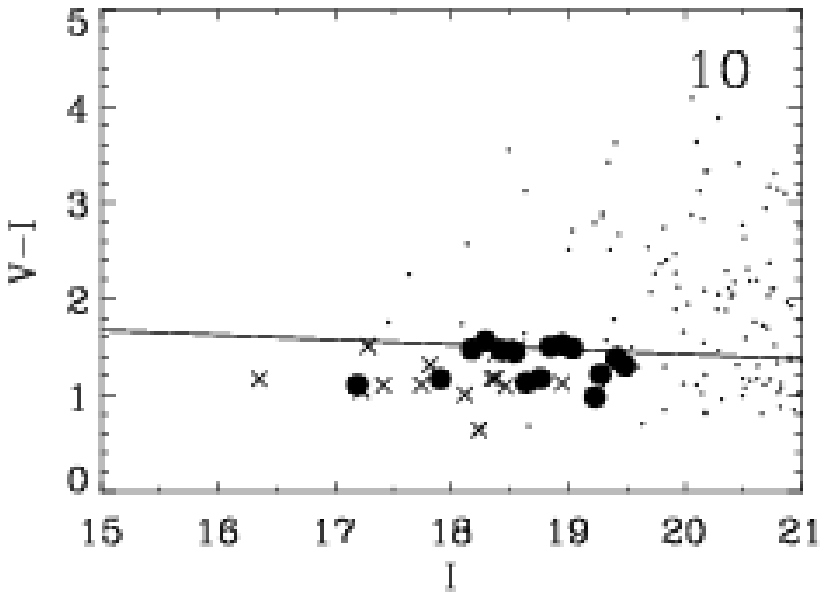}}
\resizebox{0.3\textwidth}{!}{\includegraphics{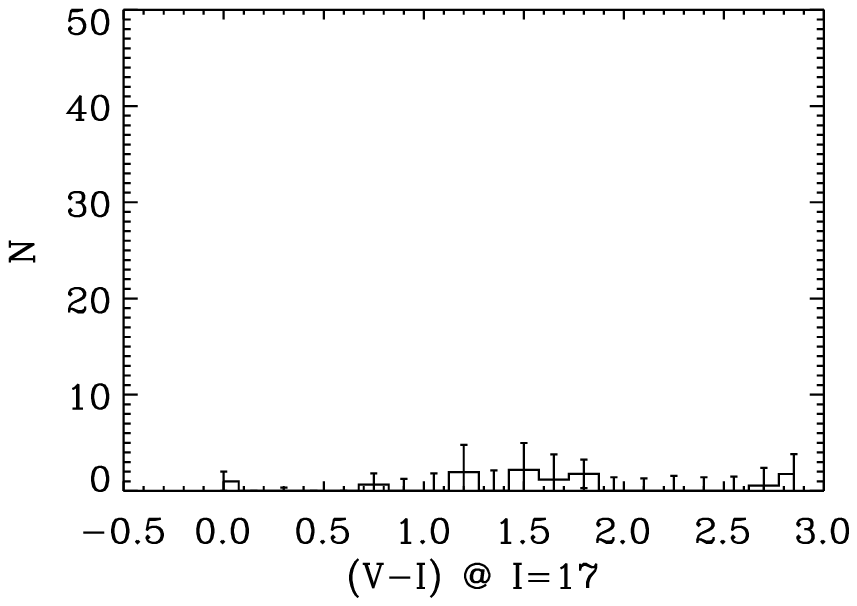}}
\resizebox{0.3\textwidth}{!}{\includegraphics{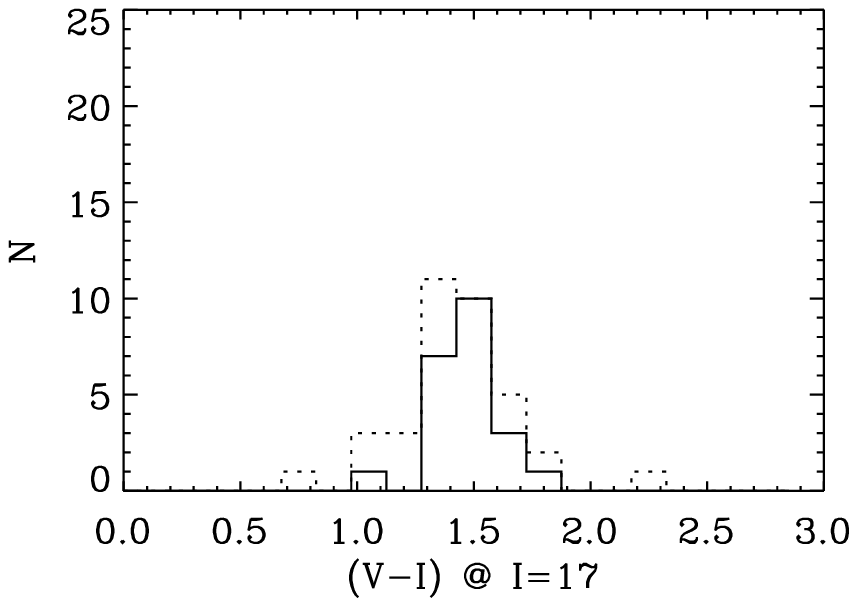}}
\resizebox{0.3\textwidth}{!}{\includegraphics{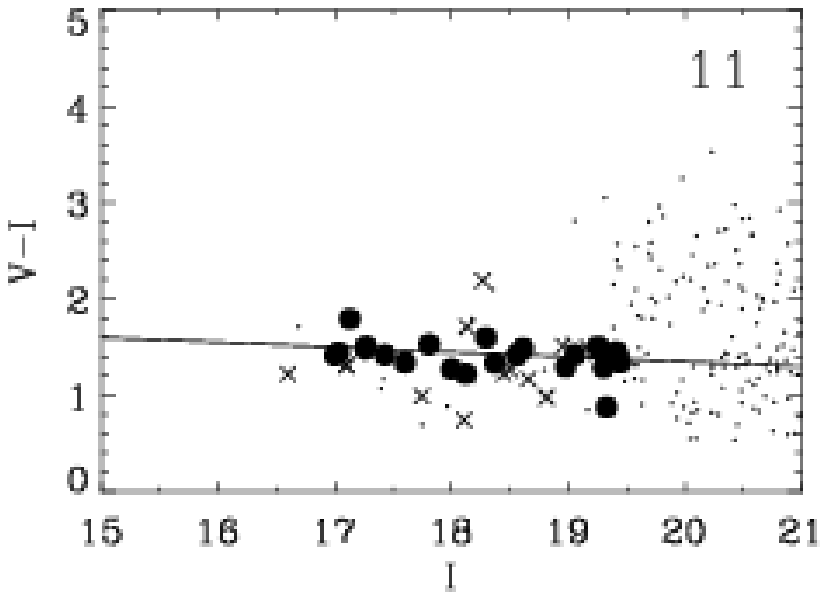}}
\resizebox{0.3\textwidth}{!}{\includegraphics{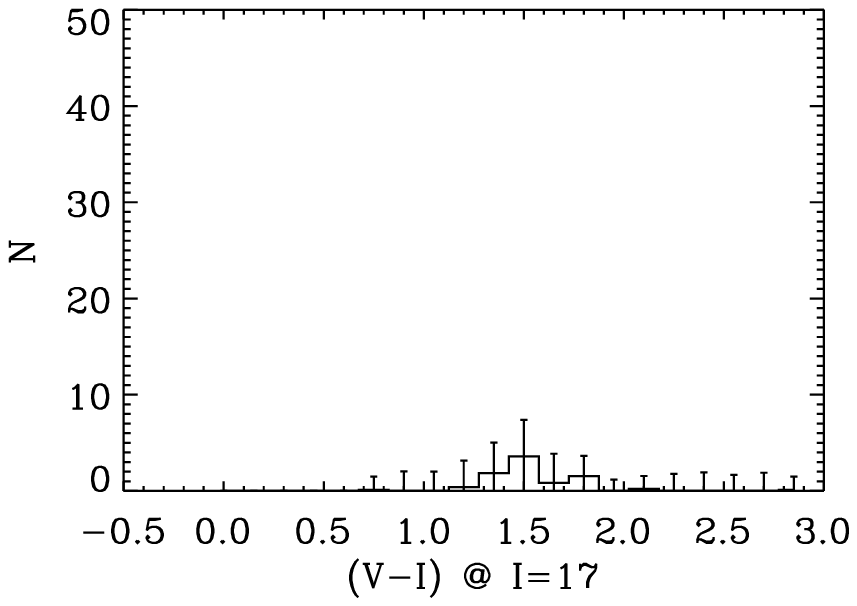}}
\resizebox{0.3\textwidth}{!}{\includegraphics{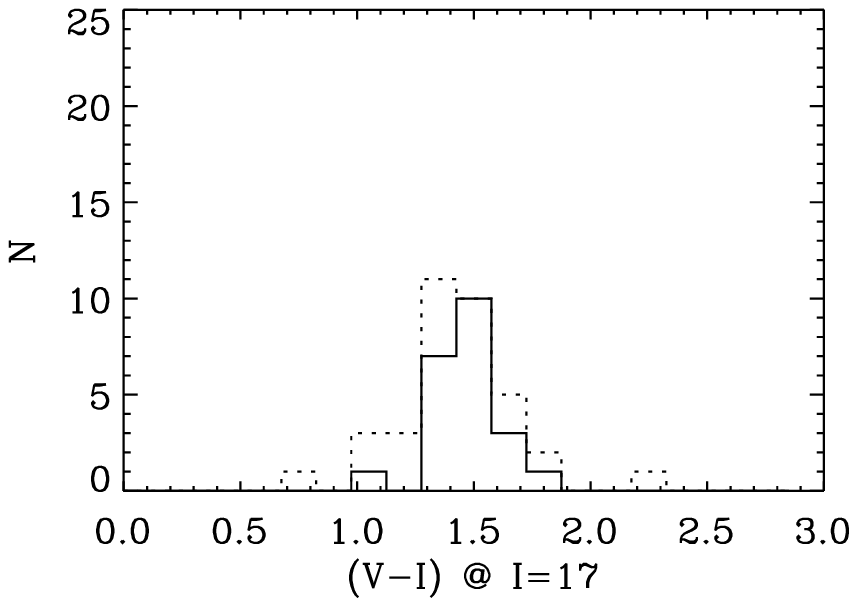}}
\end{center}
\caption{\it -- Continued }
\end{figure*}

\addtocounter{figure}{-1}

\begin{figure*}
\begin{center}
\resizebox{0.3\textwidth}{!}{\includegraphics{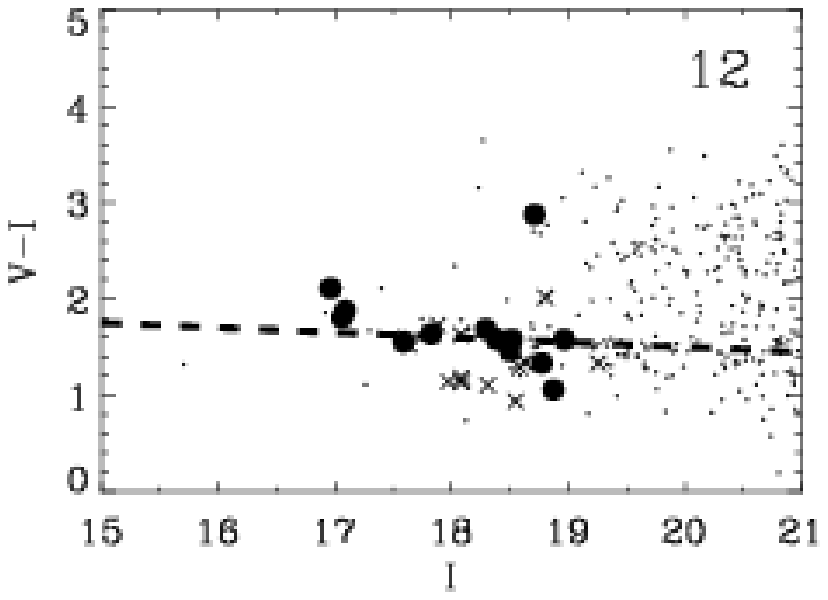}}
\resizebox{0.3\textwidth}{!}{\includegraphics{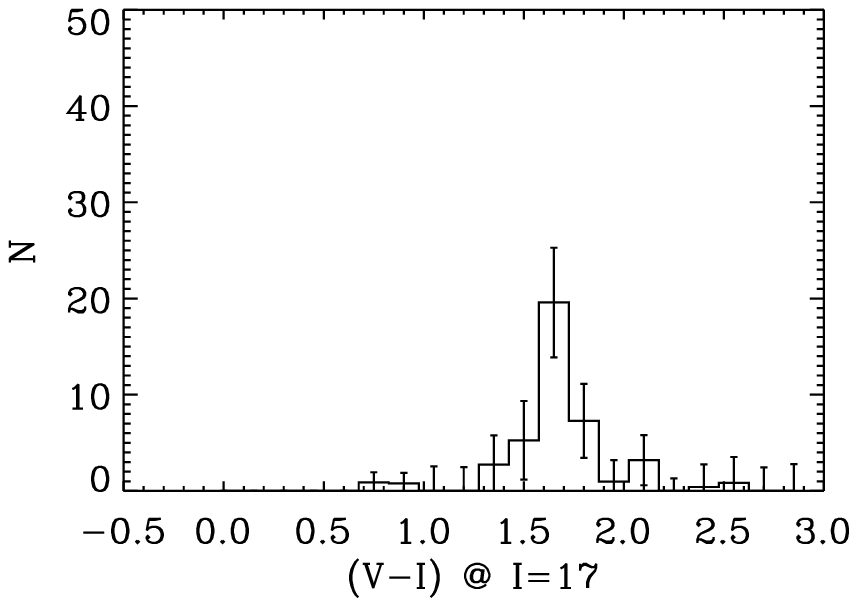}}
\resizebox{0.3\textwidth}{!}{\includegraphics{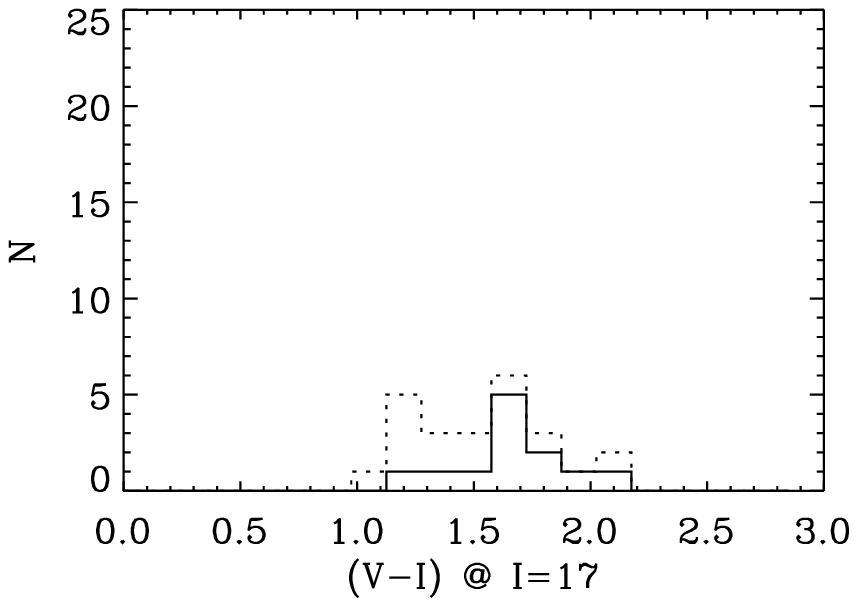}}
\resizebox{0.3\textwidth}{!}{\includegraphics{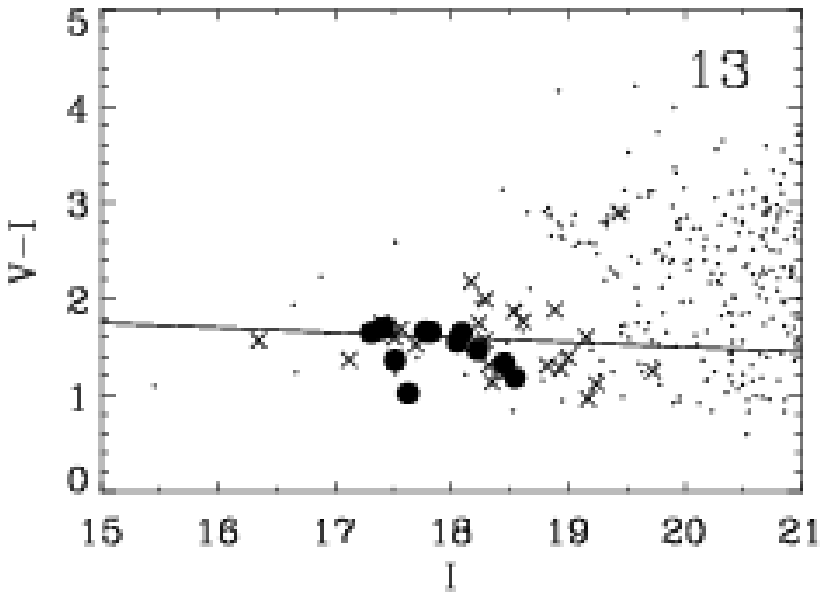}}
\resizebox{0.3\textwidth}{!}{\includegraphics{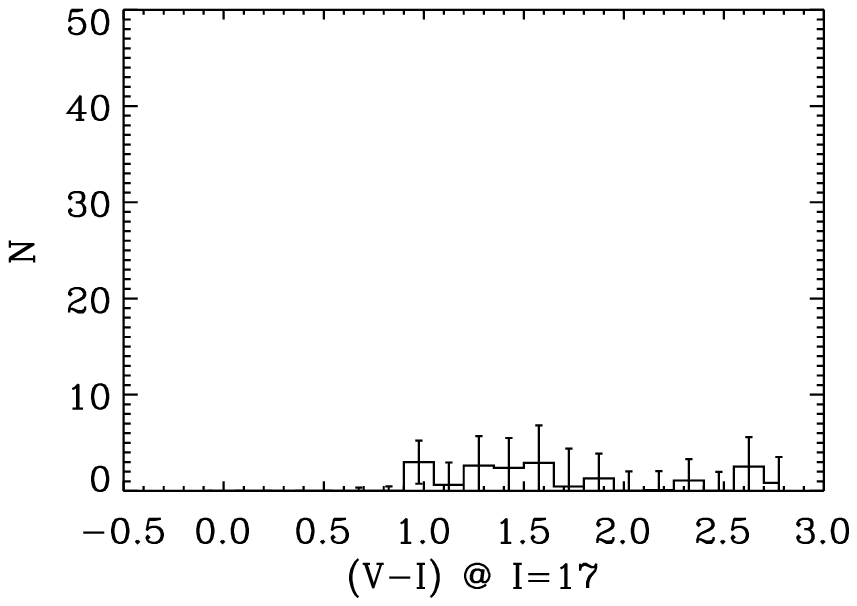}}
\resizebox{0.3\textwidth}{!}{\includegraphics{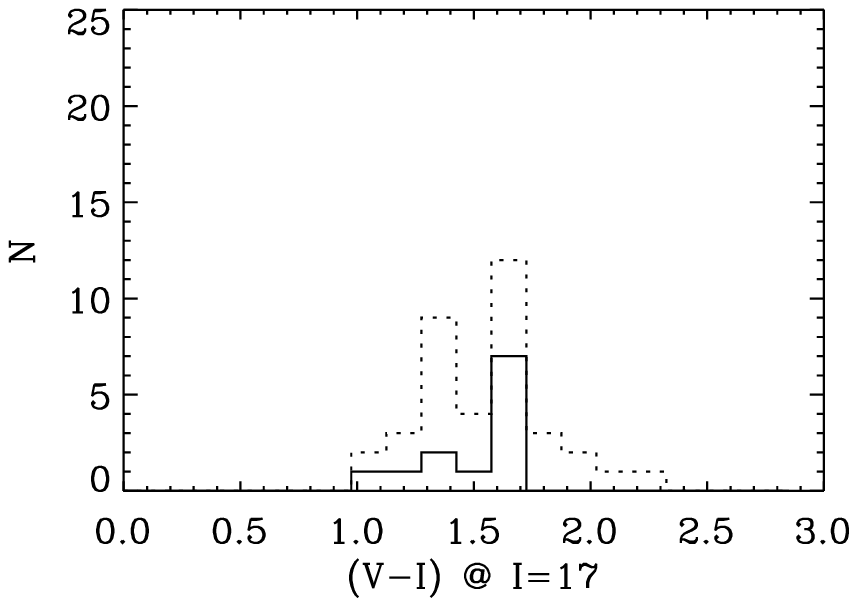}}
\resizebox{0.3\textwidth}{!}{\includegraphics{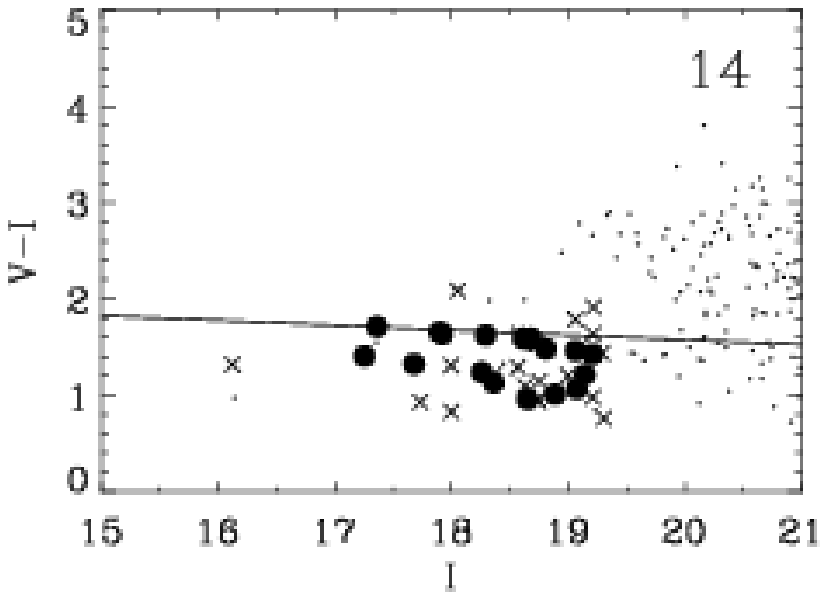}}
\resizebox{0.3\textwidth}{!}{\includegraphics{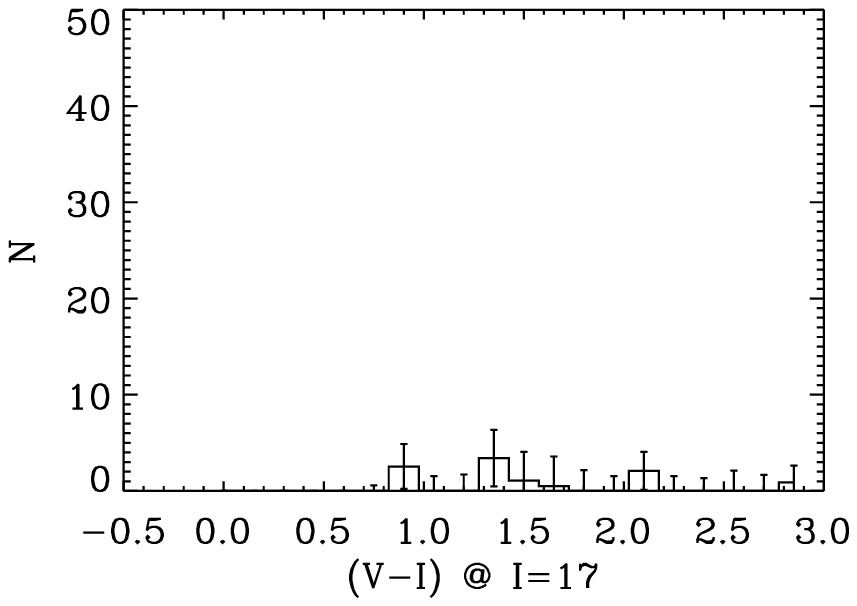}}
\resizebox{0.3\textwidth}{!}{\includegraphics{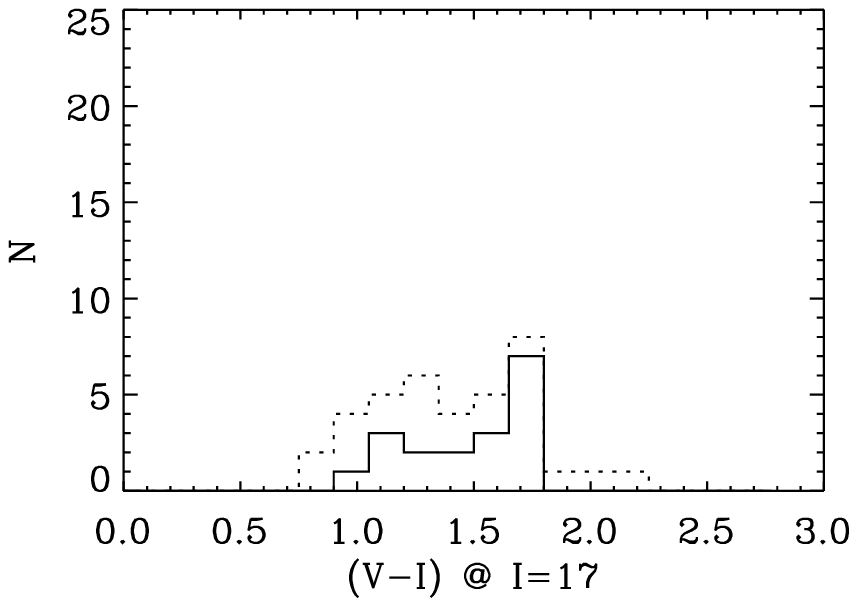}}
\resizebox{0.3\textwidth}{!}{\includegraphics{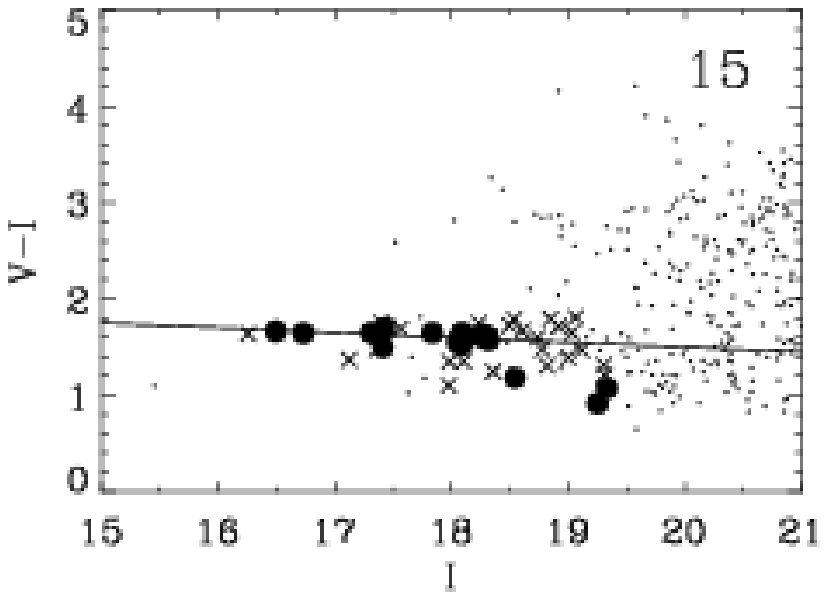}}
\resizebox{0.3\textwidth}{!}{\includegraphics{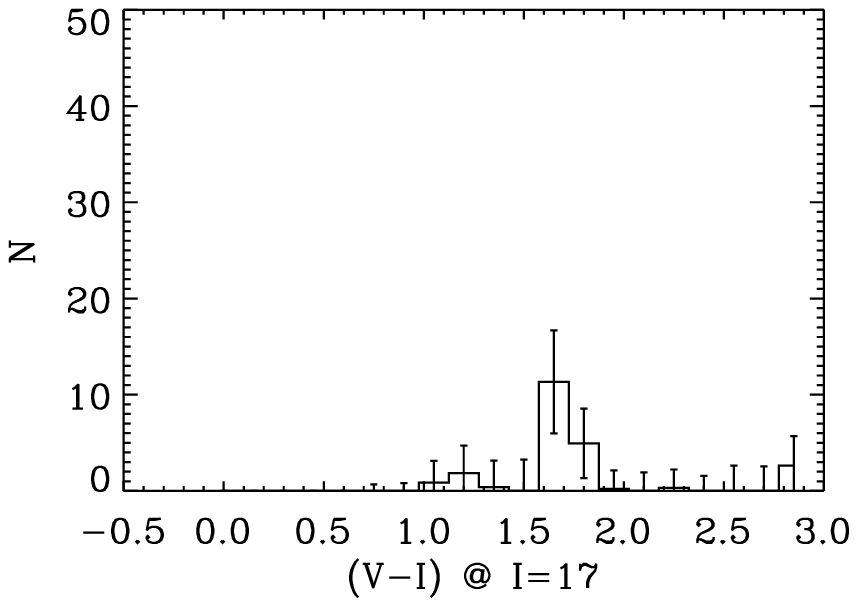}}
\resizebox{0.3\textwidth}{!}{\includegraphics{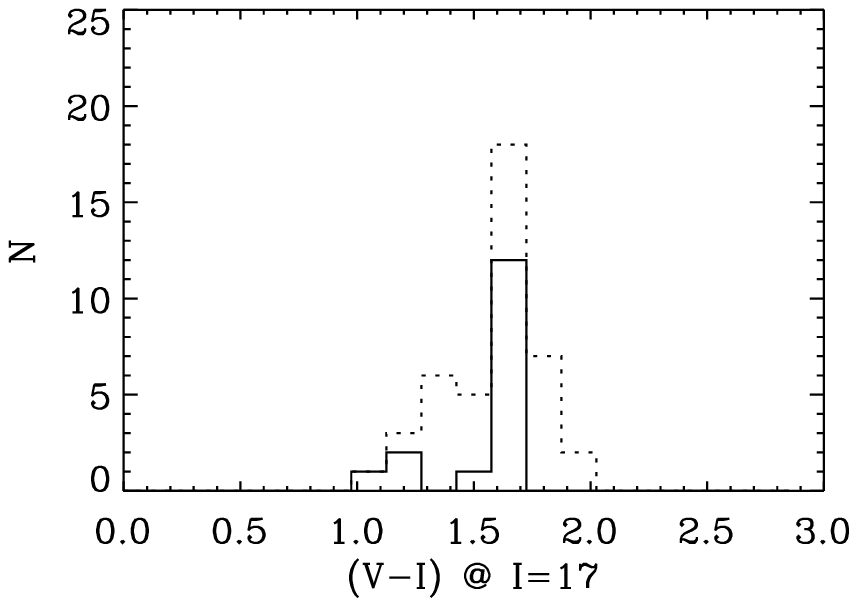}}
\resizebox{0.3\textwidth}{!}{\includegraphics{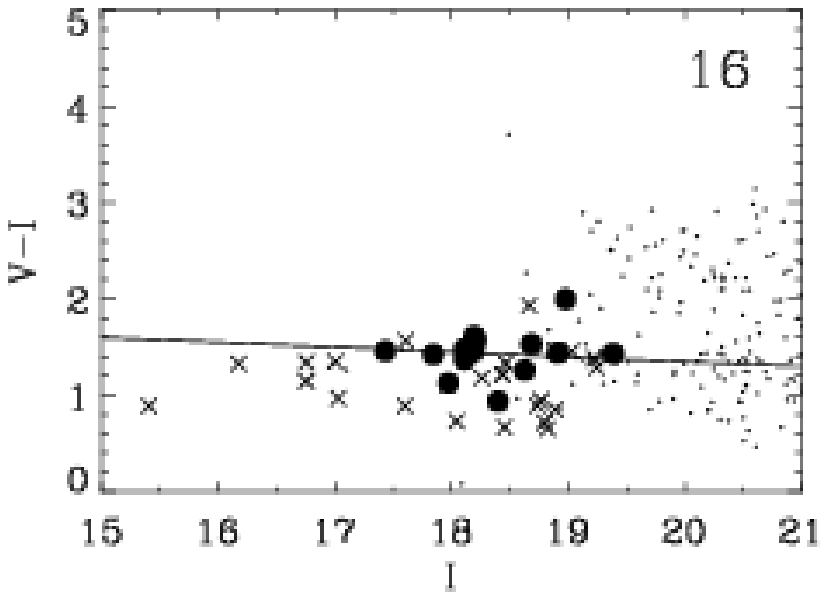}}
\resizebox{0.3\textwidth}{!}{\includegraphics{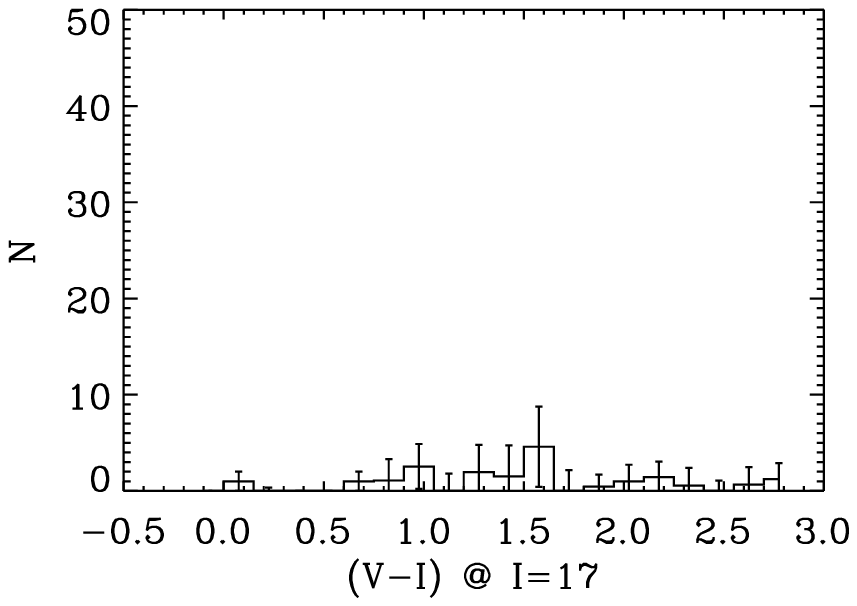}}
\resizebox{0.3\textwidth}{!}{\includegraphics{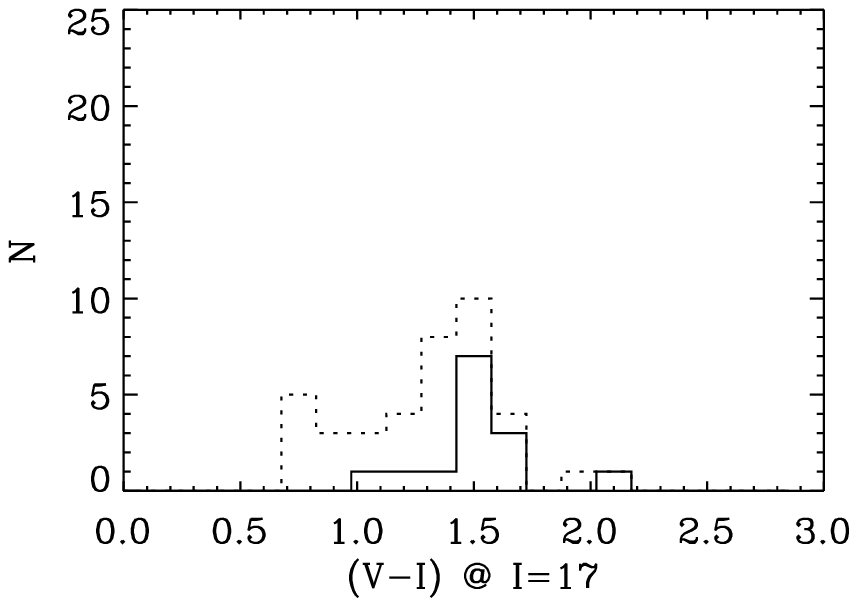}}
\resizebox{0.3\textwidth}{!}{\includegraphics{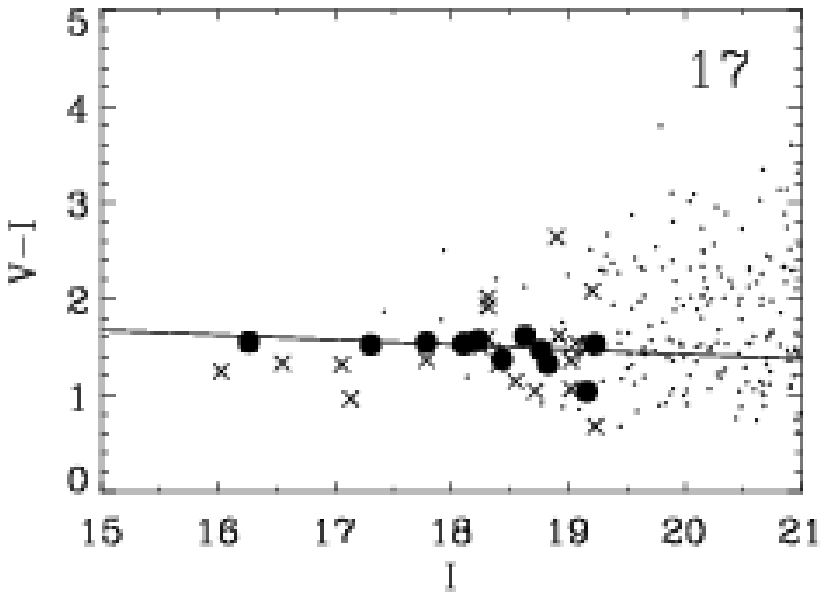}}
\resizebox{0.3\textwidth}{!}{\includegraphics{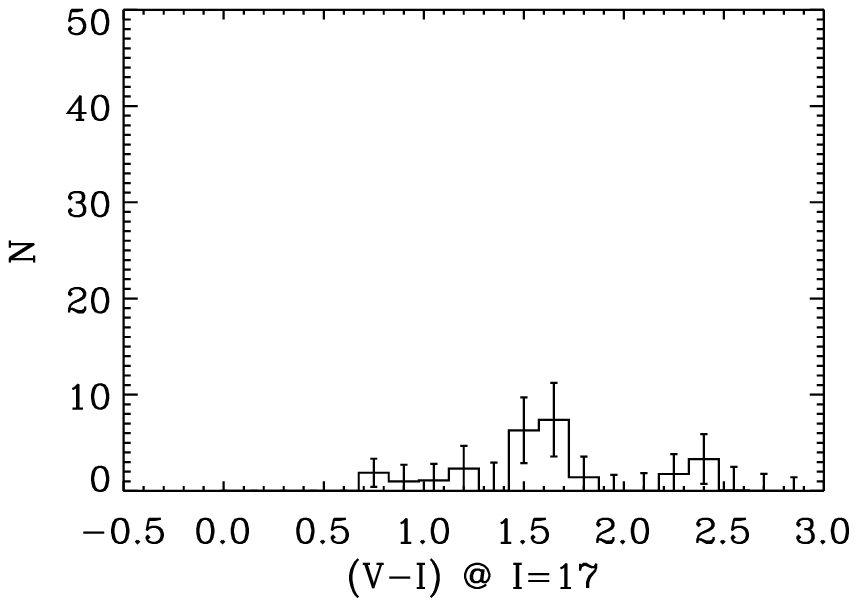}}
\resizebox{0.3\textwidth}{!}{\includegraphics{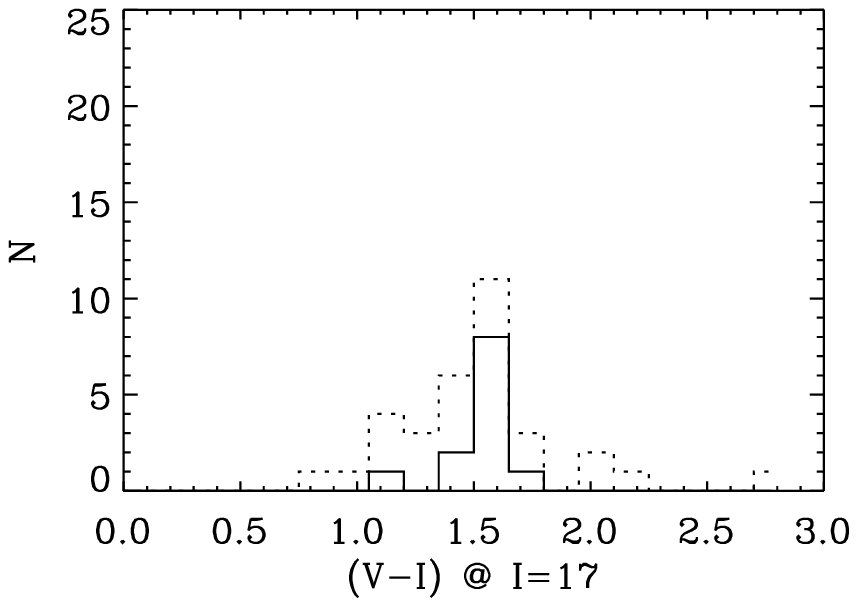}}
\end{center}
\caption{\it -- Continued }
\end{figure*}

\addtocounter{figure}{-1}

\begin{figure*}
\begin{center}
\resizebox{0.3\textwidth}{!}{\includegraphics{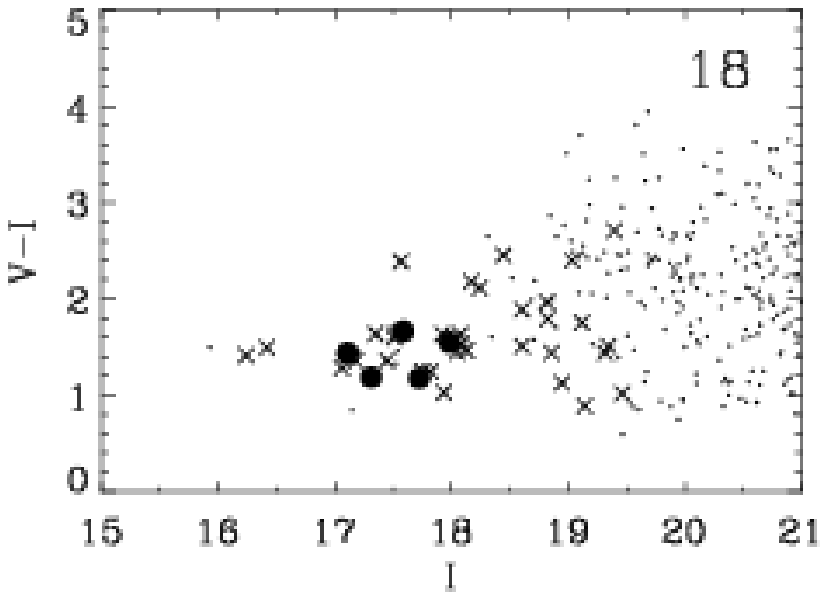}}
\resizebox{0.3\textwidth}{!}{\includegraphics{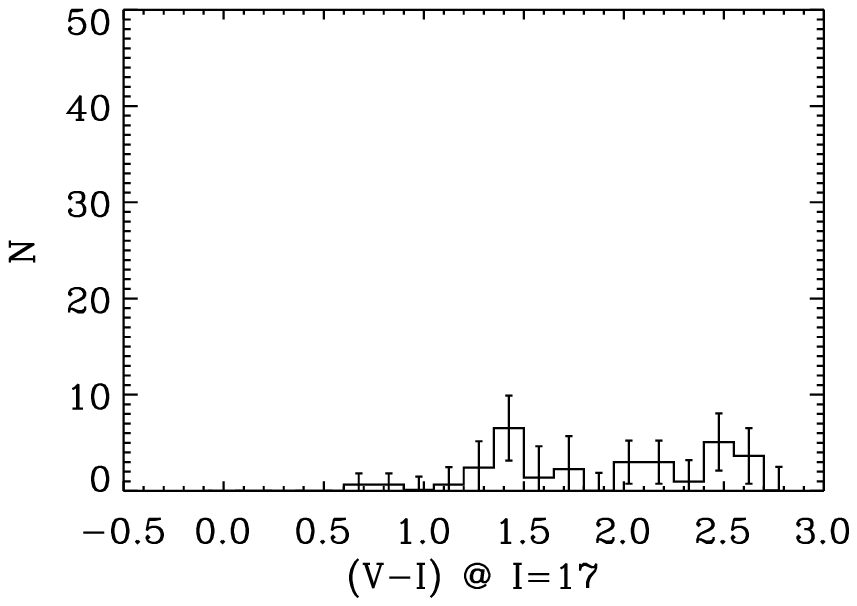}}
\resizebox{0.3\textwidth}{!}{\includegraphics{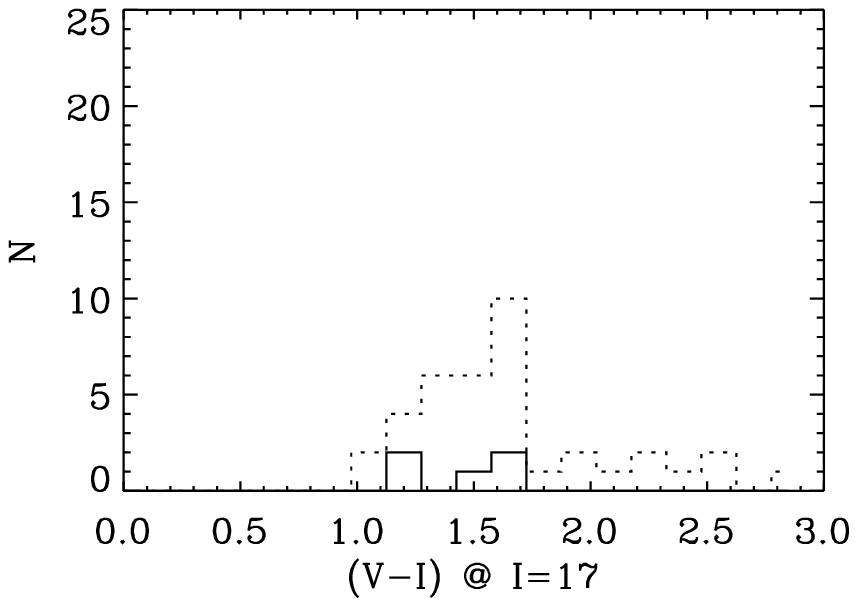}}
\resizebox{0.3\textwidth}{!}{\includegraphics{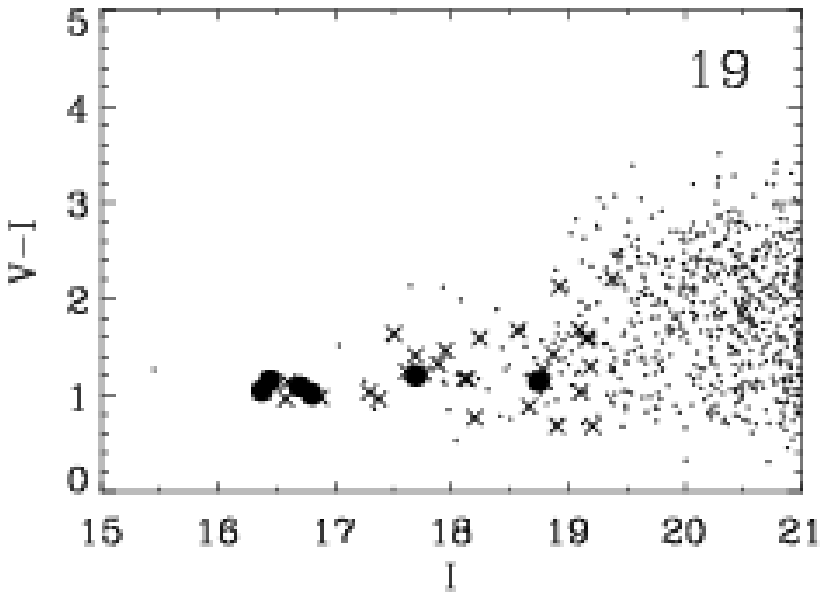}}
\resizebox{0.3\textwidth}{!}{\includegraphics{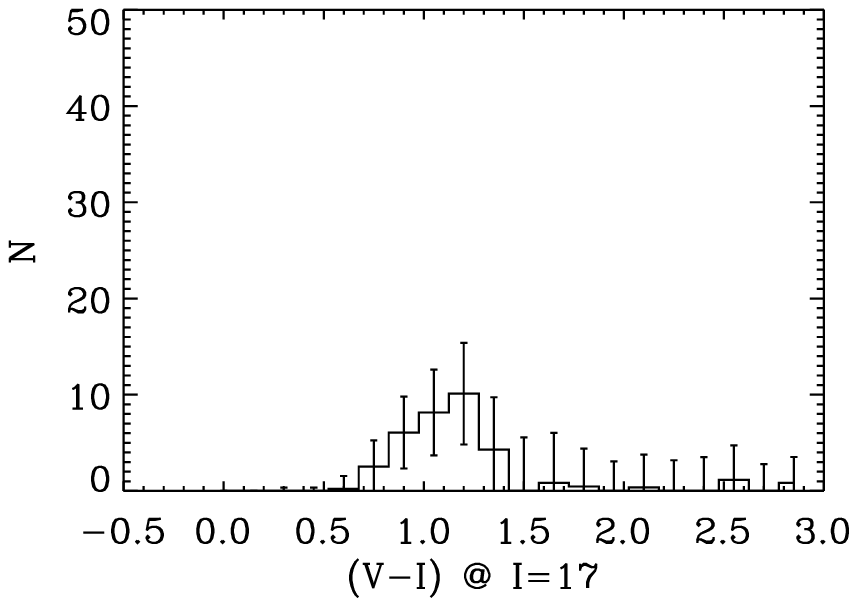}}
\resizebox{0.3\textwidth}{!}{\includegraphics{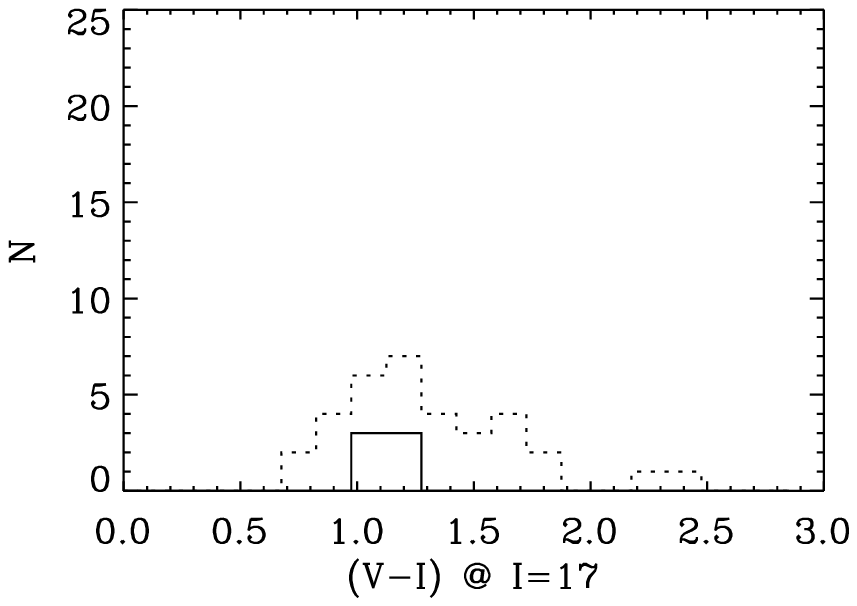}}
\resizebox{0.3\textwidth}{!}{\includegraphics{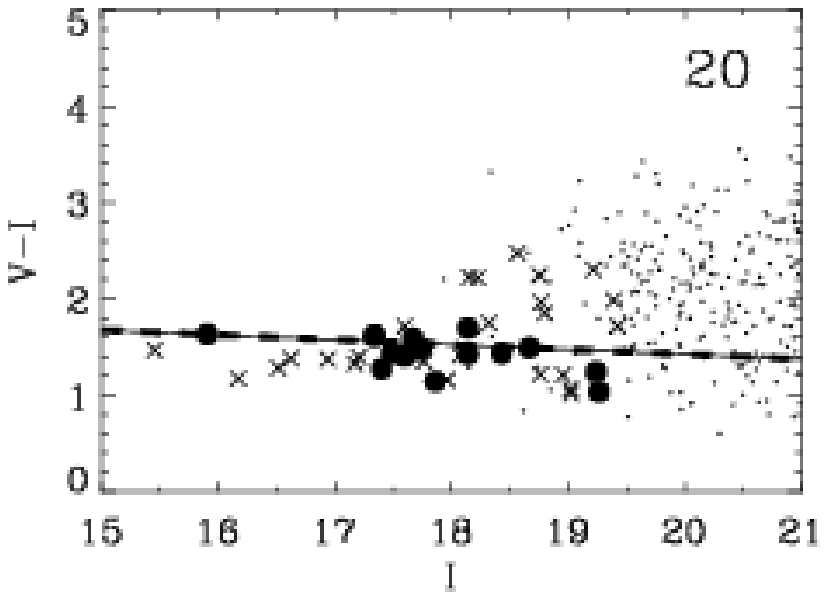}}
\resizebox{0.3\textwidth}{!}{\includegraphics{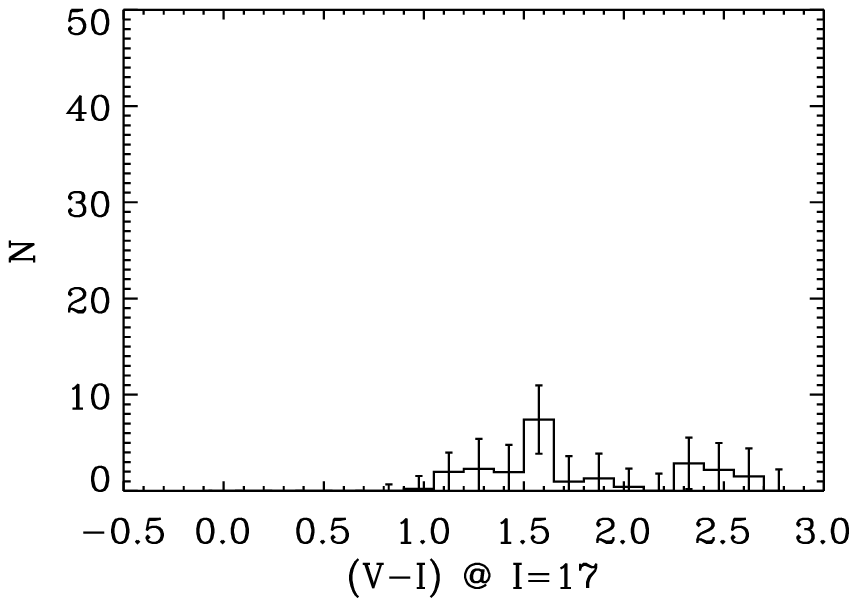}}
\resizebox{0.3\textwidth}{!}{\includegraphics{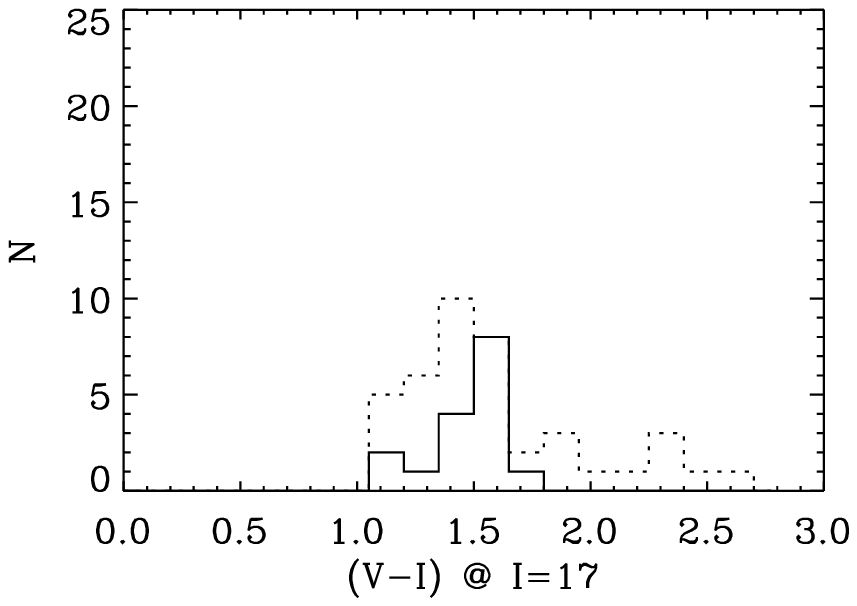}}
\resizebox{0.3\textwidth}{!}{\includegraphics{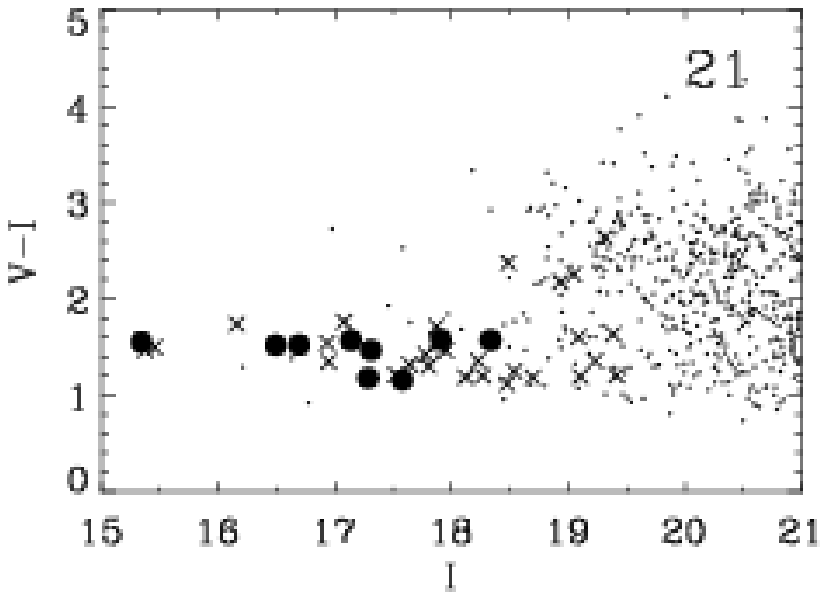}}
\resizebox{0.3\textwidth}{!}{\includegraphics{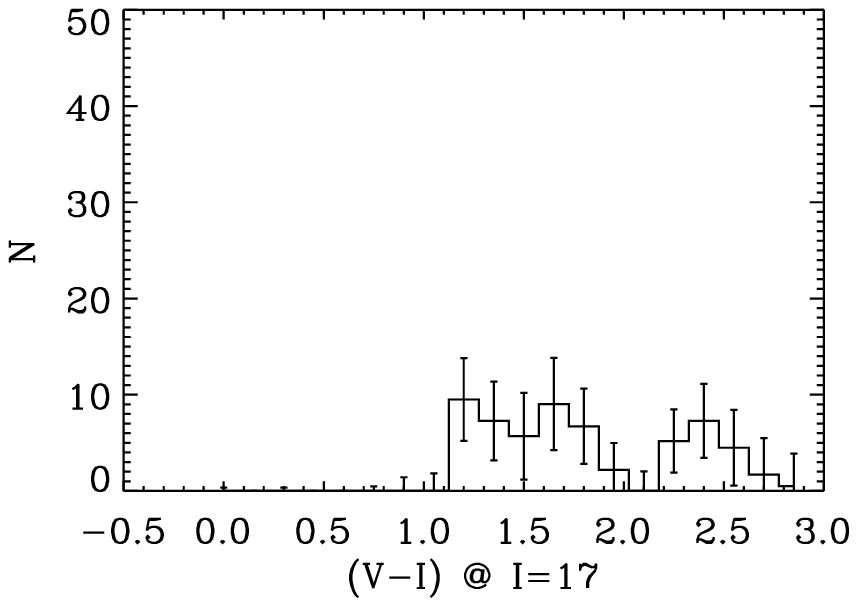}}
\resizebox{0.3\textwidth}{!}{\includegraphics{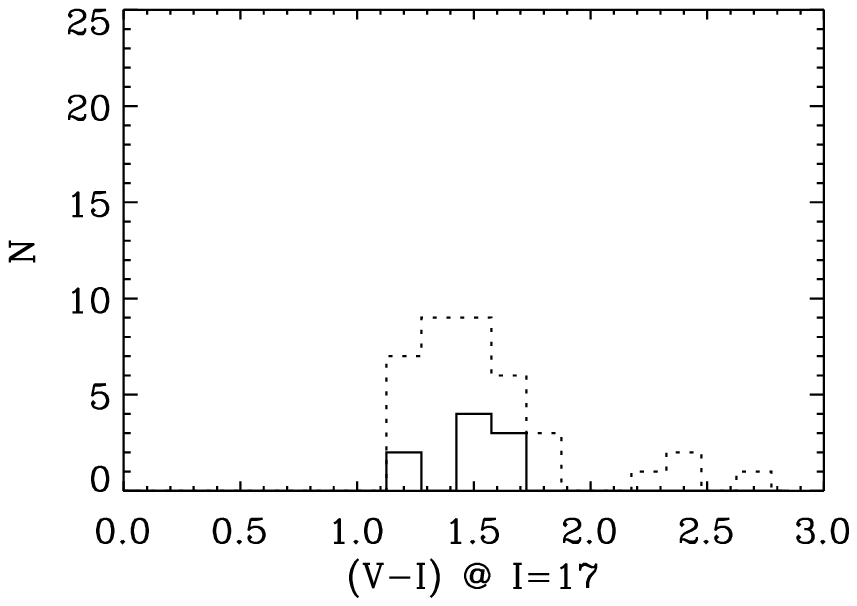}}
\resizebox{0.3\textwidth}{!}{\includegraphics{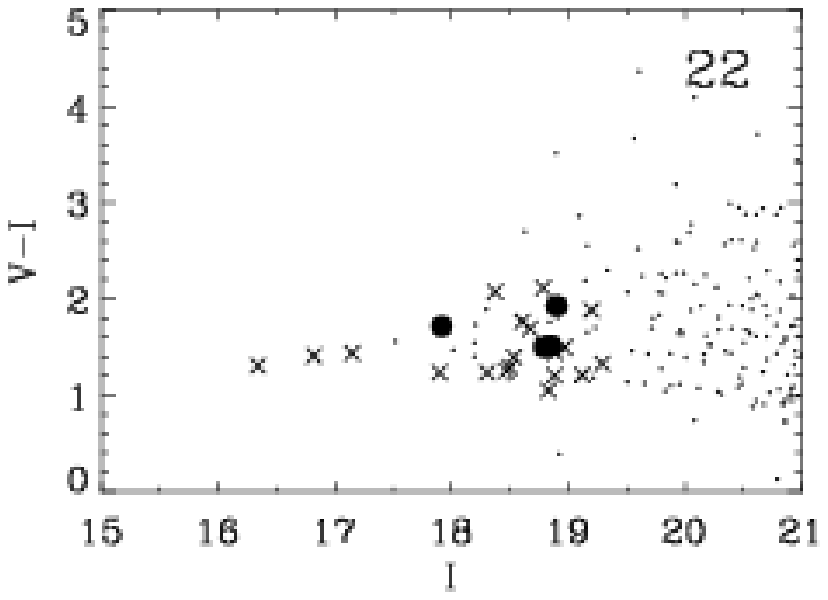}}
\resizebox{0.3\textwidth}{!}{\includegraphics{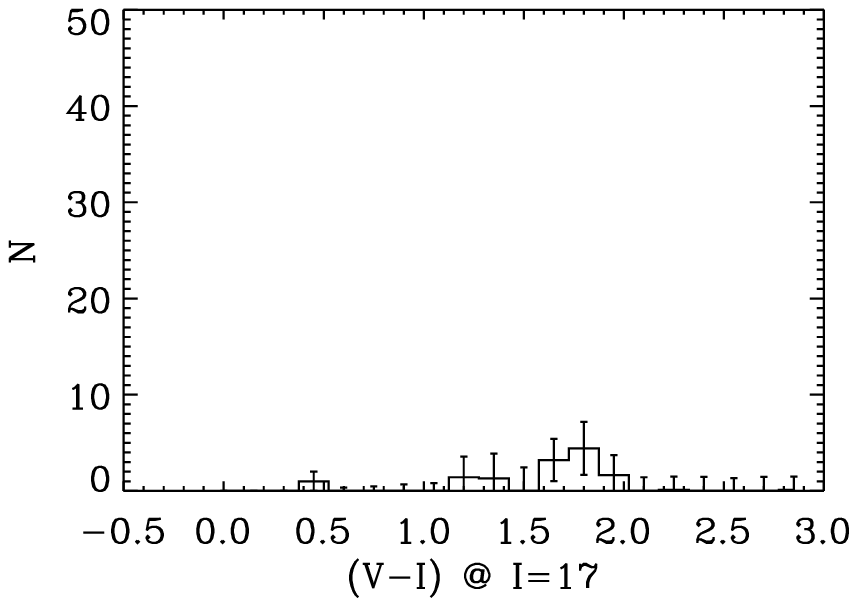}}
\resizebox{0.3\textwidth}{!}{\includegraphics{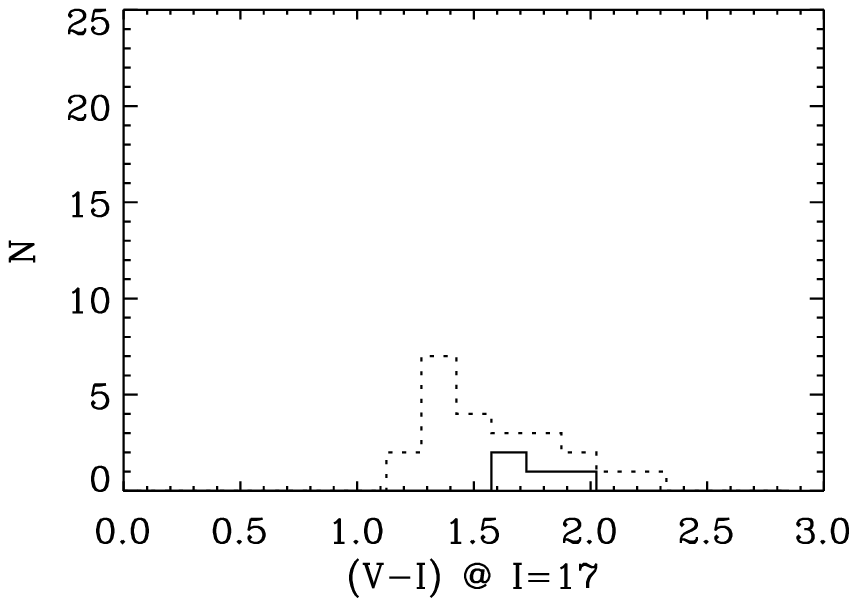}}
\resizebox{0.3\textwidth}{!}{\includegraphics{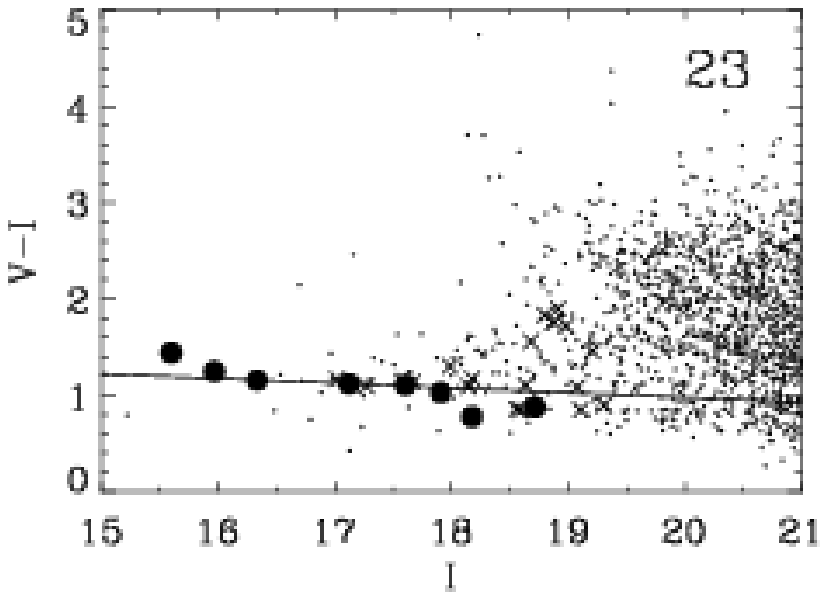}}
\resizebox{0.3\textwidth}{!}{\includegraphics{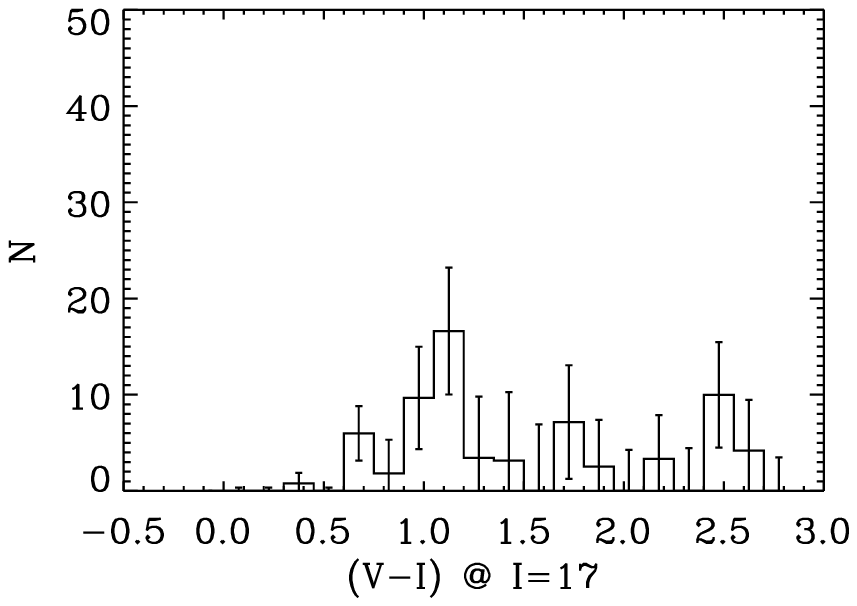}}
\resizebox{0.3\textwidth}{!}{\includegraphics{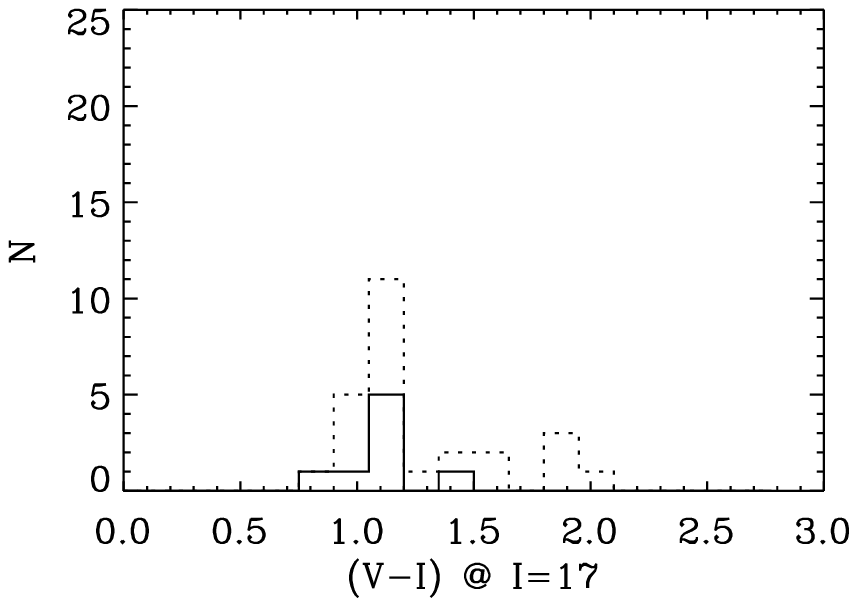}}
\end{center}
\caption{\it -- Continued }
\end{figure*}

\addtocounter{figure}{-1}

\begin{figure*}
\begin{center}
\resizebox{0.3\textwidth}{!}{\includegraphics{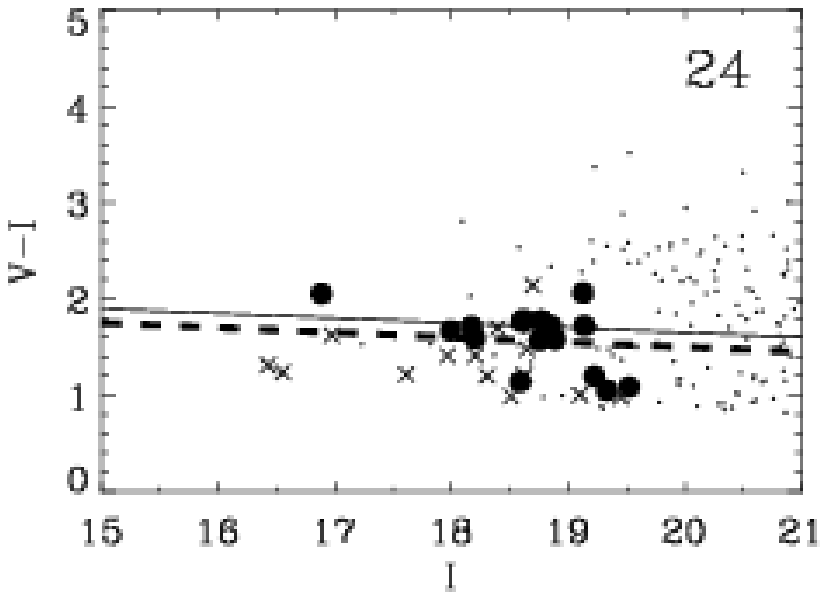}}
\resizebox{0.3\textwidth}{!}{\includegraphics{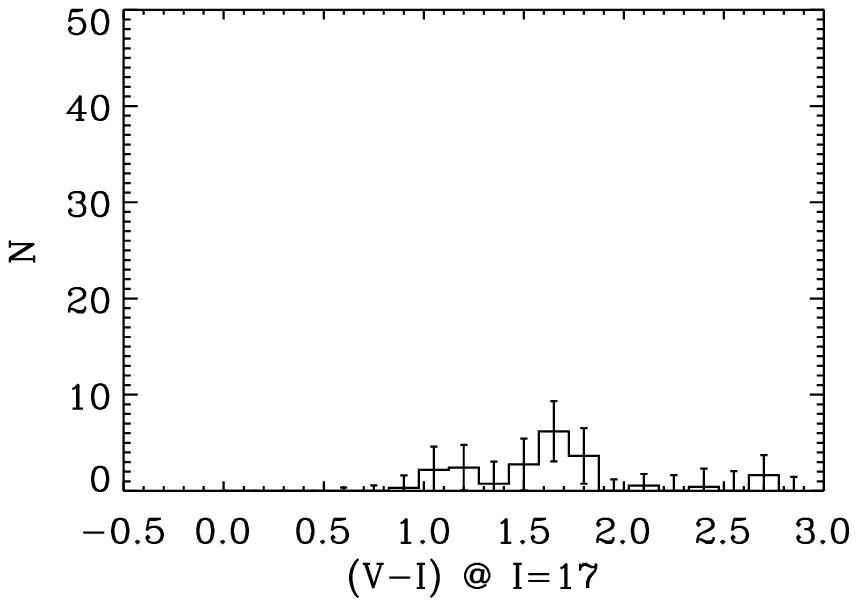}}
\resizebox{0.3\textwidth}{!}{\includegraphics{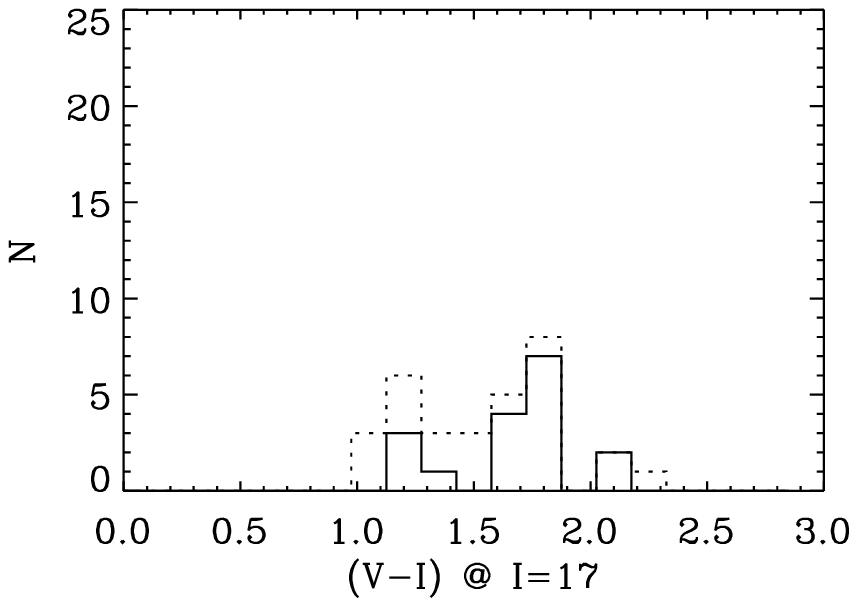}}
\resizebox{0.3\textwidth}{!}{\includegraphics{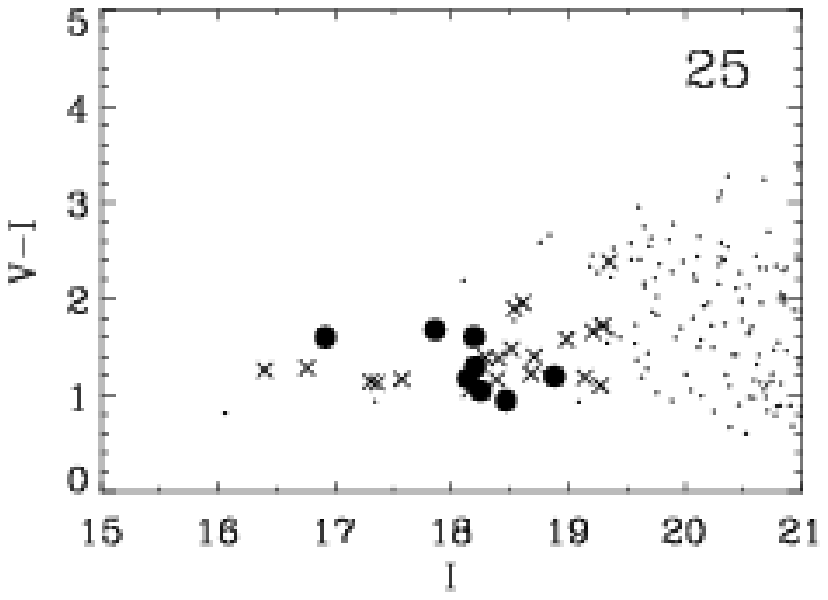}}
\resizebox{0.3\textwidth}{!}{\includegraphics{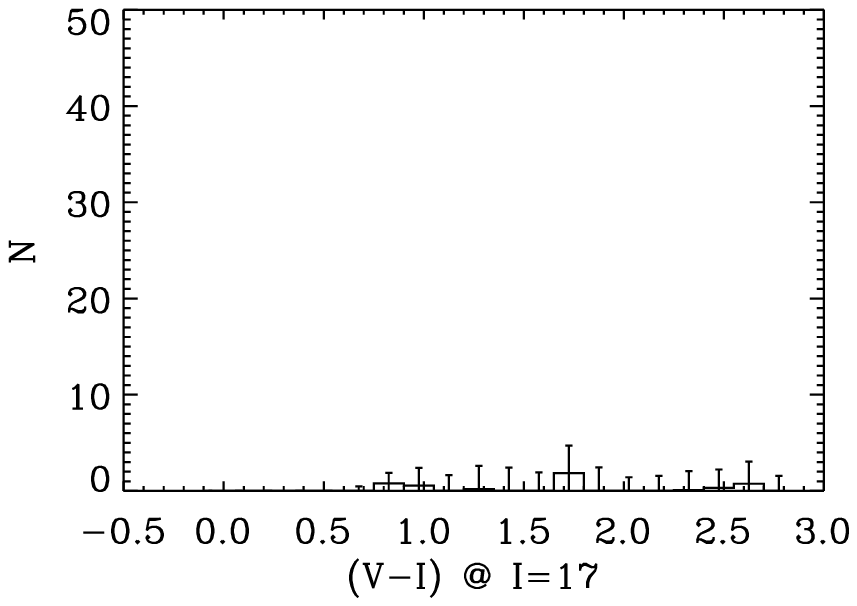}}
\resizebox{0.3\textwidth}{!}{\includegraphics{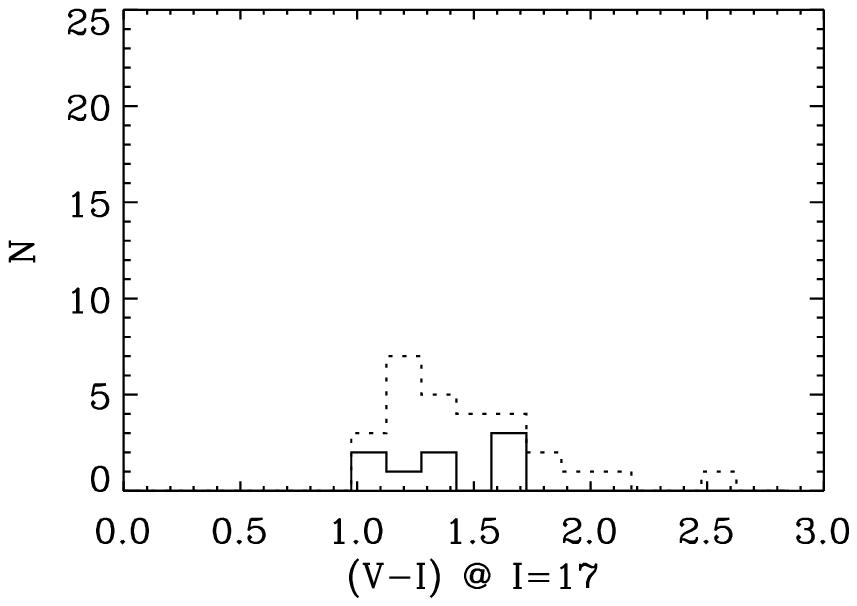}}
\resizebox{0.3\textwidth}{!}{\includegraphics{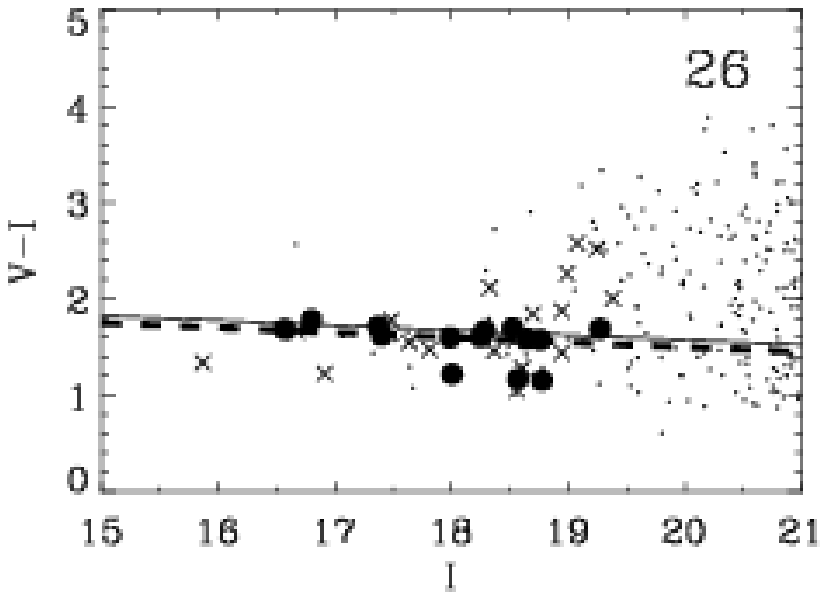}}
\resizebox{0.3\textwidth}{!}{\includegraphics{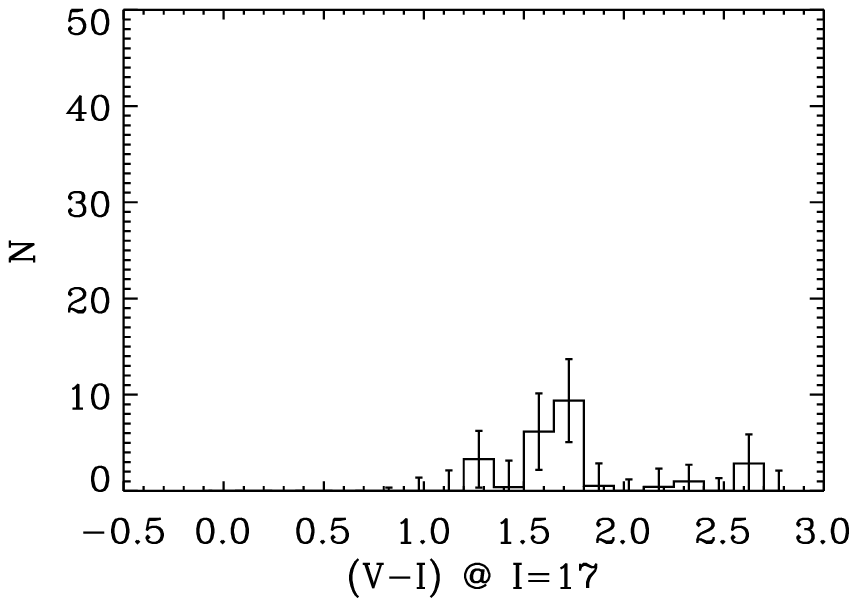}}
\resizebox{0.3\textwidth}{!}{\includegraphics{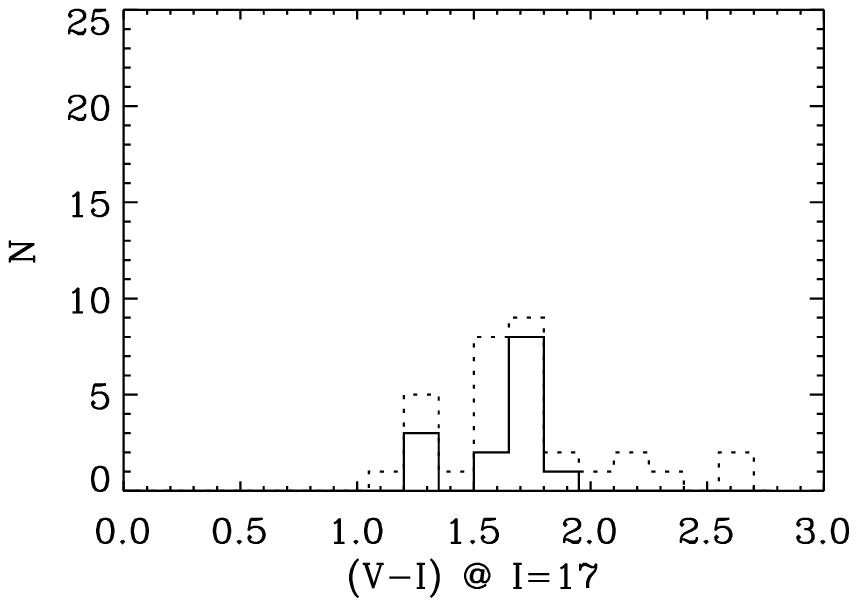}}
\resizebox{0.3\textwidth}{!}{\includegraphics{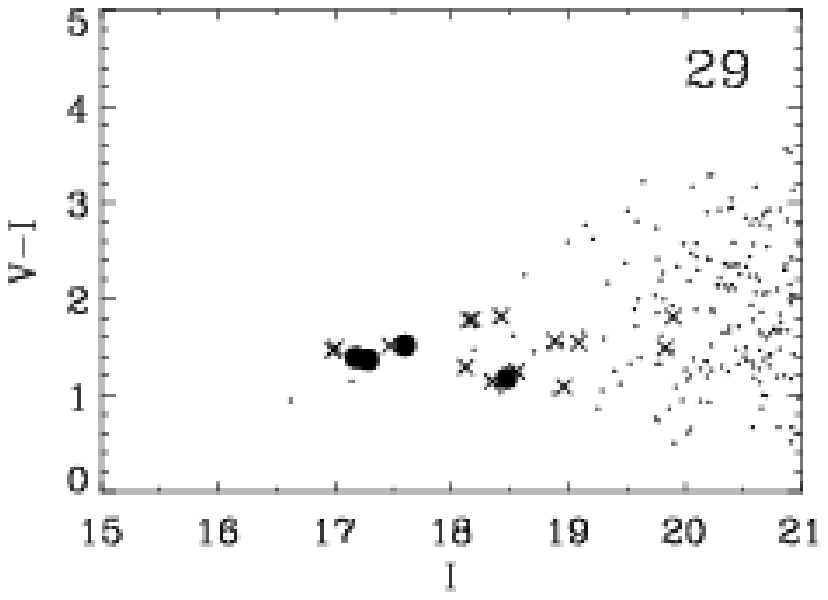}}
\resizebox{0.3\textwidth}{!}{\includegraphics{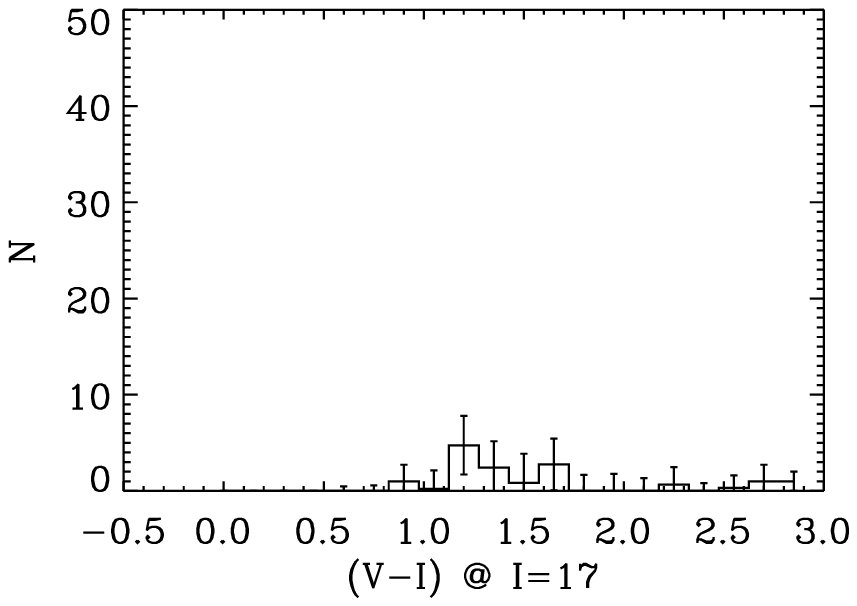}}
\resizebox{0.3\textwidth}{!}{\includegraphics{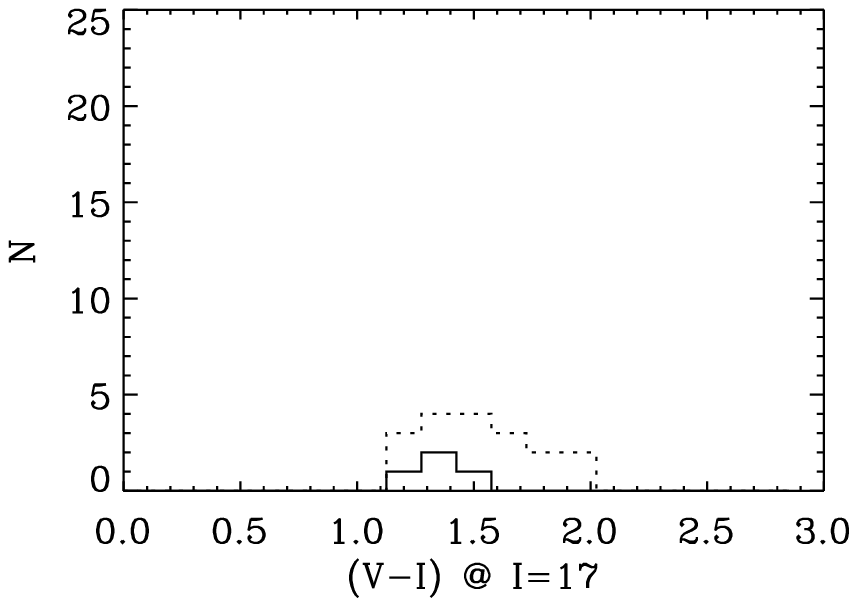}}
\resizebox{0.3\textwidth}{!}{\includegraphics{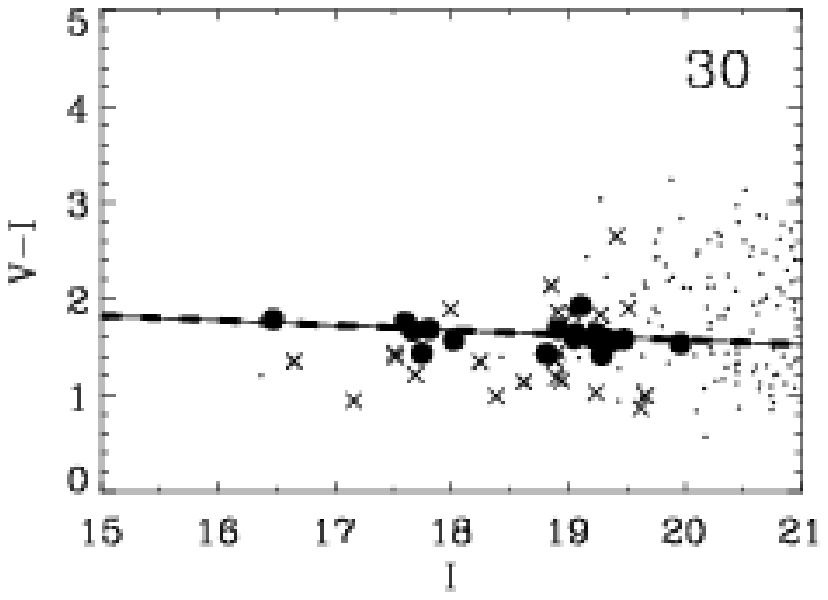}}
\resizebox{0.3\textwidth}{!}{\includegraphics{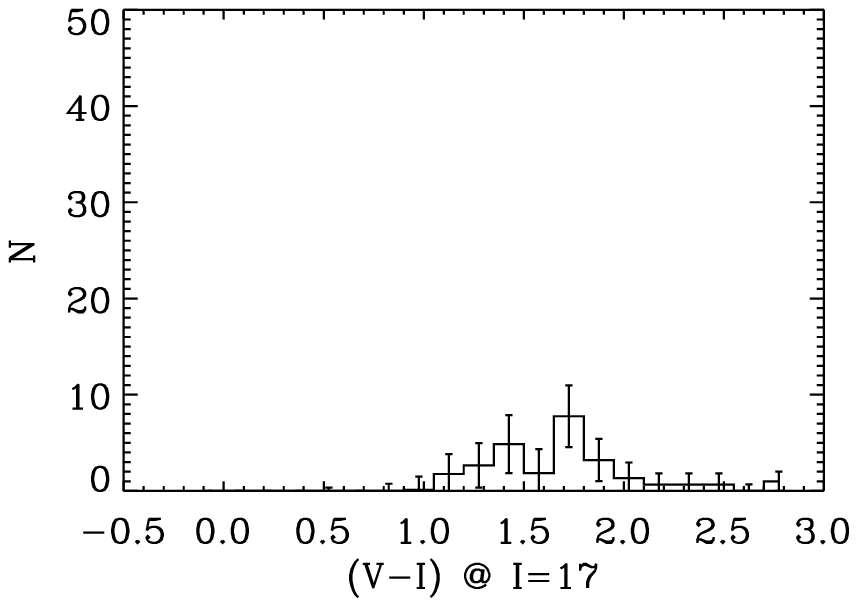}}
\resizebox{0.3\textwidth}{!}{\includegraphics{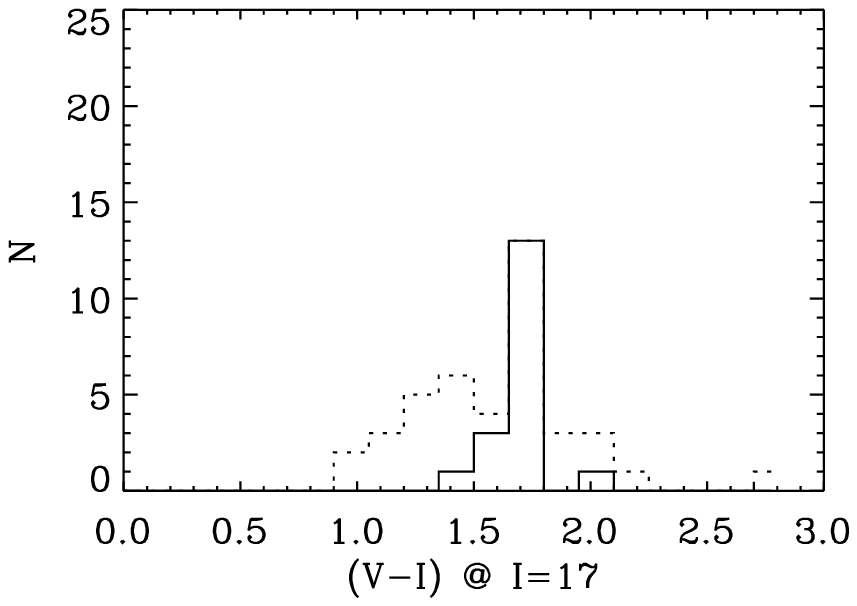}}
\resizebox{0.3\textwidth}{!}{\includegraphics{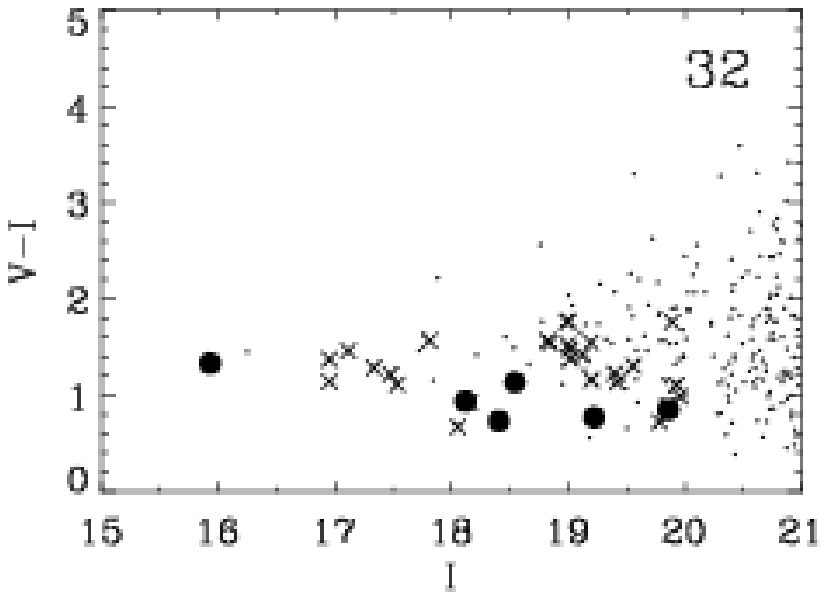}}
\resizebox{0.3\textwidth}{!}{\includegraphics{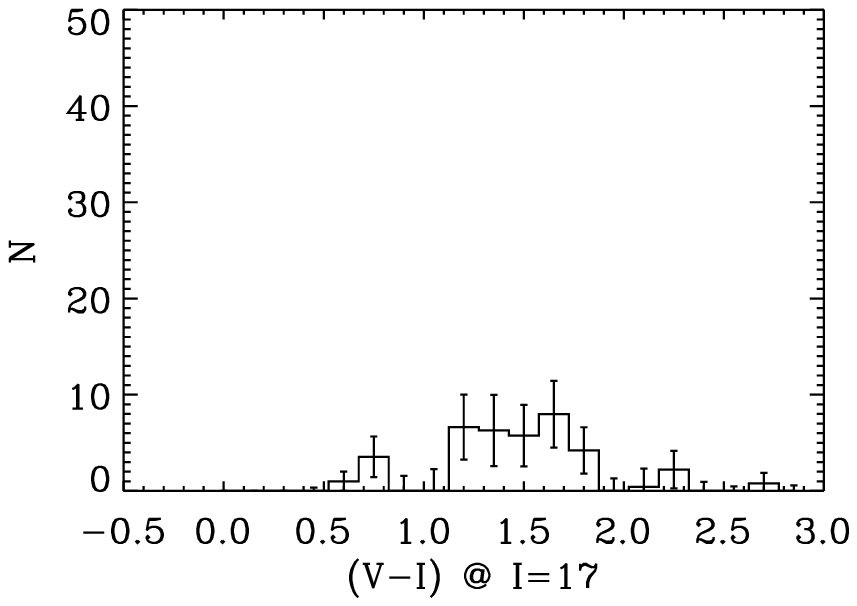}}
\resizebox{0.3\textwidth}{!}{\includegraphics{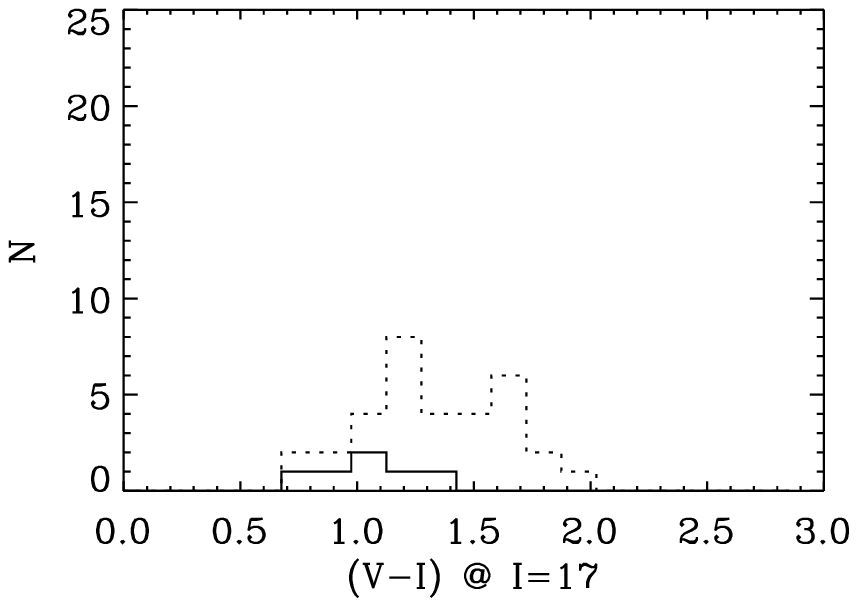}}
\end{center}
\caption{\it -- Continued}
\end{figure*}

Another important characteristic of a galaxy cluster is the colour
properties of its population. In particular, the colour-magnitude
diagram of cluster members normally reveals the presence of a narrow
sequence of bright, early-type galaxies known as the ``red sequence''
\citep[e.g. ][]{gladders98, stanford98, holden04, lopez-cruz04}. The
presence of this colour-magnitude (CM) relation serves as unambiguous
evidence for the presence of a real physical system. Furthermore, its
characteristics, such as colour, slope and scatter, have been
extensively used to constrain galaxy evolution models
\citep[e.g. ][]{gladders98}. While most of the previous studies
focused on relatively rich systems, the presence of red sequences in
very poor clusters and groups has also been reported
\citep[e.g.][]{andreon03}.

The properties of the colour-magnitude relation are found to be very
homogeneous, even though they evolve with redshift
\citep[e.g. ][]{aragon-salamanca93, stanford98}. Below we use this
fact and the measured slope for clusters at redshifts close to the
ones being considered here as the basis for an objective method
for detecting CM relations.

For 29 of the 32 systems analysed here, $V$-band images were also
available, thus allowing us to construct and investigate the CM
diagram of the ``cluster'' members.  The $(V-I)\times I$ CM diagrams
were constructed considering galaxies within a radius of
$0.75h_{75}^{-1}\mathrm{Mpc}$ and are shown in Fig.~\ref{fig:cmd}. For
each system: 1) the left panel gives the aforementioned
colour-magnitude diagram for galaxies (dots) brighter than $I=21$;
filled circles indicate spectroscopically confirmed members and crosses
other galaxies with measured redshifts. Also shown in this panel are
the best-estimated loci characterizing the so-called red sequence (see
below) as determined from the photometric data alone (dashed line) and
that obtained considering only the confirmed spectroscopic members
(solid line); 2) the middle panel shows the background-corrected
colour distribution of galaxies brighter than $I=19.5$ within the same
radius. The background correction is estimated from the colour
distribution of galaxies lying between radii
$2h_{75}^{-1}\mathrm{Mpc}$ and $3h_{75}^{-1}\mathrm{Mpc}$; and 3) the
right panel shows the colour distribution of spectroscopic members
(solid histogram) and that of all galaxies with measured redshifts
(dotted histogram).

From these diagrams it is clear that some of the clusters do have a
well-defined red sequence, while for others the red sequence is either
poorly defined or completely absent. Here we use the properties of the
CM relation mentioned above to develop an objective method for
determining its presence and properties.

As mentioned above, the red sequence consists of early-type cluster
galaxies.  Therefore, to identify the presence of a red
sequence and thus  distinguish between localized density
enhancements and physically bound systems, one would require both
morphological information and complete spectroscopic data. In the
present work we do not have morphological data nor complete
spectroscopic data and therefore we carry out a statistical analysis
to identify red sequences in the colour distribution of galaxies
in the cluster fields using the following method.  First, we build a
``tilted colour histogram'',  counting galaxies within slices of a
given width and characterized by a slope taken to be comparable to
that typically observed for the CM relations of nearby clusters.  We
then construct two such histograms with a relative shift of half a bin
width. This is done in order to avoid splitting a sequence between two
bins, thereby artificially decreasing its significance.  To obtain the
colour histogram only for the cluster galaxies, the histograms are
background corrected, as described below. We determine the colour of
the red sequence by separately analyzing both histograms.
First, we identify the highest peak considering only those that are
narrower than a certain value, taking into account the fact that the
scatter around the CM relation in nearby clusters is very small
\citep[$\leq0.1\mathrm{mag}$, e.g. ][]{lopez-cruz04}.   Next, based on
simulations, we compute the confidence level of the peaks and assign
the colour of the most significant peak to be that of the red
sequence.

We apply this method to two datasets. First, we consider all available
photometric data, which has the advantage of good statistics but is
susceptible to projection effects possibly leading to contamination by
non-cluster members, thereby diluting a possible red sequence. Then,
we consider the spectroscopic data.  While these are not affected by
projection effects, the statistics are usually poor due to both the low
richness of the systems being analysed and the incompleteness of the
spectroscopic data. There may also be false detection of a red
sequence due to incompleteness/selection effects.
 
In the analysis described below and illustrated in the
middle and right panels of Fig.~\ref{fig:cmd} we use the following
parameters. The bin width is chosen to be
$\Delta(V-I)=0.15\mathrm{mag}$. Because of the tilted nature of the
histograms we arbitrarily define the center of the bins at $I=17$ and
in the range from $(V-I)=-0.075$ and $(V-I)=2.85$.  We adopt a slope
of $-0.05$ as found by \cite{lopez-cruz04} for the $(B-R)\times R$ CM
relation at $z\sim0.05$. The $(B-R)$ at that redshift roughly
corresponds to $(V-I)$ at $z\sim0.2$. We restrict the red sequence
colours to be in the range $1.0\leq(V-I)\leq2.0$. This colour range
was chosen to avoid the tails of the colour distributions, where poor
statistics may artificially increase the peak
significances. Furthermore, it is well-matched with the colours
expected for elliptical galaxies at these redshifts
\citep[e.g. ][]{bruzual93}. We only consider peaks narrower than
$0.3\mathrm{mag}$.  This choice is rather conservative since the
estimated uncertainties of the $(V-I)$ colour at the faintest end
considered is $\sigma_{(V-I)}\sim0.04$ \citep{prandoni99}.

\subsubsection{Photometric data}

In this case the ``tilted colour histogram'' is constructed for
galaxies brighter then $I=19.5$ within the area inside a radius of
$0.75h_{75}^{-1}\mathrm{Mpc}$, hereafter referred to as the cluster
area. The background contribution is estimated from the ``tilted
colour histogram'' of galaxies in the same magnitude range but at a
distance ranging from $2h_{75}^{-1}\mathrm{Mpc}$ to
$3h_{75}^{-1}\mathrm{Mpc}$ and then scaled to the cluster area.  For
bins for which the expected  field contribution is larger than
the number of galaxies in the cluster area, we set the number of
cluster galaxies to zero. In order to compute the significance of the
identified peaks we use the signal-to-noise ratio
($S/N=N_{cl}/\sqrt{N_{tot}+N_{bkg}}$). Here $N_{cl}$ is the
number of cluster galaxies in the bin, $N_{tot}$ is the number of
galaxies in the bin in the original histogram before field correction
and $N_{bkg}$ is the expected number of background galaxies in that
bin for the cluster area.

\begin{figure}
\resizebox{\columnwidth}{!}{\includegraphics{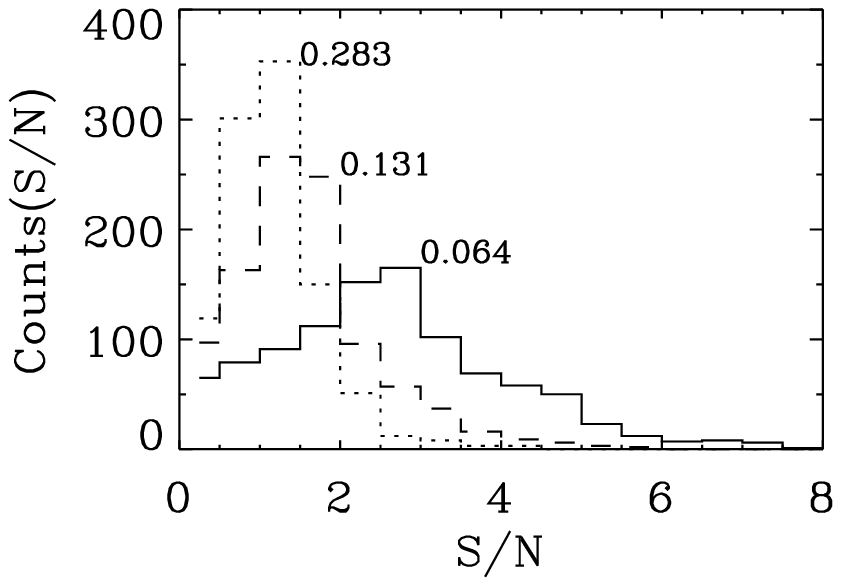}}
\caption{Distribution of red sequence S/N values for 1000 random
galaxy samples for three different redshifts.}
\label{fig:sn_significance}
\end{figure}

In order to quantify the significance ($\sigma_{S/N}$) of a given
$S/N$ we have carried out a series of simulations. For each measured
cluster redshift we have selected 1000 random positions within our
$(V-I)$ galaxy catalogue and constructed similar ``tilted colour
histograms'' for those positions. From these histograms we construct
the distribution of the measured $S/N$ and determine the probability
of finding a peak with a similar $S/N$ as the one found for the
cluster. Since we limit ourselves to a quite narrow colour range
($1.0\leq(V-I)\leq2.0$), the field contribution is roughly constant
and thus we do not separate the significances by colour. In
Fig.~\ref{fig:sn_significance} we show the distribution of the
measured $S/N$ for three different redshifts. It is clear that one
cannot rely on simply taking the same $S/N$ threshold for all
redshifts. This is mainly due to the larger sky area for the nearer
clusters.  In order to determine the significance for a given system
we use the simulations carried out using the same redshift and
determine how frequently a peak with similar or larger $S/N$ occur in
the field samples. The significance is computed as this frequency
subtracted from unity. We define the threshold for considering a red
sequence real to be $\sigma_{S/N} \geq 90\%$.  In the middle panels of
Fig~\ref{fig:cmd} we show that of the two histograms with the highest
peaks.

\subsubsection{Spectroscopic data}

The ``tilted colour histograms'' of the spectroscopic members do
not need any background subtraction. The significance of the peak is
assessed by constructing random galaxy samples. For each confirmed
system we select 1000 galaxy samples from the entire galaxy catalogue
with the same size as the confirmed system. The ``tilted histograms''
of these randomly selected galaxy samples are analyzed in the same way
as the data and we determine how frequently we encounter a peak with
at least the same number of galaxies as found in the peak of the data.
The definition of the significance, $\sigma_{spec}$, is analagous to
that above. In this case, we define the threshold for considering a
red sequence real to be $\sigma_{spec} \geq 99\%$.  The histograms
with the highest peaks are shown in the right panels of
Fig~\ref{fig:cmd}.

\subsubsection{Results}

From the application to the photometric data we find 10 systems
($\sim35\%$) that show signs of a red sequence, while we identify 17
systems ($\sim55\%$) when the spectroscopic sample is used.
Interestingly, all but  one of the systems for which a red
sequence was identified from the photometric data are confirmed by the
spectroscopic analysis. Moreover, the colours determined by the two
methods yield comparable results. The colours obtained for the red
sequences (see Fig.~\ref{fig:zvi} below) are consistent with the
passive evolution model by \cite{bruzual93}.  The only system which
was not confirmed by the spectroscopic analysis is EISJ0951-2145
(\#12) which is one of the cluster fields with the lowest completeness
(0.17) in terms of targeted galaxies.

On the other hand, we find 8 systems where a red sequence was found in
the spectroscopic sample but not from the photometric sample
alone. Curiously, in only one case (EISJ0955-2020, \#23)  have we
found, in Sect.~\ref{sec:z_identification},  that the system was
significantly affected by projection effects.

In summary, a total of 18 systems show evidence of having a red
sequence. We find 10 systems from the photometric data alone and 17 from
the spectroscopic analysis with 9 systems being in common.

\begin{figure}
\resizebox{\columnwidth}{!}{\includegraphics{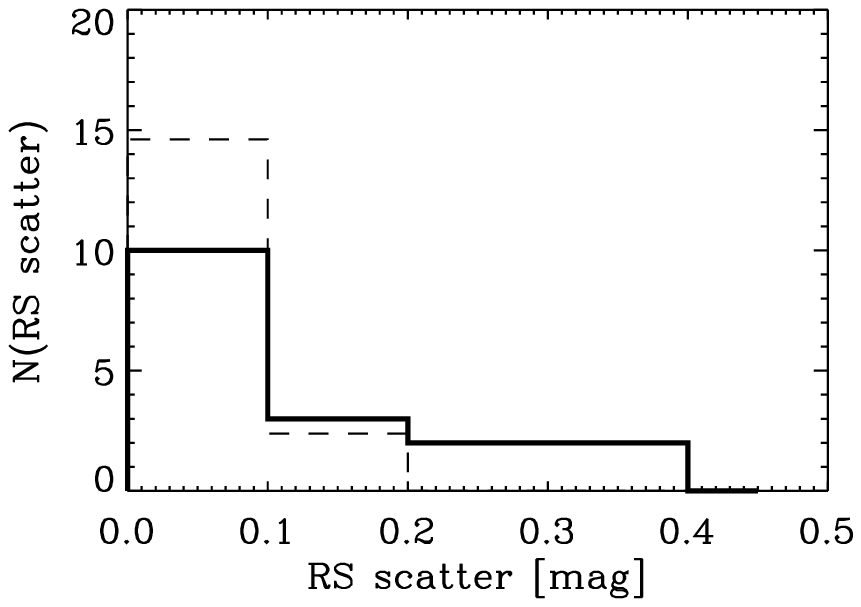}}
\caption{Distribution of the measured scatter around the red sequences
(solid line) compared to that of \cite{lopez-cruz04} (dashed line).}
\label{fig:scatter_dist}
\end{figure}

For the systems with red sequences detected from the spectroscopic
analysis we computed the scatter of the galaxy colours using an
iterative sigma-clipping method. The scatter is given in Col.~12 of
Table~\ref{tab:colour} and shown in Fig.~\ref{fig:scatter_dist} (solid
line). It ranges from 0.031 to $0.35\;\mathrm{mag}$. For comparison,
we also show the distribution of scatter measured in (B-R) by
\cite{lopez-cruz04} scaled to the same total number of clusters. This
will be discussed in more  detail below.

\subsection{Comparison with other authors}


In Fig.~\ref{fig:veldisp_dist} we compare the normalized distributions
of velocity dispersions of the EIS systems (solid line) with those of
the cluster samples considered by \cite{fadda96} (dotted line) and
\cite{zabludoff90} (dashed line), both drawn from the Abell cluster
catalogue.  Both of these samples cover roughly the same redshift
range $z\lesssim0.15$. The Fadda et al. sample includes 172 clusters
with richness as poor as $R=-1$, while that of Zabludoff et
al. consists of 65 clusters most of which have richness $R\geq1$. From
the figure we find that the EIS systems span about the same range of
velocity dispersion as the other samples, but apparently with a larger
fraction of low velocity dispersion systems. Applying the
Kolmogorov-Smirnoff test, we find that this difference is significant
with only a small probability that our sample is drawn from the same
parent population as the others.

\begin{figure}
\resizebox{\columnwidth}{!}{\includegraphics{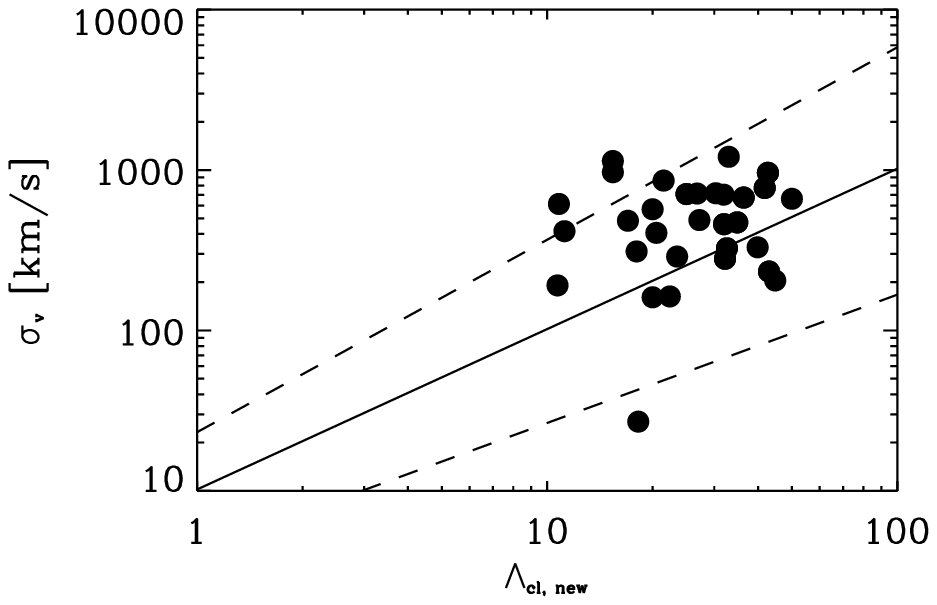}}
\caption{The relation between $\Lambda_{cl, new}$ and $\sigma_v$ for
the 31 confirmed systems with a measured velocity dispersion. The solid
line marks the scaling relation derived by \cite{bahcall03} with the
dashed lines marking its uncertainty.}
\label{fig:scaling_rel}
\end{figure}

The distribution of richness is shown in Fig.~\ref{fig:lambda_dist}.
Unfortunately no similar samples are available for
comparison. Instead, we compare, in Fig.~\ref{fig:scaling_rel}, the
relation between velocity dispersion and richness with that determined
by \cite{bahcall03}.  In the figure the solid line shows the relation
obtained by Bahcall et al., with the dashed lines indicating the
estimated uncertainty interval.  This relation was derived using 20
well-sampled clusters drawn from the Sloan Digital Sky Survey data by
a matched filter algorithm.  From the figure we find that, except for
a few cases at the low richness end, the EIS systems populate the same
region of the plot.  However, considering all systems, we find only a
weak correlation between these global parameters, perhaps indicating
that effects of projection and outliers contaminate the measurements
of the richness and velociy dispersion.  We point out  that, in their
analysis, \cite{bahcall03} only considered richer systems
($\Lambda\geq30$).

In Fig.~\ref{fig:concentration_bo} we show the distribution of
concentration indices for our clusters and for those of
\cite{butcher84}. It can be seen that the EIS sample covers a larger
range of concentration indices than that of Butcher and Oemler.  We
find a mean concentration of $\langle C\rangle_{EIS}= 0.56$ with a
standard deviation of 0.21, larger than the one found by Butcher \&
Oemler ($\langle C\rangle_{BO}=0.47$ with a standard deviation of
0.10), indicating that our sample includes more concentrated systems.

\begin{figure}
\resizebox{\columnwidth}{!}{\includegraphics{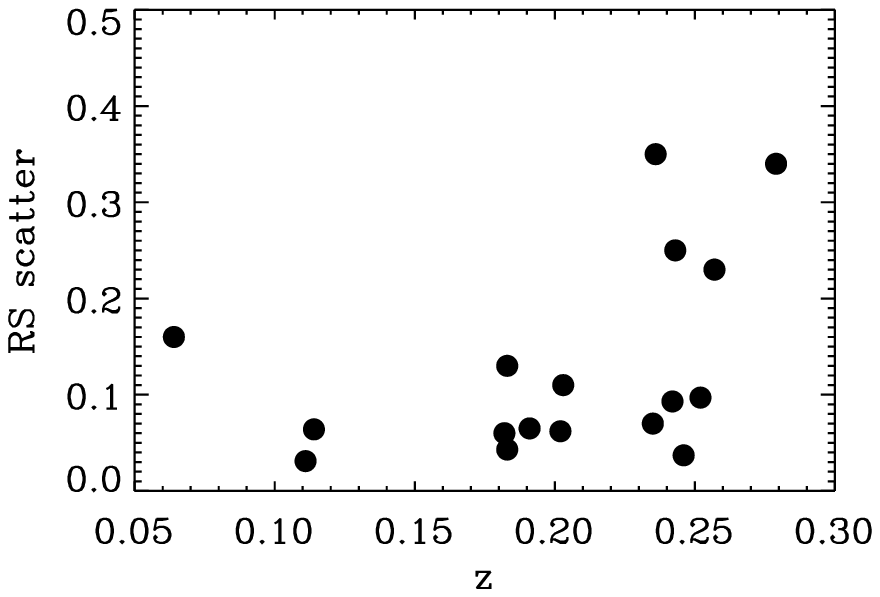}}
\caption{ The colour dispersion as function of redshift for all systems
with a red sequence identified in the spectroscopic analysis.}
\label{fig:scatter}
\end{figure}

\begin{figure*}
\resizebox{0.3\textwidth}{!}{\includegraphics{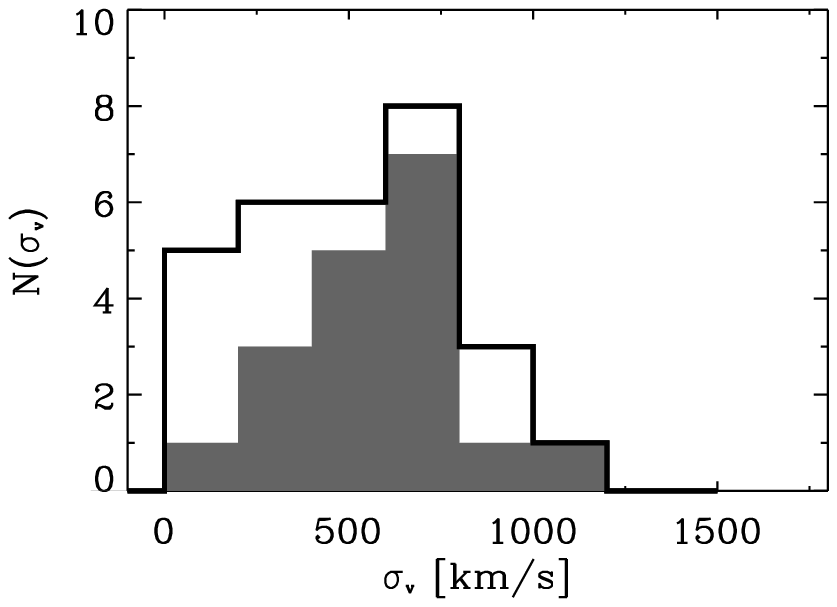}}
\resizebox{0.3\textwidth}{!}{\includegraphics{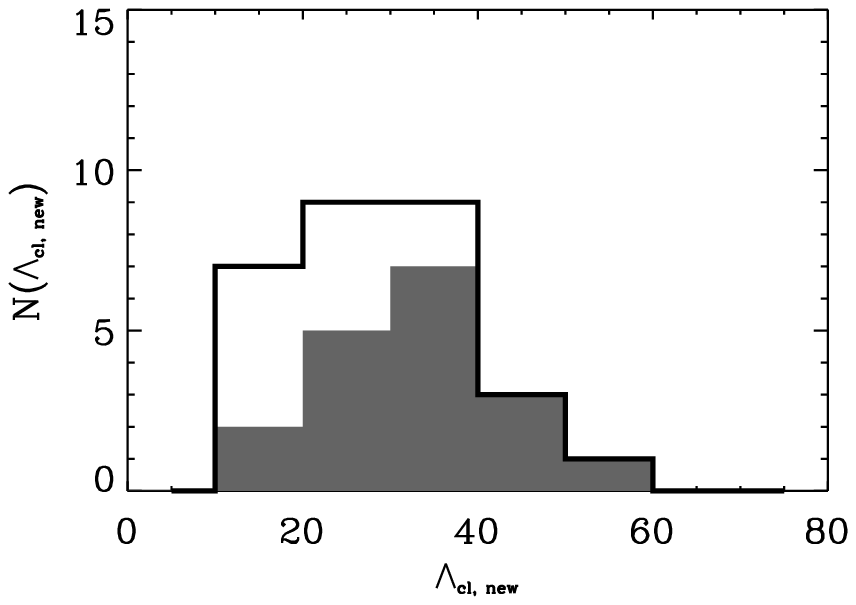}}
\resizebox{0.3\textwidth}{!}{\includegraphics{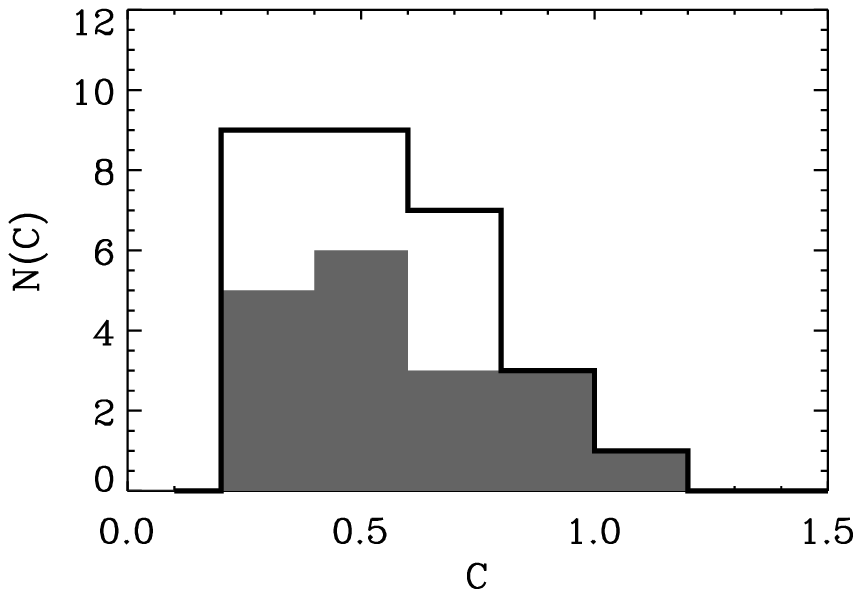}}
\caption{Global properties of the red-sequence systems (gray region)
compared with those of the entire sample (solid line). The left panel
shows the distribution of velocity dispersion, the middle panel that
of richness and the right panel that of concentration index.}
\label{fig:comp}
\end{figure*}

In Fig.~\ref{fig:scatter_dist} the distribution of the scatter about
the CM relation for the 17 systems having a red sequence detected
using the spectroscopic sample is shown and compared to that of
\cite{lopez-cruz04}.  Considering all our systems we find a mean
scatter of $0.13\pm0.10$, where the error is the standard deviation.
Discarding the four systems with measured scatters larger than
$0.2\mathrm{mag}$ we obtain a mean scatter of $0.079\;\mathrm{mag}$,
in excellent agreement with the results of \cite{lopez-cruz04}, who
found a scatter of $0.074\pm0.026\;\mathrm{mag}$.  With the exception
of one of the four outlying systems all have highly complete
spectroscopic data.  In Fig.~\ref{fig:scatter} we show the measured
scatter as function of redshift. It can be seen that the four outlying
systems are all found at redshifts $z\geq0.23$.  All the systems but
the one with incomplete spectroscopic data have concentration indices
$C>0.5$, which corresponds to concentrated systems using the
definition of \cite{butcher84}.  A possible interpretation for this
larger scatter for these higher redshift systems is that they have
bright blue cluster members, reminiscent of the Butcher-Oemler effect
\citep{butcher84}. The onset of this is thought to happen at redshift
$z\sim0.1$ with the fraction of blue galaxies increasing from
$f_B\sim0.03$ to $f_B\sim0.25$ at $z=0.5$ in compact, concentrated
clusters. Without morphological information, the origin of this
scatter cannot be determined.

\section{Discussion}
\label{sec:discussion}

To better understand the relation between the existence of a red
sequence and the global properties of the galaxy cluster/~group, here
we focus on the 17 clusters with red sequences and compare their
properties to those found for the entire sample in
Fig.~\ref{fig:comp}. The left panel gives the distribution of velocity
dispersion, the middle panel that of richness and the right panel
that of concentration index. In all panels the distribution for the
entire sample is shown (solid line), while that for the red-sequence
systems are represented by the gray histogram. From the figure, we
conclude that the red sequences are found in rich, high-velocity
dispersion systems, but their existence is independent of the
concentration. 

\begin{figure}
\resizebox{\columnwidth}{!}{\includegraphics{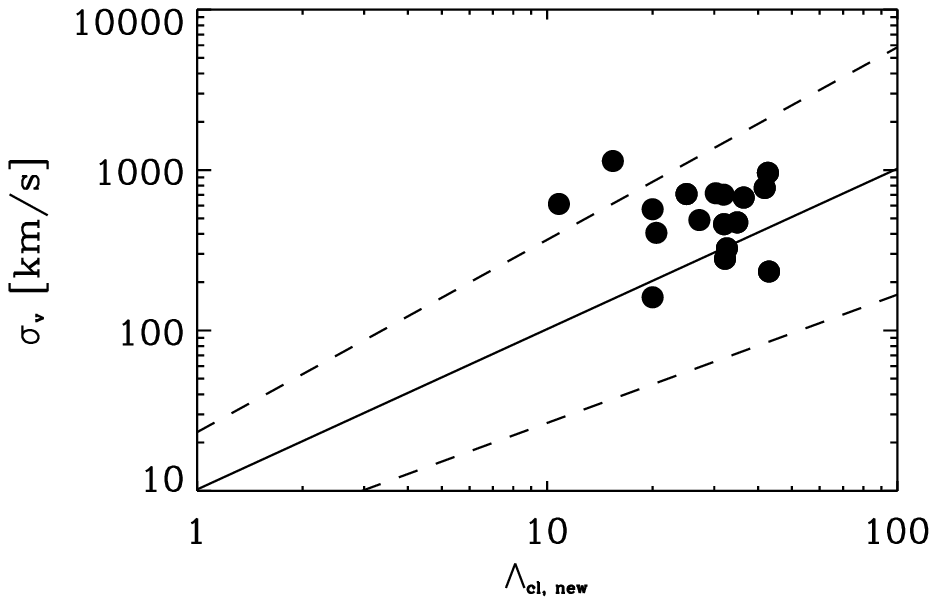}}
\caption{The relation between $\Lambda_{cl, new}$ and $\sigma_v$ for
the 17 systems for which we identify red sequences from the
spectrocopic analysis. The solid line marks the scaling relation
derived by \cite{bahcall03} with the  dashed lines marking its
uncertainty.}
\label{fig:scaling_seq}
\end{figure}

In Fig.~\ref{fig:scaling_seq} we show the richness-velocity dispersion
relation of \cite{bahcall03}, discussed in the previous section,
considering only the red-sequence systems.  Except for two low-richness
systems, all of them fall within the region indicated by the Bahcall
et al. relation. The correlation found for the 15 systems with
$\Lambda_{cl, new} \geq 20$ is 0.29, with a fitted relation of
$\sigma_v=73\Lambda_{cl, new}^{0.55}$.

\begin{figure}
\resizebox{\columnwidth}{!}{\includegraphics{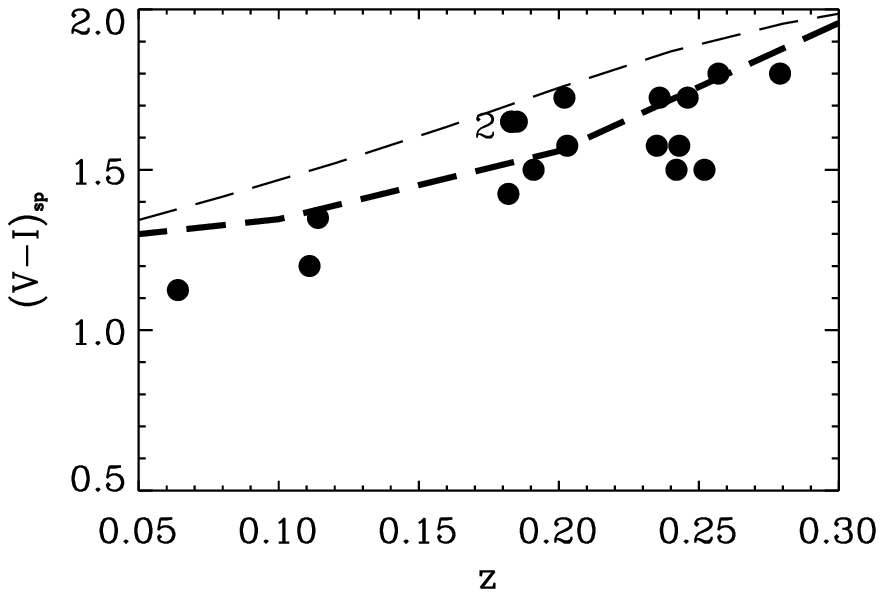}}
\caption{ The relation between the redshift of the systems and the
colour of the red sequence.  The number 2 indicates that there are two
symbols falling almost on top of each other. The lines mark the
no-evolution (thin line) and  the passive evolution (thick line)
predictions.}
\label{fig:zvi}
\end{figure}

Finally, in Fig.~\ref{fig:zvi} we compare our measured red sequence
colours as a function of redshift with those predicted by two
different models. The two models represent passively evolving (thick
line) and non-evolving (thin line) elliptical galaxies.  The passive
evolution model is based on the model by \cite{bruzual93} with a
formation redshift $z_f=10$ with a short starburst followed by passive
evolution. The colours for the non-evolving elliptical galaxy are
computed from the composite elliptical spectrum from \cite{kinney96}.
The data points seem to be consistent with the passive
evolution model, as has also been found by previous work
\citep[e.g.][]{aragon-salamanca93, gladders98, stanford98, olsen01}.
The scatter around the passive evolution model is computed to be
$0.14$, comparable to the bin width utilized to determine the colour
of the red sequence.

\section{Summary}
\label{sec:conclusions}

In this paper we report new redshifts for 738 galaxies in the field of
21 low-redshift ($z_{MF}=0.2$) candidate clusters drawn from the EIS
candidate cluster sample.  These data were used to search for
overdensities in redshift space, thus confirming the presence of a
bound group/cluster of galaxies in the direction of candidates
identified by the matched-filter analysis. The photometric and
spectroscopic data available for the 20 new confirmed clusters out of
21 candidate systems, as well as those listed in the previous papers
of this series \citep{hansen02,olsen03}, were used to compute the
properties of these systems and member galaxies. Our main results can
be summarized as follows:

\begin{enumerate}

\item For 32 (94\%) of 34 systems considered we identify significant
density enhancements in redshift space. We find that the measured
redshifts are in the range $0.06 \leq z \leq 0.28$, with a mean value of
$z=0.18$. This is in excellent agreement with the redshift estimated
by the matched filter technique.

\item The systems have a broad range of properties. The velocity
  dispersions range from small groups ($130\mathrm{km/s}$) to clusters
  ($1200\mathrm{km/s}$), the richness tends to be low and the
  concentration varies from uniform, typical of spiral-dominated
  systems to highly concentrated, typical of early-type dominated
  systems. The fact that the systems are predominantly poor is not
  surprising given the relatively small volume of space probed by the
  survey.

\item We estimate that 13 out of the 32 systems may suffer from
projection effects either due to other superposed galaxy systems along
the line-of-sight or to field galaxies, which may impact the
calculation of the system's global properties and explain the weak
correlation observed between velocity dispersion and richness.

\item We find that 17 (60\%) out of 29 systems with both photometric
  and spectroscopic data available show evidence of a red sequence in
  the colour-magnitude diagram, with colours consistent with those
  predicted from passively evolving stellar populations. Only one
  system with a detected red sequence using photometric data was not
  confirmed when only spectroscopic members were considered.

\item The systems where red sequences have been detected tend to have
higher velocity dispersions and richnesses than those
without. However, these systems seem to be independent of the
 concentration index.

\end{enumerate}

  The presence of a red sequence is consistent with the interpretation
  that we have detected bound systems containing elliptical galaxies
  that obey a scaling mass-metallicity relation  which gives rise
  to the observed CM-relation.  Therefore, these results taken
  together with the detection of density enhancements in redshift
  space provide further evidence that the systems detected by the
  matched-filter technique at z=0.2 are nearly all real. Extending the
  sample of measured redshifts would greatly help in further
  characterizing these systems, typically at the low end of the Abell
  richness.


\begin{acknowledgements}
We would like to thank the anonymous referee for thorough comments
which have greatly improved the manuscript.  We thank John Pritchard,
Lisa Germany and Ivo Saviane for making the pre-imaging observations
of the fields. We would also like to thank the 2p2 team, La Silla, for
their support at all times during the observations.  We are also
indebted to Morten Liborius Jensen for preparing the slit masks.  This
work has been supported by The Danish Board for Astronomical Research.
LFO thanks the Carlsberg Foundation for financial support.
\end{acknowledgements}

\bibliographystyle{aa}
\bibliography{/home/lisbeth/tex/lisbeth_ref}

\begin{thebibliography}{38}
\expandafter\ifx\csname natexlab\endcsname\relax\def\natexlab#1{#1}\fi

\bibitem[{Andreon(2003)}]{andreon03}
Andreon, S. 2003, A\&A, 409, 37

\bibitem[{Arag\'{o}n-Salamanca {et~al.}(1993)Arag\'{o}n-Salamanca, Ellis,
  Couch, \& Carter}]{aragon-salamanca93}
Arag\'{o}n-Salamanca, A., Ellis, R., Couch, W., \& Carter, D. 1993, MNRAS, 262,
  764

\bibitem[{Bahcall {et~al.}(2003)Bahcall, McKay, Annis, Kim, Dong, Hansen, Goto,
  Gunn, Miller, Nichol, Postman, Schneider, Schroeder, Voges, Brinkmann, \&
  Fukugita}]{bahcall03}
Bahcall, N., McKay, T., Annis, J., {et~al.} 2003, ApJS, 148, 243

\bibitem[{Beers {et~al.}(1990)Beers, Flynn, \& Gebhardt}]{beers90}
Beers, T., Flynn, K., \& Gebhardt, K. 1990, AJ, 100, 32

\bibitem[{Benoist {et~al.}(2002)Benoist, da~Costa, J{\o}rgensen, Olsen,
  Bardelli, Zucca, Scodeggio, Neumann, Arnaud, Arnouts, Biviano, \&
  Ramella}]{benoist02}
Benoist, C., da~Costa, L., J{\o}rgensen, H., {et~al.} 2002, A\&A, 394, 1

\bibitem[{Benoist {et~al.}(1999)Benoist, da~Costa, Olsen, Deul, Erben,
  Guarnieri, Hook, Nonino, Prandoni, Scodeggio, Slijkhuis, Wicenec, \&
  Zaggia}]{benoist99}
Benoist, C., da~Costa, L., Olsen, L.~F., {et~al.} 1999, A\&A, 346, 58

\bibitem[{Bruzual \& Charlot(1993)}]{bruzual93}
Bruzual, G. \& Charlot, S. 1993, ApJ, 405, 538

\bibitem[{Butcher \& Oemler(1984)}]{butcher84}
Butcher, H. \& Oemler, A. 1984, ApJ, 285, 426

\bibitem[{Fadda {et~al.}(1996)Fadda, Girardi, Giurcin, Mardirossian, \&
  Mezzerri}]{fadda96}
Fadda, D., Girardi, M., Giurcin, G., Mardirossian, F., \& Mezzerri, M. 1996,
  ApJ, 473, 670

\bibitem[{Gladders {et~al.}(1998)Gladders, L\'{o}pez-Cruz, Yee, \&
  Kodama}]{gladders98}
Gladders, M., L\'{o}pez-Cruz, O., Yee, H., \& Kodama, T. 1998, ApJ, 501, 571

\bibitem[{Gladders \& Yee(2001)}]{gladders01}
Gladders, M. \& Yee, H. 2001, in ASP Conf. Series, Vol. 232, The New Era of
  Wide Field Astronomy (Astronomical Society of the Pacific), 126

\bibitem[{Gonzalez {et~al.}(2001)Gonzalez, Zaritsky, Dalcanton, \&
  Nelson}]{gonzalez01}
Gonzalez, A., Zaritsky, D., Dalcanton, J., \& Nelson, A. 2001, ApJS, 137, 117

\bibitem[{Goto {et~al.}(2002)Goto, Sekiguchi, Nichol, Bahcall, Kim, Annis,
  Ivezic, Brinkmann, Hennessy, Szokoly, \& Tucker}]{goto02}
Goto, T., Sekiguchi, M., Nichol, R., {et~al.} 2002, AJ, 123, 1807

\bibitem[{Gunn {et~al.}(1986)Gunn, Hoessel, \& Oke}]{gunn86}
Gunn, J., Hoessel, J., \& Oke, J. 1986, ApJ, 306, 30

\bibitem[{Hansen {et~al.}(2002)Hansen, Olsen, \& J{\o}rgensen}]{hansen02}
Hansen, L., Olsen, L., \& J{\o}rgensen, H. 2002, A\&A, 388, 1

\bibitem[{Holden {et~al.}(2000)Holden, Adami, Nichol, Castander, Lubin, Romer,
  Mazure, Postman, \& Ulmer}]{holden00}
Holden, B., Adami, C., Nichol, R., {et~al.} 2000, AJ, 120, 23

\bibitem[{Holden {et~al.}(1999)Holden, Nichol, Romer, Metevier, Postman, Ulmer,
  \& Lubin}]{holden99}
Holden, B., Nichol, R., Romer, A., {et~al.} 1999, AJ, 118, 2002

\bibitem[{Holden {et~al.}(2004)Holden, Stanford, Eisenhardt, \&
  Dickinson}]{holden04}
Holden, B., Stanford, S., Eisenhardt, P., \& Dickinson, M. 2004, AJ, 127, 2484

\bibitem[{Katgert {et~al.}(1996)Katgert, Mazure, Perea, {den Hertog}, Moles,
  {Le F\`{e}vre}, Dubath, Focardi, Rhee, Jones, Escalera, Biviano, Gerbal, \&
  Giuricin}]{katgert96}
Katgert, P., Mazure, A., Perea, J., {et~al.} 1996, A\&A, 310, 8

\bibitem[{Kim {et~al.}(2002)Kim, Kepner, Postman, Strauss, Bahcall, Gunn,
  Lupton, Annis, Nochol, Castander, Brinkmann, Brunner, Connolly, Csabai,
  Hindsley, Ivezic, Vogeley, \& York}]{kim02}
Kim, R., Kepner, J., Postman, M., {et~al.} 2002, AJ, 123, 20

\bibitem[{Kinney {et~al.}(1996)Kinney, Calzetta, Bohlin, McQuade,
  Storchi-Bergmann, \& Schmitt}]{kinney96}
Kinney, A., Calzetta, D., Bohlin, R., {et~al.} 1996, ApJ, 467, 38

\bibitem[{Lopes {et~al.}(2004)Lopes, de~Carvalho, Gal, Djorgovski, Odewahn,
  Mahabal, \& Brunner}]{lopes04}
Lopes, P., de~Carvalho, R., Gal, R., {et~al.} 2004, AJ, 128, 1017

\bibitem[{L\'opez-Cruz {et~al.}(2004)L\'opez-Cruz, Barkhouse, \&
  Yee}]{lopez-cruz04}
L\'opez-Cruz, O., Barkhouse, W., \& Yee, H. 2004, ApJ, in press,
  astro-ph/0407630

\bibitem[{Nonino {et~al.}(1999)Nonino, Bertin, da~Costa, Deul, Erben, Olsen,
  Prandoni, Scodeggio, Wicenec, Wichmann, Benoist, Freudling, Guarnieri, Hook,
  Hook, M\'{e}ndez, Savaglio, Silva, \& Slijkhuis}]{nonino99}
Nonino, M., Bertin, E., da~Costa, L., {et~al.} 1999, A\&AS, 137, 51

\bibitem[{Olsen {et~al.}(2001)Olsen, Benoist, da~Costa, Scodeggio,
  J{\o}rgensen, Arnouts, Bardelli, Biviano, Ramella, \& Zucca}]{olsen01}
Olsen, L., Benoist, C., da~Costa, L., {et~al.} 2001, A\&A, 380, 460

\bibitem[{Olsen {et~al.}(2003)Olsen, Hansen, J{\o}rgensen, Benoist, da~Costa,
  \& Scodeggio}]{olsen03}
Olsen, L., Hansen, L., J{\o}rgensen, H., {et~al.} 2003, A\&A, 409, 439

\bibitem[{Olsen(2000)}]{olsen00}
Olsen, L.~F. 2000, PhD thesis, Copenhagen University Observatory

\bibitem[{Olsen {et~al.}(1999{\natexlab{a}})Olsen, Scodeggio, da~Costa,
  Benoist, Bertin, Deul, Erben, Guarnieri, Hook, Nonino, Prandoni, Slijkhuis,
  Wicenec, \& Wichmann}]{olsen99a}
Olsen, L.~F., Scodeggio, M., da~Costa, L., {et~al.} 1999{\natexlab{a}}, A\&A,
  345, 681

\bibitem[{Olsen {et~al.}(1999{\natexlab{b}})Olsen, Scodeggio, da~Costa,
  Benoist, Bertin, Deul, Erben, Guarnieri, Hook, Nonino, Prandoni, Slijkhuis,
  Wicenec, \& Wichmann}]{olsen99b}
Olsen, L.~F., Scodeggio, M., da~Costa, L., {et~al.} 1999{\natexlab{b}}, A\&A,
  345, 363

\bibitem[{Postman {et~al.}(2002)Postman, Lauer, Oegerle, \&
  Donahue}]{postman02}
Postman, M., Lauer, T., Oegerle, W., \& Donahue, M. 2002, ApJ, 579, 93

\bibitem[{Postman {et~al.}(1996)Postman, Lubin, Gunn, Oke, Hoessel, Schneider,
  \& Christensen}]{postman96}
Postman, M., Lubin, L., Gunn, J., {et~al.} 1996, AJ, 111, 615

\bibitem[{Prandoni {et~al.}(1999)Prandoni, Wichmann, da~Costa, Benoist,
  M\'{e}ndez, Nonino, Olsen, Wicenec, Zaggia, Bertin, Deul, Erben, Guarnieri,
  Hook, Hook, Scodeggio, \& Slijkhuis}]{prandoni99}
Prandoni, I., Wichmann, R., da~Costa, L., {et~al.} 1999, A\&A, 345, 448

\bibitem[{Ramella {et~al.}(2000)Ramella, Biviano, Boschin, Bardelli, Scodeggio,
  Borgani, Benoist, da~Costa, Girardi, Nonino, \& Olsen}]{ramella00}
Ramella, M., Biviano, A., Boschin, W., {et~al.} 2000, A\&A, 360, 861

\bibitem[{Rizzo {et~al.}(2004)Rizzo, Adami, Bardelli, Cappi, Zucca, Guideroni,
  Chincarini, \& Mazure}]{rizzo04}
Rizzo, D., Adami, C., Bardelli, S., {et~al.} 2004, A\&A, 413, 453

\bibitem[{Scodeggio {et~al.}(1999)Scodeggio, Olsen, da~Costa, Slijkhuis,
  Benoist, Deul, Erben, Hook, Nonino, Wicenec, \& Zaggia}]{scodeggio99}
Scodeggio, M., Olsen, L.~F., da~Costa, L., {et~al.} 1999, A\&AS, 137, 83

\bibitem[{Stanford {et~al.}(1998)Stanford, Eisenhardt, \&
  Dickinson}]{stanford98}
Stanford, S., Eisenhardt, P., \& Dickinson, M. 1998, ApJ, 492, 461

\bibitem[{Yee {et~al.}(2000)Yee, Morris, Lin, Carlberg, Hall, Sawicki, Patton,
  Wirth, Ellingson, \& Shepherd}]{yee00}
Yee, H., Morris, S., Lin, H., {et~al.} 2000, ApJS, 129, 475

\bibitem[{Zabludoff {et~al.}(1990)Zabludoff, Huchra, \& Geller}]{zabludoff90}
Zabludoff, A., Huchra, J., \& Geller, M. 1990, ApJS, 74, 1

\end{thebibliography}





\end{document}